\documentclass[11pt]{article}
\pdfoutput = 1
\usepackage[utf8]{inputenc}
\usepackage{color,graphicx}
\usepackage{epsfig}
\usepackage{amsmath}
\usepackage{verbatim}
\usepackage{amssymb}
\usepackage{mathabx}
\usepackage{stmaryrd}
\usepackage{empheq}
\usepackage{esint}
\usepackage{physics} 
\usepackage{amsfonts}
\usepackage{cite}
\usepackage{array}
\usepackage{setspace}
\usepackage{braket}
\usepackage{float}
\usepackage{url}
\usepackage{mathtools}
\usepackage{tensor}
\usepackage{bbold}
\usepackage{slashed}
\usepackage{tikz}
\usepackage{tikz,pgfplots}
\usepackage{epstopdf}
\usepackage{subfigure}
\usepackage[labelfont=bf,font={sf}]{caption}
\usepackage{pgfplots}
\usepackage{mathrsfs}
\usepackage{fancybox}
\usepackage{eurosym}
\usepackage{tcolorbox}
\usepackage{tensor}
\allowdisplaybreaks
\usepackage{changepage}

\usepackage[mathcal]{eucal}

\usepackage[margin = 2.2cm]{geometry}
\usepackage[ragged]{footmisc}
\usepackage[bookmarks=true,bookmarksnumbered=false,hyperindex=true,bookmarksopen=true,hyperfigures=true,colorlinks=true,linkcolor=webblue,citecolor=webgreen,urlcolor=webblue,breaklinks]{hyperref}
\definecolor{webgreen}{rgb}{0, 0.5, 0}
\definecolor{webblue}{rgb}{0, 0, 0.5}
\definecolor{webred}{rgb}{0.5, 0, 0}
\definecolor{darkgreen}{rgb}{0,0.5,0}

\def\darkblue{blue!85!black}
\def\darkred{red!60!black}
\def\darkgreen{green!50!black}

\def\ben{\begin{equation}}
\def\een{\end{equation}}

   \let\d=\delta 
   
   \let\x=\xi  \let\r=v

\let\X=\Xi  \let\U=\Upsilon

\def\be{\begin{equation}}
\def\ee{\end{equation}}
\def\ba{\begin{array}}
\def\ea{\end{array}}

\def\dalemb#1#2{{\vbox{\hrule height .#2pt
       \hbox{\vrule width.#2pt height#1pt \kern#1pt
               \vrule width.#2pt}
       \hrule height.#2pt}}}

\newcommand{\bea}{\begin{eqnarray}}
\newcommand{\eea}{\end{eqnarray}}

\def\Z{{{\mathbb{Z}}}}

\def\V{{\mathcal{V}}}

\let\tilde=\widetilde

\def\smpc{\hspace{.5pt}}

\renewcommand{\d}{\mathrm{d}}
\renewcommand{\i}{\mathrm{i}}

\numberwithin{equation}{section}

\def\lket{\ket}
\def\lbra{\bra}

\newcommand{\cls}{CLS}
\newcommand{\gs}{$G \Sigma$}

\newcommand{\de}{\partial}
\newcommand{\ca}{\mathcal}
\newcommand{\LL}{\left}
\newcommand{\RR}{\right}
\newcommand{\TT}{\text}

\newcommand{\ishiket}[1]{\lket{#1}  \hspace{-.08cm}\rangle \hspace{-.05cm}}
\newcommand{\ishibra}[1]{\hspace{-.05cm} \langle \hspace{-.08cm} \lbra{#1}}
\DeclareMathOperator{\arccosh}{arccosh}

\def\is{\!&\! = \! & \!}
\def\hh{M}
\def\ll{L}
\def\U{\textsc{U}}
\def\V{\textsc{V}}
\def\T{{\mbox{\small \textsc{T}}}}
\def\X{x}%{\mbox{\small \textsc{\smpc \hat{\phi}}}}}
\newcommand{\dmub}{\tilde{\mu}_{\textB}}

\def\ptau{{{\tau}}}
\def\sfb{{\mathsf b}}
\def\sfq{{\mathsf q}}
\def\sfbb{b}
\def\lgray{gray}
\def\textB{{\smpc\text{B}}}
\def\spc{\hspace{1pt}}
\begin{document}

\thispagestyle{empty}
\begin{adjustwidth}{-1cm}{-1cm}
\begin{center}
    ~\vspace{9mm}
    
     {\LARGE \bf 
    %From CLS to SYK via the crosscap
    %\\~\\
    SYK collective field theory as complex Liouville gravity
   }
   \end{center}
    \end{adjustwidth}
    \begin{center}
   \vspace{0.4in}
    
    {\bf Andreas Blommaert$^1$, Damiano Tietto$^2$, Herman Verlinde$^2$ }

    \vspace{0.4in}
    {$^1$School of Natural Sciences, Institute for Advanced Study, Princeton, NJ 08540, USA\\
    $^2$Department of Physics, Princeton University, Princeton, NJ 08544, USA
    }
    \vspace{0.1in}
    
    {\tt blommaert@ias.edu, dtietto@princeton.edu, verlinde@princeton.edu}
\end{center}

\vspace{0.4in}

\begin{abstract}
\noindent We establish a precise relationship between the $G\Sigma$ collective field theory of the double scaled SYK model and the worldsheet theory of the complex Liouville string a.k.a. sine dilaton gravity. The relationship is similar to the lightcone gauge  in critical string theory, and to what transpires when we gravitationally dress to an observer in gravity: one of the Liouville fields plays the role of a dynamical  clock with respect to which the second Liouville field evolves. This other Liouville field is identified with the collective field of SYK, which thus acquires a direct gravity interpretation.

The relevant 2D worldsheet geometry is that of a disk with specific crosscap and FZZT boundary conditions, as deduced from the $G\Sigma$ formulation. We compute the CLS amplitude on this geometry and find that this coincides with the DSSYK partition function. We indicate how our results can be lifted to 3D gravity, previewing upcoming work. An outflow of our results is that physical operators of DSSYK are mapped to holonomy operators (Verlinde lines) of complex Liouville theory on the crosscap geometry, which in turn have a 3D representation in terms of line operators in 3D de Sitter gravity. We show that the partition function of SYK can be represented as the expectation value of a circular gravitational Wilson line on $\mathbb{RP}^3$ (a.k.a. elliptic 3D de Sitter space).

\end{abstract}

\def\nspc{\hspace{-1pt}}
\pagebreak
\setcounter{page}{1}
\tableofcontents

\pagebreak

% %%%%%%%%%%%
% % SECTION %
% %%%%%%%%%%%
\section{Introduction}\label{sect:intro}

Solvable models of low-dimensional holography such as the SYK model are an important testing ground for studying the underlying dynamical mechanisms of holography and for exploring potential generalizations of holography to spacetimes with positive cosmological constant. The SYK model is a quantum mechanics of $N$ Majorana oscillators $\psi_i$ with the following Hamiltonian
\begin{equation}
    H_\text{SYK}=\i^{p/2}\sum_{i_1<\dots <i_p}J_{i_1\dots i_p}\psi_{i_1}\dots \psi_{i_p}\,,\quad \lambda =\frac{2 p^2}{N}\,.\label{1.1}
\end{equation}
At low energies, large $N$ and small $\lambda$, the SYK model is holographically dual to 2D JT gravity on the Lorentzian strip or euclidean disk, where the SYK temperature determines the size of the boundary \cite{Maldacena:2016upp,Engelsoy:2016xyb,Jensen:2016pah} (see \cite{Mertens:2022irh} for a review). In the double scaling limit $p\to\infty$ and $N\to\infty$ with $0<q= e^{-\lambda}<1$ finite, SYK is exactly solvable \cite{Cotler:2016fpe,Berkooz:2018jqr,Berkooz:2018qkz}.\footnote{Some interesting recent work on DSSYK includes \cite{Okuyama:2025hsd,Cui:2025sgy,Aguilar-Gutierrez:2024oea,Berkooz:2024ifu,Lin:2023trc,Belaey:2025kiu,Bossi:2024ffa}. Additional relevant references can be found in \cite{Berkooz:2024lgq}.} Quickly, it was recognized that DSSYK is a good candidate for a microscopic description of quantum gravity in a cosmological de Sitter spacetime \cite{Narovlansky:2023lfz,Verlinde:2024znh,Susskind:2022bia,Lin:2022nss,Rahman:2022jsf}. Any microscopic dual for a cosmological spacetime would be a valuable step forward, and obtaining a detailed understanding of the gravitational dual of DSSYK is therefore a well warranted goal.

Progress on pinpointing the relations between DSSYK and pure dS$_3$ quantum gravity was made in \cite{Gaiotto:2024kze,Verlinde:2024znh,Narovlansky:2023lfz,Verlinde:2024zrh,Tietto:2025oxn,Narovlansky:2025tpb}. However, while some quantitative links have been uncovered, a full holographic dictionary is lacking and many geometric aspects are yet to be understood. A related dual description of dS$_3$ quantum gravity was recently found \cite{Collier:2025lux,Verlinde:2024zrh} in the form of the worldsheet theory of the complex Liouville string  \cite{Collier:2024kmo_base}, a theory of 2D quantum gravity that consists of two Liouville fields with complex central charges adding up to $c_++c_-=26$. The worldsheet action of the CLS has some similarities with the G$\Sigma$ theory --- the bilocal collective field theory of DSSYK \cite{Cotler:2016fpe,Lin:2023trc}. However, a precise correspondence between the two systems has not been established yet. One of the goals of this paper is to fill this gap.

In a parallel development, much recent progress was made in developing a duality between DSSYK and a theory of 2D quantum cosmology called sine dilaton gravity \cite{Blommaert:2023wad,Blommaert:2023opb,Blommaert:2024whf,Blommaert:2024ydx,Blommaert_spheres:2025rgw,blommaert_wormholes_2025}. The action of the sine dilaton model coincides with that of the complex Liouville string \cite{Blommaert:2024ydx,Verlinde:2024zrh,Collier:2025pbm_sinedil,mertens_liouville_2021}. However, unlike the~CLS, sine dilaton gravity was designed with particular ``initial'' boundary conditions, whose goal in life is to ensure that its quantization yields the exact amplitudes of DSSYK \cite{Blommaert:2023wad,Blommaert:2023opb,Blommaert:2024whf,Blommaert:2024ydx,Blommaert_spheres:2025rgw,blommaert_wormholes_2025}. A second goal of this work, directly related to the first, is to give a natural geometric characterization of these boundary conditions and thereby put the duality between DSSYK and 2D sine dilaton gravity on firmer footing.

From this, it may perhaps sound like our aim is to fill in some technical details in an already well-understood story. Neither is true. First, our direct aim is to establish a precise equivalence between the $G\Sigma$ theory of double scaled SYK and gauge-fixed sine dilaton gravity, thus giving a direct embedding of DSSYK into a gravitational system. This is a significant step forward. Secondly,
there is a deep and fundamental distinction between the spectral properties of DSSYK and those of conventional 2D CFTs, as well as of most well-understood low-dimensional gravity models such as Schwarzian QM, JT gravity and AdS$_3$ gravity. The DSSYK energy $E(\theta)$ and spectral density $\cal{Z}_\text{spec}(\theta)$ are conventionally parametrized by an angle $\theta$ via
\bea
E(\theta) = -\frac{2\cos\theta}{\lambda}\,,\qquad{\cal{Z}}_\text{spec}(\theta) \is \big( e^{\pm 2 \i \theta}; q\big)_\infty \,, \qquad q=e^{-\lambda}. \label{1.2}
\eea
The shape of the DSSYK spectrum is depicted in figure 1. 
We see that the spectrum runs over a finite range and that $\cal{Z}_\text{spec}(\theta)$ has a smooth maximum at the symmetry point $\theta=\pi/2$, where $E=0$. This behavior should be contrasted with the unbounded spectrum of all known models of AdS holography, where energy can become arbitrarily large and entropy always increases with energy (because AdS black holes can become arbitrarily large and their horizon area grows with mass). Clearly, the gravitational dual of a system with the spectrum shown in figure~1 needs to be very different. Indeed, there are many indications that the dual description of DSSYK is related with a closed dS$_3$ universe \cite{Narovlansky:2023lfz, Susskind:2021esx, Tietto:2025oxn,Narovlansky:2025tpb}. 
With this motivation, our third aim is to identify the properties and boundary conditions of complex Liouville gravity responsible for truncating its energy spectrum to a finite range. 

%We believe that deriving boundary conditions for the gravitational path integral from a microscopic hologram is an important problem.  
%This is quite remarkable, because the $G\Sigma$ theory is inherently a boundary description, which acquires a direct gravitational interpretation.\footnote{It would be interesting to understand if there is a relation to the role played by the collective fields of vector models in \cite{Aharony:2022feg}.} 
%In particular, a quite important boundary condition is to be imposed at the  ``beginning of time''. One proposal to describe a big-bang at the quantum level is the ``no-boundary'' initial condition \cite{hartle1983wave}. We will argue in section \ref{sect:threedee} that DSSYK does implement a smooth capping off of spacetime in dS$_3$ gravity. But it does \emph{not} correspond with the usual implementation of the no-boundary state.

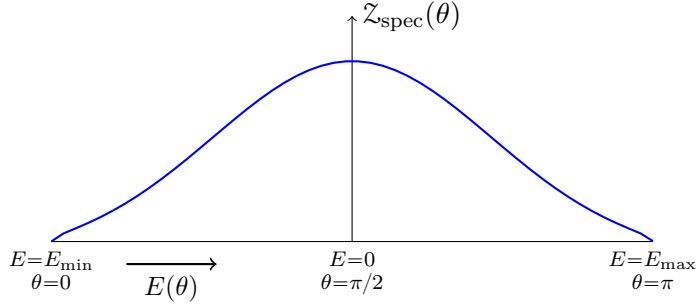
\begin{figure}[t]
\centering
\begin{tikzpicture}[scale=1]
  % Draw x-axis
  \draw (-4,0) -- (0,0) node[below] {{${E=0}\atop{\theta=\pi/2}$}} -- (4,0) node[below] {${E=E_{\rm max}}\atop{\theta=\pi}$} ;
  \draw[thick,->] (-3,-.3) -- (-2.4,-.3) node[below] {\small $E(\theta)$} -- (-1.8,-.3) ;
  \draw (-4,0)  node[below] {${E=E_{\rm min}}\atop{\theta=0}$};
  \draw[->] (0,0) -- (0,3) node[right] { ${\cal{Z}}_\text{spec}(\theta)$};
  % Draw Gaussian center at x=0
  \draw[thick, blue] (0,0) plot[domain=-4:4, samples=50] (\x,{.3*exp(-\x*\x/8)*sqrt(4-\x)*(2*sqrt(4+\x))});
\end{tikzpicture}
\caption{The SYK spectral density is bounded: it runs over a finite range and has a maximum at $E=0$. }
\end{figure}

Complex Liouville gravity comes with precise rules of quantization and a well-studied assortment of conformal boundary conditions \cite{Zamolodchikov:2001ah,Fateev:2000ik,Collier:2024kmo_base,Collier:2024mlg_zz}. The bilocal G$\Sigma$-theory, on the other hand, lives on a specific 2D geometry given by the kinematic space (the space of pairs of points) of the thermal SYK circle \cite{Maldacena:2016hyu,Cotler:2016fpe,Lin:2023trc}. This space is topologically a disk with a crosscap \cite{Lin:2023trc}. The collective field $g$ has been suggested to have a direct bulk gravity interpretation \cite{Goel:2023svz,Berkooz:2024evs,Berkooz:2024ofm}. However, the boundary conditions on $g$ are neither diff invariant nor conformal. How can we view the $G\Sigma$ theory as a gravity theory if it is not generally covariant? As we will show, it is nevertheless possible to recast the G$\Sigma$ theory in the form of a covariant CLS worldsheet theory, placed on a disk geometry (depicted in figure 2) with conformal boundary conditions on one side and with a suitable crosscap boundary condition on the other side.

\begin{figure}[t]
\begin{center}
\begin{tikzpicture}[thick,xscale=0.56,yscale=0.69, every node/.style={scale=0.9}]
\draw[very thick,fill=blue!05!white] (3,-2)--(3,2)--(-3,2)--(-3,-2)--cycle;
\draw[very thick,darkgreen,fill=white] (0,2)  ellipse (3cm and 1cm);
\draw[very thick,blue,fill=blue!10!white] (0,-2)  ellipse (3cm and 1cm);
\draw[very thick, <<->>] (-1.1,-2.45)--(1.1,-1.55);
\draw[very thick, <<->>] (1.1,-2.45)--(-1.1,-1.55);
\draw (0,1.3) node[color=blue] {\footnotesize $\T=0$};
\draw[->,blue] (-.9,-.7)--(-.9,.3) node[midway, right] {\footnotesize $\T=\frac{\U-\V}{2}$};
\draw (0,-3.5) node[color=white] {$\tau=-1/4$};
\end{tikzpicture}~~~~~~~~~~
\begin{tikzpicture}[thick,scale=.85, every node/.style={scale=0.9}]
\draw[blue,fill=blue!02, thick] (1.75,0.2)--(1.75,1.8)--(5.75,1.8)--(5.75,0.2) --cycle ;
\draw[blue,fill=blue!02, thick] (-1.75,0.2)--(-1.75,1.8)--(-5.75,1.8)--(-5.75,0.2) --cycle ;
\draw[darkgreen,fill=cyan!02, thick] (-1.1,0.25)--(-1.1,1.25)--(1.1,1.25)--(1.1,0.25) --cycle ;
\draw (3.75,1) node[color=blue] {\large $\varphi_-$ Liouville CFT};
\draw (0,.75) node[color=darkgreen] {\large $bc$ ghosts};
\draw (-3.75,1) node[color=blue] {\large $\varphi_+$ Liouville CFT};
\draw[darkgray,fill=darkgray!02, thick] (2,-3.8)--(2,-2.2)--(5.5,-2.2)--(5.5,-3.8) --cycle ;
\draw[\darkred,fill=purple!02, thick] (-2,-3.8)--(-2,-2.2)--(-5.5,-2.2)--(-5.5,-3.8) --cycle ;
\draw (3.8,-3) node[color=darkgray] {\large CLOCK};
%\draw (3.8,-3.4) node[color=darkgray] {\small $\T=\frac{\U-\V}{2}$};
\draw[darkgray,thick,->] (3.5,0) -- (3.5,-2) node[midway, right] {\ $e^{\varphi_-}\! = \frac{b}{\sin
(\frac{b}{2}(\U-\V))}$};
\draw[darkgray,thick,->] (-3.5,0) -- (-3.5,-2) node[midway, left] {$e^{\varphi_+} =e^{g/2}$\quad};
\draw (-3.8,-3) node[color=\darkred] {\large  $G\Sigma$-theory};
\draw (0,-1) node[color=black] {\large lightcone gauge fixing};
\draw[darkgray,<->]  (-1.7,-2.8) -- (1.7,-2.8);
\draw[darkgray] (-0,-2.5) node {\small WDW constraint};
%\draw[black] (-0,-3.3) node {\small constraint};
\draw[blue] (-0,2) node {\Large CLS};
\draw[\darkred] (-0,-3.55) node {\Large SYK};
\end{tikzpicture}
\end{center}
\caption{Overview of our main results. We place the CLS on a disk with a crosscap and perform a lightcone gauge fixing by setting the physical worldsheet coordinates equal to the classical uniformizing coordinates of the $\varphi_-$ Liouville field. The other Liouville field $\varphi_+(\U,\V)$ then becomes identified with the SYK collective field $g(\U,
\V)$, while $\varphi_-$ takes on the role of a clock. We match the partition functions on both sides. \label{fig2}}
\end{figure}
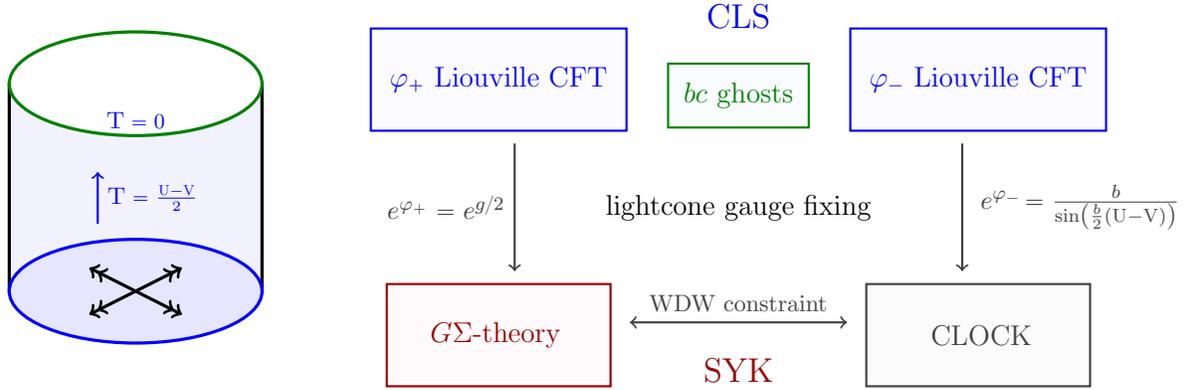

The main idea is summarized in figure \ref{fig2}. We consider the CLS worldsheet theory on the disk with a crosscap (a.k.a.\,the  M\"obius strip). Next we perform a ``lightcone gauge'' fixing by setting the physical worldsheet coordinates equal to the classical uniformizing coordinates $(\U,\V)$ of the $\varphi_-$ Liouville field. The other Liouville field $\varphi_+(\U,\V)$ then becomes identified with the SYK collective field $g(\U,\V)$, while $\varphi_-$ takes on the role of a clock that measures the physical time. 
This construction is similar to lightcone gaug quantization method in the ordinary bosonic string \cite{mandelstam1973interacting}. From a Hamiltonian point of view, our construction mirrors recent discussions of dressing to dynamical observers in quantum gravity \cite{Chandrasekaran:2022cip,Witten:2023xze}, with the difference that our clock is automatically provided by the gravitational theory. In that sense, our construction is morally more akin to \cite{Chen:2024rpx} where one dresses to the inflaton field.

After introducing the main actors in our story, we describe the CLS light-cone gauge quantization and the equivalence with the G$\Sigma$ theory in \textbf{section \ref{sec: lightcone quantization}}. As a decisive quantitative check, we compute the M\"obius band amplitude in complex Liouville gravity in \textbf{section \ref{sec: liouville crosscap amplitude}}, using the crosscap boundary state motivated by the discussion on sine dilaton gravity in \textbf{section \ref{sec: hamiltonian analysis}}. The main input in this calculation are the standard boundary wavefunctionals of Liouville CFT. As a crucial intermediate step, we show that the integral over Liouville momentum $p$ in the intermediate channel reduces to a discrete sum of residues at an infinite set of poles, indicating that the spectrum of momenta in the closed channel is discrete. In the open channel, this means that the boundary cosmological constant of the FZZT brane, which becomes the DSSYK energy, runs over a finite range. The calculated spectrum exactly matches that of DSSYK \eqref{1.2}, thus establishing the claimed equivalence.

An important theme in our work is that the CLS worldsheet does not support any transverse fluctuations and that its physical phase space is therefore exactly described by the zero mode sector. We investigate the quantization of this zero mode sector in two  complementary ways. First, we consider a minisuperspace quantization. Secondly, we identify the zero modes with holonomies of the dynamical coordinates that uniformize the classical solutions to the Liouville equations of motion. The latter identifies the spectral angle $\theta$ as the label of an elliptic SU(1,1) holonomy, which naturally explains why the energy spectrum runs over a finite range.

The remaining sections are organized as follows.

In \textbf{section \ref{sect:2actors}} we review necessary background material on the collective field theory formulation of DSSYK, on the complex Liouville string, and on the relation between sine dilaton gravity and DSSYK.

In \textbf{section \ref{sec: lightcone quantization}} we explain the embedding of the collective field $g$ in CLS (as one of the Liouville fields expressed in the gauge where the second Liouville field measures time) from the classical point of view. The physics is analogous to the usual lightcone gauge \cite{mandelstam1973interacting}. We also motivate the crosscap boundary conditions in CLS that will lead to a match with the partition function of DSSYK.

In \textbf{section \ref{sec: hamiltonian analysis}} we study the minisuperspace description of complex Liouville (or sine dilaton) gravity and explain from a Hamiltonian perspective how one of the Liouville fields provides the ``clock'' relative to which the dynamics of the second Liouville field takes place. An important outflow of our study is that it gives a starting point for identifying the dual gravity interpretation of DSSYK operators. Our results indicate that DSSYK operators are mapped to Verlinde lines of complex Liouville theory, which in turn have a 3D representation in terms of gravitational line operators in dS$_3$.

In \textbf{section \ref{sec: liouville crosscap amplitude}} we describe the calculation of the complex Liouville crosscap amplitude. The match between this amplitude and the DSSYK partition function is the main validation of the stated equivalence between the G$\Sigma$ theory and complex Liouville gravity.

Finally, in \textbf{section \ref{sect:threedee}} we describe elements of dS$_3$ implementation of the DSSYK partition function. In particular, we argue that a Wilson line amplitude of SU$(1,1)$ CS theory in $\mathbb{RP}^3$ computes the DSSYK spectral density \eqref{1.2}. The Wilson line represents the worldline of an observer in the static patch. The SU$(1,1)$ CS theory describes the quantization of non-rotating SdS black holes in dS$_3$ and describes the ``chiral half" of full dS$_3$ quantum gravity $-$ an SL$(2,\mathbb{C})$ CS theory. The embedding in $\mathbb{RP}^3$ indicates that the cosmological horizon satisfies an unusual smoothness condition akin to elliptic de Sitter space \cite{Parikh:2002py}. More details on this SYK-3D de Sitter correspondence will be presented elsewhere.

In the \textbf{appendices} we compute some other CLS amplitudes for comparison.

%%%%%%%%%%%
% SECTION %
%%%%%%%%%%%
\section{Main characters}\label{sect:2actors}
\vspace{-1mm}

In this section we present some background material on the $G\Sigma$ formulation of DSSYK and on complex Liouville string theory (sine dilaton gravity). In section \ref{sec: lightcone quantization} and section \ref{sec: hamiltonian analysis} we will then clarify the relation between these two bulk dual descriptions of DSSYK. One may read this as explaining the precise sense in which the collective field $g$ has a gravity interpretation. The $G\Sigma$ SYK collective field theory will be reviewed in section \ref{sect2.1gsigma}. CLS/sine dilaton gravity is reviewed in section \ref{sect:actor2}.
%%%%%%%%%%%%%%%%%%%
\subsection{Actor 1. SYK collective field theory}\label{sect2.1gsigma}

\vspace{-1mm}

\begin{figure}[t]
    \centering
\begin{tikzpicture}[thick,scale=0.69, every node/.style={scale=1}]
\draw[black,fill=blue!10, thick] (-4,0)--(0,4)--(4,0)--cycle ;
\draw[black, thick] (-4,0)--(0,4)--(4,0)--(0,-4)--cycle ;
\draw[green,very thick] (-4,0)--(4,0);
\draw[thick,->] (-2.1,1.9)--(-2.05,1.95);
\draw[thick,->] (-2,2)--(-1.95,2.05);
\draw[thick,<-] (2.1,1.9)--(2.05,1.95);
\draw[thick,<-] (2,2)--(1.95,2.05);
\draw (-0.2,-0.4) node[color=blue] {$\tau_1=\tau_2$};
\draw[->] (-4.5,0.5)--(-1.5,3.5);
\draw[->] (-4.5,-0.5)--(-1.5,-3.5);
\draw (-4.5,0) node[color=black] {$0$};
\draw (0,4.5) node[color=black] {$\beta$};
\draw (0,-4.5) node[color=black] {$\beta$};
\draw (-3.5,2.0) node {$\tau_1$};
\draw (-3.5,-2.0) node {$\tau_2$};
\end{tikzpicture}~~~~~~~~~~~
\begin{tikzpicture}[thick,scale=0.69, every node/.style={scale=1}]
\draw[black,fill=blue!10, thick] (-4,0)--(4,0)--(4,-2)--(-4,-2) --cycle ;
\draw[black, thin,lightgray] (-4,0)--(0,4)--(4,0)--(0,-4)--cycle ;
\draw[green,very thick] (-4,0)--(4,0);
\draw[blue,thick] (-4,-2)--(4,-2);
\draw[thin,gray,->] (-2.1,1.9)--(-2.05,1.95);
\draw[thin,gray,->] (-2,2)--(-1.95,2.05);
\draw[thin,gray,<-] (2.1,1.9)--(2.05,1.95);
\draw[thin,gray,<-] (2,2)--(1.95,2.05);
\draw[thick,<<-] (-1.95,-2)--(-2,-2);
\draw[thick,<<-] (2.05,-2)--(2,-2);
\draw[thick,<-] (4,-.9)--(4,-1);
\draw[thick,<-] (-4,-.9)--(-4,-1);
\draw (-0.2,0.4) node[color=blue] {$\tau=0$};
\draw (-0.2,-2.4) node[color=blue] {$\tau=-1/4$};
\draw (-4.8,0) node[color=blue] {\small $x=0$};
\draw (4.8,0) node[color=blue] {\small $x=1$};
\draw[->] (-4.5,-4.25)--(-3.25,-4.25);
\draw (-3,-4.7) node {$x$};% = \frac{\tau_1+\tau_2}{2\beta}$};
\draw[->] (-4.5,-4.25)--(-4.5,-3.2);
\draw (-4.5,-2.9) node {$\tau$};% = \frac{\tau_1-\tau_2}{2\beta}$};
\end{tikzpicture}
  \caption{\textbf{Left:} the \gs{} model lives in the blue shaded crosscap, in the region $0 \le \tau_1 \le \tau_2 \le \beta$ of the $(\tau_1,\tau_2)$ plane. \textbf{Right:} the equivalence between the crosscap and the M\"obius strip can be easily seen by ``doubling'' the crosscap and then folding it in half again in a different way.} 
  \label{FIGURE crosscap g sigma}
\end{figure}
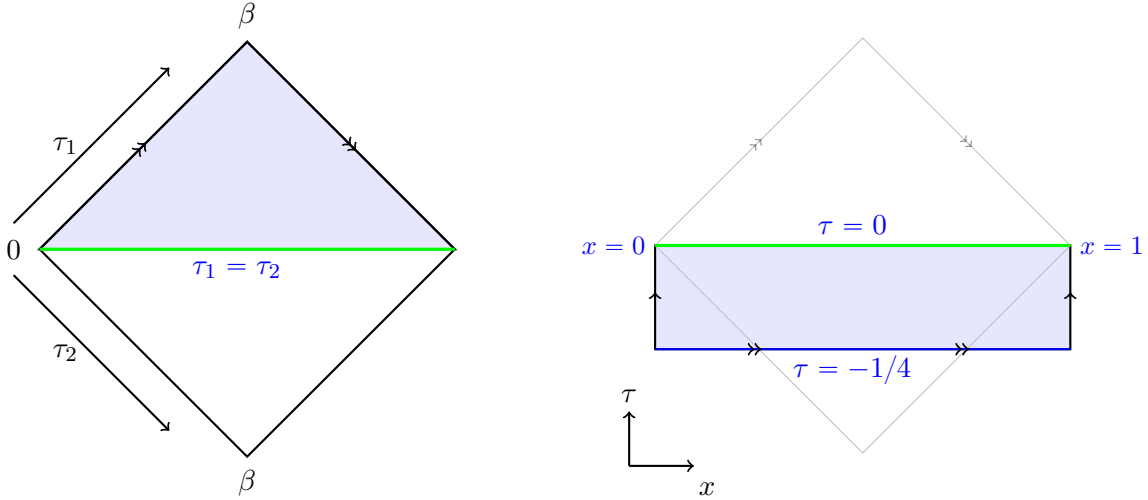

The start by reviewing the collective field theory formulation of DSSYK \cite{Cotler:2016fpe, Lin:2023trc,Maldacena:2016hyu,Sachdev:2015efa}. The $G\Sigma$ field theory is designed to capture the large $N$ collective behavior of the two-point function
\bea
\label{twopointG}
    G(\tau_1,\tau_2)\is \frac{1}{N}\sum_{i=1}^N\psi_i(\tau_1)\psi_i(\tau_2)\,,
\eea
In the double scaling limit $p,N\to \infty$ with finite $\lambda$ in \eqref{1.1} held fixed, $G(\tau_1,\tau_2)$ scales as
\bea
    G(\tau_1,\tau_2)\is \frac{\text{sgn}(\tau_1\!-\nspc\tau_2)}{2}\Bigl(1+\frac{g(\tau_1,\tau_2)}{p}\Bigr)\,. %,\quad\Sigma(\tau_1\tau_2)=\frac{\sigma(\tau_1,\tau_2)}{p}\,.
\eea
In these equations $\tau$ denotes euclidean time. We will work at finite inverse temperature $\beta$, so the collective field $g(\tau_1,\tau_2)$ is $\beta$ periodic in both coordinates. It is further subject to the boundary condition $g(\tau,\tau)=0$ and to the identification \cite{Lin:2023trc}
\begin{equation} \label{g sigma g field boundary conditions}
    g(\tau_1,\tau_2) = g(\tau_2,\tau_1)   \, .
\end{equation}
Hence we may restrict its fundamental domain to the   $\tau_2<\tau_1$ half of the $\beta$-periodic torus. For the discussion in this paper, it will be useful to transition the Euclidean cartesian coordinates $(x,\tau)$ restricted to the finite annulus (see figure \ref{FIGURE crosscap g sigma})
\begin{equation}
    x=\frac{\tau_2+\tau_1}{2\beta}\,,\quad \tau=\frac{\tau_2-\tau_1}{2\beta}\,,\qquad 
    -1/4<\tau<0\,,\quad x\sim x+1\,.
    \label{xtaudomain}
\end{equation}

The derivation of the collective field theory starts by writing the Euclidean path integral for the thermal partition function of the SYK model \eqref{1.1} while imposing the field identification \eqref{twopointG}  by means of a bilocal Lagrange multiplier field $\Sigma(\tau_1,\tau_2)$. After performing the Gaussian average over the coupling $J_{i_1\dots i_p}$ and integrating out the fundamental SYK fermionic modes,  one obtains the $G\Sigma$ theory \cite{Sachdev:2015efa,Maldacena:2016hyu,Cotler:2016fpe}.
In the large $p$ double scaling limit, the $G\Sigma$ action undergoes a further drastic simplification: the action for $\Sigma$ becomes Gaussian. Performing this Gaussian path integral results in the following surprisingly simple and familiar looking collective field theory \cite{Cotler:2016fpe,Lin:2023trc}
\bea
\label{2.3gintegral}
%   && Z(\beta) \, = \,  2^{N/2}  \int\! \mathcal{D}g\exp\Bigl(\frac{\i}{\lambda}\spc I \Bigr), \\[2.5mm]
I  \is  \frac \i 4  \int_{-1/4}^0\!\!\!\!\d \tau \int_0^{1}\!\! \d x \spc \Bigl(\frac{1}{4}\smpc \bigl(\de_x g\bigr)^{2}- \frac{1}{4}\smpc \bigl(\de_\tau g\bigr)^{2} - e^g  \Bigr)  \,.
%I  \is   \i   \int_0^\beta\!\!\d \tau_1\!\int_0^{\tau_1}\!\!\! \d \tau_2\Bigl(\frac 1 4\spc \de_{\tau_1} g\spc \de_{\tau_2} g-\mathcal{J}^2e^g  \Bigr)  \,.
\label{2.4gaction}
\eea
Here we set $\mathcal{J}=\frac{1}{2}$ and performed a constant shift $g \to g - 2\log \beta$ of the collective field $g(x,\tau)$ relative to the standard conventions. 
Hence the inverse temperature $\beta$ is now encoded in the boundary conditions at $\tau=0$ 
\begin{equation}
   e^{g(x,0)/2}= \beta\,.
 \label{2.8gbc}
\end{equation}
The action $I$ in \eqref{2.4gaction} is, up to the prefactor of $\i$, identical to that of a Liouville CFT in Lorentzian signature, though with a negative bulk cosmological constant. This relationship with a familiar CFT will play an important role in the following. Note, however, that the extra factor $\i$ in the action implies that, when viewed as a CFT, the central charge would be complex. Note further that the conformal symmetry requires that $e^{g(\tau_1,\tau_2)}$ transforms as a (1,1) form and that this symmetry is in fact broken by the Dirichlet boundary condition \eqref{2.8gbc}.

To specify the boundary condition at the other spacelike boundary at $\tau=-1/4$, we observe that the reflection symmetry
$g(\tau_1,\tau_2)=g(\tau_2,\tau_1)$ implies an antipodal spatial identification 
\begin{equation}
g(x,-1/4) = g(x+1/2,-1/4)
\end{equation}
Such an antipodal identification on a circle is the definition of a ``crosscap'' boundary condition. Thus, the SYK collective field $g$ lives on a spacetime which is topologically a disk with a crosscap. This is topologically equivalent to the M\"obius strip.  In the following we will think of the partition function of the $G\Sigma$ theory as a transition amplitude between an initial crosscap state at $\tau=-1/4$ and the final boundary state at $\tau=0$. The domain \eqref{xtaudomain} is translation invariant in $x$. So classical vacuum solutions are expected to be $x$-independent.

\subsubsection*{Partition function from minisuperspace}

\vspace{-1mm}

Our main object of study is the SYK partition function $Z(\beta)$. By computing the collective field theory path integral  \eqref{2.4gaction} via perturbation theory in $\mathcal{J}$ one obtains  \cite{berkooz2019towards,Lin:2023trc}
 \bea
  &&\ \  Z  = \int_0^\pi \! \d\theta \,
{\mathcal{Z}}_{\rm spec}(\theta) \,e^{\beta E(\theta)}\, 
\label{2.13zsyk}\\[2.5mm]
E(\theta) \!\! &\! =\! & \!\! -\frac{2\cos(\theta)}{\lambda},
\quad
{\mathcal{Z}}_{\rm spec}(\theta) =  \big( e^{\pm 2 i \theta}; e^{-\lambda}\big)_\infty\,.
\label{zspecz}
\eea
The symbol $(a,b)_n$ is a q-Pochhammer symbol.
Our goal is to give a geometric derivation of this result by relating the $G\Sigma$ collective field theory to the complex Liouville string, or equivalently, sine dilaton gravity. The following heuristic derivation of the result \eqref{zspecz} via the minisuperspace treatment of the collective field theory will be useful.

Restricting to $x$-independent configurations, the  action \eqref{2.4gaction} reduces to Liouville quantum mechanics
\begin{align}
 I&= {\i}\int_{-1/4}^{0}\!\!\!\!\d\tau \,\bigl(\pi_g \spc \dot{g} - H \bigr)\label{liouac2}\,,\quad 
H =   {\pi_g{\!\!}^2} + e^g\,. 
\end{align}
Here we introduced a dual momentum $\pi_g$ to $g$. A general solution to the Hamilton equations of motion $\dot{g} = 2\pi_g$, $\dot{\pi}_g = e^g$ takes the form
\begin{equation}
\label{cantrafo}
    e^{g(\tau)/2}=\frac{\sfbb}{\sin(\theta - b\spc \tau)},\,\quad\ \  \pi_g(\tau) = b \cot(\theta - \sfbb\spc \tau).
\end{equation}
The boundary condition $e^{g/2}=\beta$ at $\tau=0$ implies the relation $\beta  = {b}/{\sin(\theta)}$, whilst the identification $g(\tau_1,\tau_2)=g(\tau_2,\tau_1)$ implies smoothness at the crosscap location
\begin{equation}
    \frac{\d g}{\d \tau}(-1/4) = % -2\beta b^2 \cot(\theta+ \sfbb/4) = 
    0\,  \quad \to \quad \boxed{\ \cos(\theta +\sfbb/4) = 0\ \raisebox{-.5mm}{\Large $\strut$}}\label{2.10gbccrosscap}
\end{equation}
This uniquely fixes the parameters of the solution
\bea
    \beta=\frac{b}{\sin(\theta)}\,,\quad %e^{g/2}=\frac{b}{\sin(\theta-b \tau)}\,, 
    & & \quad b = 2\pi - 4\theta\,,
    \label{gsigmasol2.12}
\eea
where the angle $\theta$ runs over the physical range $[0,\pi]$. 
At $\tau=0$, the phase space coordinates $(\pi_g,g)$ and   $(\sfbb,\theta)$ are related via a canonical transformation specified by setting $\tau=0$ in equation \eqref{cantrafo}.
%\bea
%\label{cantrafo}
%e^{g/2} =  \frac{b}{\sin(\theta)}\,,\quad \pi_g = \frac{b}{2} \cot(\theta)\,.
%\eea
Hence in the quantum theory, both sets of variables form canonically conjugate pairs
\bea
 [\pi_g,g] = {\hbar}, \quad \ \ 
 [b,\theta] ={\hbar}, \quad \ \hbar = \lambda\,.
\eea
Note that the spectrum of $b$ is expected to be discrete: $b = \i \hbar m$ with $ m\in \mathbb{Z}$.

Motivated by the above description of the domain and boundary conditions on the $G\Sigma$ theory, we wish to reinterpret the functional integral \eqref{2.3gintegral} as computing the overlap
\bea
Z(\beta) = \langle \psi_{\rm final}|\psi_{\rm initial}\rangle 
%= \int \! d\theta \, \psi^*_{\rm final}(\theta)\spc \psi_{\rm initial}(\theta)\,,
\label{zoverlap}
\eea
between a suitable initial ``crosscap state'' at $\tau=-1/4$, 
and a final state at $\tau=0$ that implements the finite temperature boundary condition \eqref{2.8gbc}. This overlap can be computed as follows. The initial and final states implement the boundary conditions highlighted in  \eqref{2.8gbc} and \eqref{2.10gbccrosscap}
\bea
 \langle \psi_{\rm final} | \bigl( e^{g/2}-\beta\bigr)  = 0, \quad & & \quad \cos(\theta +\sfbb/4)|\psi_{\rm initial}\rangle = 0.
\eea
Expressing these quantum mechanical conditions as operator equations on wavefunction of $\theta$ gives rise to a differential equation for the final state and a difference equation for the initial state:
\bea
\bigg(-\frac{\hbar}{\sin \theta} \frac{\d}{\d\theta} -\beta\bigg) \psi_{\rm final}(\theta) = 0\,, \ & & \ 
\Bigl(e^{ \i \frac{\hbar}{2} \frac{\d\, }{\d \theta}} +  e^{\hbar  - 2 \i\theta} \Bigl) \psi_{\rm initial}(\theta)=0\, . 
\eea
Here we used that $b= \hbar\frac{d\ }{d\theta}$, and the relation \eqref{cantrafo}. In the second equation we made an inspired guess for how to treat the normal ordering ambiguity in the quantum definition of $\cos(\theta +\sfbb/4)$. From the above differential and difference equation, we deduce the explicit form of the initial and final wavefunction
\bea
\psi_{\rm final}(\theta) =  \exp\bigg( \frac{\beta \cos\theta}{\hbar}\bigg),\quad & & 
\quad
\psi_\text{initial}(\theta)= \big( e^{\pm 2 i \theta}; e^{-\hbar}\big)_\infty.
\eea
Plugging these expressions into  \eqref{zoverlap}, we reproduce the exact SYK partition function \eqref{2.13zsyk}-\eqref{zspecz}.
%%%%%%%%%%%%%%%%

\subsection{Actor 2. Complex Liouville and sine dilaton gravity}\label{sect:actor2}

\vspace{-1mm}

A second key actor in our discussion is the complex Liouville string \cite{Collier:2024kmo_base} or its reformulation as sine dilaton quantum gravity \cite{Blommaert:2024ydx}. We review some aspects of CLS and what was known about the relation between DSSYK and sine dilaton gravity. In addition, we provide initial evidence for the relation with the $G\Sigma$ formulation that we introduced in section \ref{sect2.1gsigma}.

\subsubsection*{Complex Liouville gravity}
\vspace{-1.5mm}
 The complex Liouville string (CLS) \cite{Collier:2024kmo_base} is a string theory whose worldsheet description consists of two Liouville scalar fields $\varphi_+$ and $\varphi_-$ with complex conjugate central charges, adding up to 26
\begin{equation}
    c_+=13+6 \bigg(\i \sfb^2+\frac{1}{\i \sfb^2}\bigg)\,,\quad c_+=13-6 \bigg(\i \sfb^2+\frac{1}{\i \sfb^2}\bigg)\,.\label{2.14cc}
\end{equation}
The parameter $\sfb^2$ is real, and will correspond in DSSYK with 
\begin{equation}
    2\pi \sfb^2= \lambda = {2p^2}/{N}\,.
\end{equation}
The Euclidean action of the complex Liouville string reads
\begin{align}
    I&=-\frac{\i}{4\pi \sfb^2}\int\d^2x \,\Bigr( \de^\mu \varphi_+ \de_\mu \varphi_++  e^{2\varphi_+}\Bigr)-\frac{\i}{2\pi \sfb^2}\,{\mu_{\text{B}}}_+\int \d x \,e^{\varphi_+}\nonumber\\[2mm]&\qquad\qquad\qquad +\frac{\i}{4\pi \sfb^2}\int\d^2x \,\Bigr( \de^\mu \varphi_- \de_\mu \varphi_- +  e^{2\varphi_-} \Bigr)+\frac{\i}{2\pi \sfb^2}\,{\mu_{\text{B}}}_-\int \d x \,e^{\varphi_-}\,.\label{2.17liouac}
\end{align}
The action \eqref{2.17liouac} may \emph{wrongly} lead one to suspect that the two Liouville theories are decoupled. But, the complex Liouville string worldsheet theory starts out as a fully covariant 2D gravity theory invariant under two-dimensional diffeomorphisms and Weyl transformations. Following the usual Polyakov path integral treatment of string theory, we can eliminate the gauge redundancy by
fixing the (euclidean) worldsheet metric to be of the form $
    \eta_{\mu\nu}\d x^\mu \d x^\nu = \d\tau^2+\d x^2.$
This gauge fixing removes the background metric as a dynamical degree of freedom, but still leaves an imprint: the full gauge fixed action also includes a ($b,c$) ghost action. We will mostly ignore this ghost system in what follows, except to note that the physical BRST invariant observables of the CLS worldsheet theory must be invariant under the conformal symmetry group generated the sum $L_n^{\rm tot} = L_n^+ + L_n^-$ of the Virasoro generators of $\varphi_+$ and $\varphi_-$. These Virasoro constraints
\bea
    L_n^{\rm tot} = L_n^+ + L_n^- \!\is \! 0
\eea
play a crucial role in our discussion of the lightcone gauge quantization in section \ref{sec: lightcone quantization}.

In Liouville CFT, fixed $\mu_\text{B}$ boundary conditions are FZZT \cite{Fateev:2000ik, Teschner:2000md,Zamolodchikov:2001ah} boundary conditions. FZZT boundary conditions are associated with Virasoro primaries with labels $s$. The precise relation is
\begin{equation}
    {\mu_\text{B}}_\pm=\cos(2\pi \sfb \spc \mathsf{s}_\pm) =\cos(\theta_\pm)
    \quad \leftrightarrow\quad \ket{\text{FZZT}(\mathsf{s}_\pm)}\,.\label{2.22fzzt}
\end{equation}
Complex Liouville CFT has the special property that the parameter  $\mathsf{s}_+$ can be chosen real.\footnote{
The boundary cosmological constant is conventionally parametrized via \cite{Teschner:2000md}
\begin{equation}
    {\mu_\text{B}}_+=\cosh(2\pi b_+S_+)\,.
\end{equation}
For complex values of the central charge we have a non-standard spectrum (equations (2.9) and (2.12) in \cite{Collier:2024kwt})
\begin{equation}
    b_+=\sqrt{\i}\sfb\,,\quad S_+=\sqrt{\i}\mathsf{s}_+\,,
\end{equation}
with $\sfb$ and $\mathsf{s}_+$ real. In the notation of equation (2.12) and footnote 3 of \cite{Collier:2024kwt} indeed $S_+=\i {p_+}_\text{there}=\sqrt{\i}\mathsf{s}_+$ with real $\mathsf{s}_+$.\label{fn6}
} 
These boundary states are discussed in more detail in section \ref{sec: liouville crosscap amplitude}. Alternatively \cite{mertens_liouville_2021}, one can Legendre transform with respect to ${\mu_\text{B}}_+$ and (or) ${\mu_\text{B}}_-$. This replaces the boundary terms in the Liouville action \eqref{2.17liouac} with the constraints 
\bea
\beta_\pm \!\is \! \oint \d x\, e^{\varphi_\pm}
\eea
fixing the boundary lengths  at $\tau=0$.

To get more insight into the meaning of the boundary conditions, it is useful to consider the classical solutions of the complex Liouville equation of motion. As before, it is sufficient to consider only the  $x$-independent classical solutions of the minisuperspace Liouville action 
\begin{align}
 I&= {\i}\int_{-1/4}^{0}\!\!\!\!\d\tau \,\bigl(\pi_+ \spc \dot{\varphi}_+ + \pi_-\dot{\varphi}_- - H_+  + H_-\bigr)\label{liouac1d}, 
 %\\[1.5mm]&\ 
\qquad {H_\pm =   {\pi_\pm^2} + e^{2\varphi_\pm}}
%\,,\quad  {H_- =   {\pi_-^2} + e^{\varphi_-}}\,.
\end{align}
The general solution to the minisuperspace Hamilton equations of motion $\dot{\varphi_\pm} = 2\dot{\pi}_\pm$, $\dot{\pi}_\pm = e^{\varphi_\pm}$ now comes with four integration constants $b_\pm, \theta_\pm$ (c.f. \eqref{gsigmasol2.12})
\begin{align}
    \label{2.29solbis}
   \; e^{\varphi_+}\!=\, \frac{\sfbb_+}{\sin(\theta_+\! - \sfbb_+\spc \tau)}&\,,\quad e^{\varphi_-}\! =\, \frac{\sfbb_-}{\sin(\theta_-\! +  \sfbb_-\spc \tau)}\,,    \\[2mm]
     \   \pi_+ = b_+ \cot(\theta_+\! - \sfbb_+\spc \tau)&\,,\quad\ \spc    \pi_- = b_-\cot(\theta_-\! + \sfbb_-\spc \tau)\,.
    \label{2.20liousol}
\end{align} 
The minisuperspace phase space thus consists of two canonically conjugate pairs 
\bea
 [\pi_+,\varphi_+] =  [\pi_-,\varphi_-] = {\hbar}\,, & & \hbar = 2\pi b^2,
 \eea
subject to the WDW constraint 
\bea
H_{\rm WDW} \equiv H_+ - H_-= {\pi_+^2} + e^{2\varphi_+}-{\pi_-^2} - e^{2\varphi_-}\, =\, 0\,. 
\eea
We will describe the minisuperspace quantization in more detail in section 4.  For now, we mention that the zero mode of the WDW Hamiltonian constraint forces $b_+=b_-$ in \eqref{2.29solbis} and in \eqref{2.20liousol}.

The angles $\theta_\pm$ that parametrize the boundary cc in \eqref{2.22fzzt} equal the integration constants in \eqref{2.20liousol}. Indeed, according to the Legendre transform which we carried out in the action, $\mu_{{\rm B}_\pm}$ is conjugate to $e^{\varphi_\pm}$. Furthermore, using the relations $e^{\varphi_\pm} = {b_\pm}/{\sin(\theta_\pm)}$ and $\pi_\pm = b_\pm \cot(\theta_\pm)$ that characterize phase space (the initial conditions at some Cauchy slice, say $\tau=0$) we deduce the parameterization \eqref{2.22fzzt}.
%\bea
%\mu_{\textB_+}  = e^{-\varphi_+}\pi_+ = \cos(\theta_+)\,,\quad
%\mu_{\text{B}_-}\! = e^{-\varphi_-}\pi_- = \cos(\theta_-)_{\strut}\,.
%\eea

\subsubsection*{Sine dilaton gravity}
\vspace{-1.5mm}

Sine dilaton gravity \cite{Blommaert:2024ydx} is defined by the action
\bea
    I\is \frac{1}{4\pi \sfb^2}\int\d^2 x\sqrt{g}\,(\Phi R+2\sin(\Phi))+\frac{1}{2\pi \sfb^2}\int \d u \sqrt{h}\,\Phi K +\,\text{boundary terms}\,.\label{2.19sdaction}
\eea
This covariant dilaton gravity action is invariant under two-dimensional diffeomorphisms. Upon fixing the conformal gauge $g_{ab} = e^\rho \eta_{ab}$ and after performing the field definition \cite{Blommaert:2024ydx,Verlinde:2024zrh,Collier:2025pbm_sinedil,mertens_liouville_2021}
\begin{equation}
    \varphi_+=\rho+\i \Phi/2\,,\quad \varphi_-=\rho-\i\Phi/2\,,
\end{equation}
one discovers that the sine dilation gravity theory is equivalent to the worldsheet theory of the complex Liouville string. The Virasoro constraints associated with the conformal gauge fixing can be recast in the form of  the usual WDW Hamiltonian and momentum constraint. The role of these constraints  from the dilaton gravity point of view will be discussed in more detail in section \ref{sec: hamiltonian analysis}.

It has been understood that sine dilaton gravity has a holographic relation with DSSYK \cite{Blommaert:2024whf,Blommaert:2024ydx,blommaert_wormholes_2025}. For our present purposes, the two key elements of what has been understood are the following. Firstly, the partition function of DSSYK $Z(\beta)$ is found by considering sine dilaton with the following boundary conditions at the worldsheet's boundary $\tau=0$ \cite{blommaert_wormholes_2025}\footnote{Consider equation (2.19) in \cite{blommaert_wormholes_2025}. The first identity in equation (2.19) in \cite{blommaert_wormholes_2025} means that %the  AdS$_2$ metric 
$e^{\varphi_-}$ diverges at the boundary, while the second identity sets the $\varphi_+$ boundary length equal to $\oint \d x\, e^{\varphi_+}\rvert_\text{bdy}=\beta\,$.
Since $x\sim x+1$, this implies  $e^{\varphi_+}\rvert_\text{bdy}=\beta$.
}
\begin{equation}
   e^{\varphi_+}\rvert_\text{bdy}=\beta\,, \quad \mu_{\text{B}_-}\rvert_\text{bdy}=1\,.\label{2.26holobdy}
\end{equation}
The second boundary condition ($\mu_{\text{B}_-}\rvert_\text{bdy}=1$) implies that $e^{\varphi_-}$ diverges at the  boundary, which implies an FZZT boundary state $\ket{\text{FZZT}(0)}$ in Liouville theory.\footnote{A closely related boundary state is the ZZ state \cite{Zamolodchikov:2001ah}. 
%Originally, the ZZ boundary was introduced precisely to have the Liouville field diverge at the worldsheet's boundary \cite{Zamolodchikov:2001ah}. 
The ZZ boundary state is a linear combination of FZZT boundary states \cite{Martinec:2003ka} 
which in our normalization \eqref{2.22fzzt} satisfies (as seen from  equation (A.1) in \cite{mertens_liouville_2021} and plug in the classical values for $s_+(1,1)$ in \eqref{2.22fzzt})
\begin{equation}
    {\mu_\text{B}}_-=-1\quad \leftrightarrow \quad \ket{\text{ZZ}}\,.\label{2.28zz}
\end{equation}
Indeed, the ZZ state and $\mu_{\text{B}_-}\rvert_\text{bdy}=-1$ are both boundaries where the Liouville field diverges. Note that complex conjugation of the action \eqref{2.17liouac} flips the sign of $\mu_\text{B}$ such that $\bra{\text{ZZ}}\leftrightarrow\ket{\text{FZZT}(0)}$. This remark will be useful in section \ref{sec: liouville crosscap amplitude}. } The condition $\mu_{\text{B}_-}\rvert_\text{bdy}=1$ fixes in the classical solutions \eqref{2.20liousol} at $\tau=0$ that $\theta_-=0$. The condition $e^{\varphi_+}\rvert_\text{bdy}=\beta$ equates the inverse temperature to $\beta = b_+/\sin(\theta_+)$. %The classical solutions for general $\tau$ thus simply to $e^{\varphi_+}={b}/{\sin(\theta-\sfb \tau)}, e^{\varphi_-}= {b}/{\sin(b\tau)}$ with $ \beta = {b}/{\sin(\theta)}$.

Comparing the resulting solutions \eqref{2.29solbis} with the classical solution \eqref{gsigmasol2.12} for the collective field $g$  suggests that the holographic dictionary between DSSYK and sine dilaton gravity should identify  
\bea
\quad \label{gident}
\boxed{\ g=2\varphi_+\ \raisebox{-.5mm}{\large $\strut$}}
\eea
This leaves the question: what is the role of the other Liouville field $\varphi_-$?  As we will discuss in section 4, the physical meaning of $\varphi_-$  is that it provides the clock that keeps track of a gauge invariant notion of time $\tau$. It is well known that in quantum gravity, the coordinate system, including the time coordinate, is not diff invariant and so a priori non-physical. The invariant statement to be made is that the identification \eqref{gident} holds {\it provided} we use the $\tau$-dependence of $\varphi_-$ as ``reference'' clock, relative to which we can specify the time-dependence of the physical  field $\varphi_+$.  One of the main goals of this work is to turn this intuitive statement into an exact quantum mechanical fact.

Before proceeding, we recall a second fact about the relation between sine dilaton and DSSYK. Aside from the holographic boundary conditions \eqref{2.26holobdy} the partition function of DSSYK is computed in sine dilaton gravity by imposing a peculiar set of ``initial conditions'' \cite{Blommaert:2024ydx,Blommaert:2024whf,blommaert_wormholes_2025}. In particular, as we explain in more detail in section \ref{sec: hamiltonian analysis}, one can write the relevant sine dilaton path integral as the following transition amplitude in minisuperspace quantum mechanics
\begin{equation}
    Z(\beta)= \bra{e^{\varphi_+}\! =\beta}\otimes\bra{\text{FZZT}(0)}
    \,\int_{0}^{\infty}\!\! \d T \, e^{-T H_\text{WDW}}
    \ket{\psi_\text{initial}}
    \,.\label{2.31sdpathint}
\end{equation}
$H_\text{WDW}$ is the WDW Hamiltonian \eqref{H_WDW4.5}. In earlier work \cite{Blommaert:2024ydx,Blommaert:2024whf,blommaert_wormholes_2025} it was shown that the correspondence with SYK specifies a unique minisuperspace wavefunction for the initial state $\ket{\psi_\text{initial}}$ which differs from the usual smooth no-boundary state \cite{hartle1983wave} of dilaton gravity. However, the full geometric meaning of this initial state $\ket{\psi_{\rm initial}}$ has thus far not been identified. As we will show in section \ref{sec: hamiltonian analysis}, this initial state has the natural geometric interpretation of a crosscap boundary condition, analogous to the boundary condition \eqref{2.10gbccrosscap} of the SYK collective field. %In section \ref{sect:threedee} we show that in terms of dS$_3$ quantum gravity, $\ket{\psi_{\rm initial}}$ is smooth. Of course, a 2D crosscap is also smooth.

\bigskip

%%%%%%%%%%%%%%%
 % SECTION %
 %%%%%%%%%%%
\section{Liouville lightcone gauge} \label{sec: lightcone quantization}
\vspace{-1mm}
In this section we explain how the lightcone gauge approach to the complex Liouville string (\cls{}) \cite{Collier:2024kmo_base} allows us to make direct contact with the \gs{} collective field theory approach to the DSSYK \cite{Cotler:2016fpe}. Here we will stick to a purely classical analysis, leaving a quantum mechanical treatment to sections \ref{sec: hamiltonian analysis} and~\ref{sec: liouville crosscap amplitude}.

For comparison, we briefly recall the key steps in lightcone quantization of the bosonic string \cite{mandelstam1973interacting,Goddard:1973qh, Kiritsis:2019npv}. The philosophy behind lightcone quantization is ``constrain first, quantize later'', meaning that one gets rid of all the gauge symmetries at the classical level, and only quantizes the physical degrees of freedom. Starting from the  bosonic string action, diffeomorphism invariance can be fixed by setting $\d s^2=e^{\varphi}\eta$. This still leaves conformal transformations as unbroken symmetries. Then, one solves the classical equation of motion for the $X^+ = X^0+X^1$ field as $X^+ = X^+_L(u) + X^+_R(v)$, with $u$ and $v$ the worldsheet lightcone coordinates. The residual conformal invariance $u \to U(u), \, v \to V(v)$ can be used to set $X^+_L = \frac 12(x^+ - p^+ u)$, $X^+_R = \frac 12 (x^+ + p^+ v)$, such that $X^+ = x^+ + p^+ t$. The zero modes  $x^+$ and $p^+$ cannot be gauged away \cite{Tong:2009np}. Finally, one uses the Virasoro constraints $T_{u u} = T_{v v} = 0$ to solve $X^-$ in terms of $x^+, p^+$ and the transverse string coordinates $X^i$. Quantization of the remaining degrees of freedom is then carried out as usual.

In subsection \ref{sect3.1lc}, we will introduce an analogous lightcone gauge procedure in CLS, gauge-fixing the $\varphi_-$ field modulo zero modes. We then relate the resulting gauge fixed action to that of the SYK $G\Sigma$ theory \eqref{2.4gaction}. Next we will impose the Virasoro constraints, which eliminates all local excitations of the $\varphi_+$ field: unlike the bosonic string, only the zero modes remain as physical degrees of freedom. Their quantization is presented in section \ref{sec: hamiltonian analysis}. In section \ref{sect:3.3cross} we motivate the boundary conditions on the Liouville fields $\varphi_-$ and $\varphi_+$ that complete the match with the $G\Sigma$ collective field theory and formulate the string amplitude that computes the DSSYK partition function \eqref{2.13zsyk}
%, which will be further motivated in section \ref{sec: hamiltonian analysis}, and computed in section \ref{sec: liouville crosscap amplitude}.

%%%%%%%%%%%%%%%
\subsection{Physical phase space}\label{sect3.1cls}
\vspace{-1mm}

We start by describing the space of classical solutions to the CLS equation of motion. The worldsheet theory is given by the sum of two Liouville CFTs \eqref{2.17liouac} with total central charge 26.\footnote{After fixing diff invariance, thus introducing a $bc$ ghost system with central charge $c = -26$, the total worldsheet theory has vanishing conformal anomaly.} We consider the two Liouville theories on a Lorentzian cylinder with coordinates $(x,\tau)$ with $x \sim x +1$. Using lightcone coordinates $u=x-\tau,v=x+\tau$,
the \cls{} action \eqref{2.17liouac} reads:
\bea \label{gauge fixed action cls}
    I\!\is \! \frac{\i}{2 \pi   \sfb^2} \int \d u\,\d v \,  \Bigl( \de_u \varphi_+ \de_v \varphi_+ + \frac{1}{4}e^{2\varphi_+}   - \de_u \varphi_- \de_v \varphi_--\frac{1}{4}e^{2\varphi_-} \Bigr)  \, .
\eea
From now on, let us consider a double Wick rotation of the worldsheet coordinates $u \to \i u$, $v \to \i v$, which effectively changes the sign of the bulk cosmological constant. A general solution of the equations of motion of this action may be expressed in terms of uniformizing coordinates $U_\pm(u), V_\pm(v)$ as \cite{Seiberg:1990eb}
\bea \label{liouville metric}
    e^{2\varphi_\pm(u,v)}\is \frac{\de U_\pm(u) \spc \de V_\pm(v)}{\sin\bigl( (U_\pm(u)-V_\pm(v))/2\bigr)^2} \,  .
\eea
Unlike in the bosonic string case these are not solutions of a free wave equations; but the key analogy is that they still depend only on left moving and right moving functions $U_\pm(u), V_\pm(v)$. 
Using the solutions \eqref{liouville metric}, the stress-energy tensor of each Liouville theory reads (with $\{ \cdot, \cdot\}$ the Schwarzian derivative)
\begin{align}
    & \sfb^2 \spc T^{\pm}_{u u } = \pm \i\smpc ((\de_u \varphi_\pm)^2 -\de^2_{u} \varphi_\pm  ) = \mp \frac{ \i}{2} \LL\{e^{\i \smpc U_\pm}, u \RR\} \,,\nonumber \\[-2.5mm]
\label{stress tensors}\\[-2.5mm]
  & \sfb^2  \spc T^{\pm}_{v v } = \pm {\i }\smpc ((\de_v \varphi_\pm)^2 -\de^2_{v} \varphi_\pm  ) = \mp \frac{ \i }{2} \LL\{e^{-\i \smpc V_\pm}, v \RR\}   \, .\nonumber
\end{align}

The functions $U_\pm(u)$, $V_\pm(v)$ may be multi-valued on the worldsheet. Indeed, the right-hand side of \eqref{liouville metric} and the Schwarzian derivative in \eqref{stress tensors} are invariant under M\"obius transformations $e^{\i \smpc U_\pm} \to g_\pm \cdot e^{\i\smpc U_\pm} $, $e^{-\i \smpc V_\pm} \to g_\pm e^{-\i \smpc V_\pm}$ 
acting on the dynamical lightcone coordinates via
\bea
\label{mobiustrafo}
e^{\i\smpc U_\pm} \to g_\pm\cdot e^{\i \smpc U_\pm}= \frac{\alpha_\pm \, e^{\i\smpc U_\pm}+ \beta_\pm}{\bar{\beta}_\pm\, e^{\i\smpc U_\pm} + \bar{\alpha}_\pm}\,,\  & & \  
g_\pm = \bigg(\! \begin{array}{cc} \alpha_\pm \! &\!  \beta_\pm \\[0mm] \bar{\beta}_\pm \! &\! \bar{\alpha}_\pm \end{array}\!\bigg) \in SU(1,1)\,,
\eea
and the same transformation for $e^{-\i V_\pm}$. This in particular implies that $U_\pm$ and $V_\pm$ can have a non-trivial SU(1,1) monodromy around the spatial circle
\bea
e^{\i \smpc U_\pm (x+1)} = \hh_\pm \cdot 
e^{\i \smpc U_\pm(x)}\,,\  & & \ \  e^{-\i \smpc V_\pm(x+1)} = \hh^{-1}_\pm \cdot 
e^{-\i \smpc V_\pm(x)}\,,
\label{hmonodromy1}
\eea
whilst preserving the single-valuedness of the Liouville fields $\varphi_+$ and $\varphi_-$. Because we can still act with the global M\"obius isometries \eqref{mobiustrafo}, the monodromy transformations $\hh_\pm$ are defined up to conjugation. So we can rotate the holonomy matrices to a diagonal form
\bea
\label{hmonodromy2}
\hh_+ = \bigg(\! \begin{array}{cc} e^{\i b_+/2} \! &\!  0\\[0mm] 0 \! &\! e^{-\i b_+/2} \end{array}\!\bigg)\,,\quad
\hh_- = \bigg(\! \begin{array}{cc} e^{\i b_-/2} \! &\!  0\\[0mm] 0 \! &\! e^{-\i b_-/2} \end{array}\!\bigg)\,.
\eea
Here we assumed that $\hh_\pm$ are both in an elliptic conjugacy class of SU(1,1). This is the relevant choice for relating the CLS to the SYK collective field theory, as we will see shortly.

In our setting, we have a worldsheet boundary. Without loss of generality, we choose the boundary to be located at $\tau=0$, or, expressed in terms of the $(u,v)$ coordinates, at $
u|_{\text{bdy}}=v|_{\text{bdy}}\,.$
The boundary conditions at $u=v$ must preserve conformal invariance and reflect stress energy. Conformal invariance and the fact that the $\varphi_+$ and $\varphi_-$ Liouville CFTs  have different central charges $c_+$ and $c_-$ implies that each have their own boundary conditions: there can be no exchange of stress energy between the two systems. Hence, stress energy conservation holds independently for $\varphi_+$ and $\varphi_-$:
\bea
T_{uu}^\pm|_{\text{bdy}}  = T_{vv}^\pm|_{\text{bdy}}  \quad \to\quad \LL\{e^{\i U_\pm}, u \RR\}|_{\text{bdy}}  = \LL\{e^{-\i V_\pm}, v \RR\}|_{\text{bdy}}  
\eea
This condition demands that $U_\pm(u)$ and $V_\pm(v)$ are identical functions up to a M\"obius transformation:
\bea
e^{\i \smpc U_\pm }|_{\text{bdy}}  = \ll_\pm \cdot 
e^{-\i \smpc V_\pm}|_{\text{bdy}}\,,\quad \ll_\pm \in SU(1,1)\,.
\label{lrelation}
\eea
These transformations $\ll_\pm$ must commute with the corresponding monodromy matrices $\hh_\pm$ \eqref{hmonodromy2}
\bea
\ll_+ \hh_+ = \hh_+ \ll_+\,,\quad 
\ll_- \hh_- = \hh_- \ll_-\,.\label{3.10}
\eea
Assuming $\ll_\pm$ are both elliptic, we deduce that the most general $L_{\pm}$ are parameterized as
\bea
\label{lmonodromy}
\ll_+ = \bigg(\! \begin{array}{cc} e^{\i \theta_+} \! &\!  0\\[0mm] 0 \! &\! e^{-\i \theta_+} \end{array}\!\bigg)\,,\quad 
\ll_- = \bigg(\! \begin{array}{cc} e^{\i \theta_-} \! &\!  0\\[0mm] 0 \! &\! e^{-\i \theta_-} \end{array}\!\bigg)\,.
\eea
Inserting this into \eqref{lrelation}, this means that $U_\pm$ and $V_\pm$ are related up to a constant shift
\bea
U_\pm(u)|_{\text{bdy}} = V_\pm(v)|_{\text{bdy}} + 2 \theta_\pm\,.\label{3.12}
\eea
The four monodromy variables $b_\pm$ and $\theta_\pm$ are a sufficient to parametrize the full physical phase space.

%%%%%%%%%%%%%%%
\subsection{Lightcone gauge}\label{sect3.1lc}

\vspace{-1mm}

We are now ready to implement the lightcone gauge fixing of the CLS worldsheet theory. The action \eqref{gauge fixed action cls} still has a redundancy under $u$ and $v$ dependent conformal transformations
\begin{equation} \label{conformal transformation}
    u \to \U(u)\,,\quad v \to \V(v) \,,\quad e^{2 \varphi_\pm(\U,\V)} \d \U \d \V = e^{2\varphi_\pm(u,v)} \d u \, \d v\,.
\end{equation}
These transformations are generated by the the stress-tensor; we will discuss the associated Virasoro constraints momentarily. Following the lightcone gauge approach, we  use the conformal invariance to choose a special gauge in which the (non-dynamical) worldsheet coordinates are set equal to the dynamical coordinate functions $U_-$ and  $V_-$ that label the space of classical solutions of $\varphi_-$. After imposing this gauge, the field $\varphi_-$ becomes non-dynamical.

The presence of the worldsheet boundary at $u=v$ restricts the conformal symmetry group \eqref{conformal transformation} to the set of transformations for which
\bea
\U(u)|_{\text{bdy}}=\V(v)|_{\text{bdy}}\,.
\eea
Hence, $\U$ and $\V$ are the same functions  $\U(z) = \V(z)$. So, the conformal symmetry group with boundary is half as big as without boundary. However, as we see from \eqref{3.12}, the phase space of classical solutions to the Liouville equations is also half as without boundary. We can thus still eliminate the $\varphi_-$ dynamics, by choosing a suitable lightcone gauge coordinate system $(\U,\V)$. 
For the following, it is useful to introduce the space and time coordinate
\bea
\X = \frac{\U\spc +\spc \V}{2} \,,\quad\T = \frac{\U\spc -\spc \V}{2}\,.
\eea
Without loss of generality, we can choose the reflecting boundary to be located at $\T=0$ and impose the periodicity condition $\X\sim \X+1$.

In the CLS lightcone gauge, we choose the worldsheet coordinates $(\U,\V)$ such that the dynamical uniformizing coordinates of the $\varphi_-$ field are set equal to
\bea \label{uniformizing coordinate conformal transformationminus}
    U_-(\U) =  b_-  \U + 2 \theta_- \, \quad & & \   V_-(V) =  b_- \V  \, .
\eea 
The remaining constants $b_-$ and $\theta_-$ are dynamical zero modes.
%\footnote{The constant modes $b_-$ and $\theta_-$  are analogous to the constant zero modes $x^+$ and $p^+$ that appear in standard lightcone gauge string theory.} 
The mode $b_-$ parametrizes the Liouville momentum and specifies the multi-valuedness  \eqref{hmonodromy1}-\eqref{hmonodromy2}  of  $U_-$ and $V_-$ around the spatial circle, 
while the variable $\theta_-$ determines the location of the $T=0$ boundary in terms of the $\varphi_-$ time coordinate, or equivalently, the $L_-$ monodromy \eqref{3.12}
\bea
\begin{array}{c}{U_-(\U + 1)\, =\, U_-(\U) + b_-\,,}\\[1.5mm] 
{ V_-(\V + 1) \, =\,  V_-(\V) + b_-\,,}\end{array} \quad & &  T_- = \frac{U_- -V_-}{2} = b_- \T + \theta_-\,.
\eea
After this lightcone gauge fixing, the $\varphi_-$ Liouville field has been eliminated as an independent dynamical degree of freedom. This leaves us with one single Liouville action:
\bea \label{liouville action cls only phi_+}
    I\!\is\!\frac{\i}{2 \pi   \sfb^2} \int \d \U \, \d \V \,  \Bigl( \de_{\U} \varphi_+ \de_{\V} \varphi_+ + \frac{1}{4}  e^{2\varphi_+} \Bigr) \,.
\eea
We propose that this action should be identified with the $G\Sigma$ collective field action \eqref{2.4gaction}, with the following field identification and relation between CLS and SYK coupling and lightcone coordinates
\bea\label{gsigma coordinates}
 g = 2\varphi_+\,,\quad \lambda = 2\pi \sfb^2\,,\quad  \tau_1 =   \U \,,\quad \ \tau_2 =  \V   \, .
\eea
In other words, the collective field $g$ is identified with $2\varphi_+$ in the gauge where $\varphi_-$ is used to provide the worldsheet space and time coordinates. We make this statement more concrete at the quantum level in section \ref{sec: hamiltonian analysis}. 

Next, following the standard lightcone gauge procedure, we impose the Virasoro constraints
\bea
    T^+_{uu}+ T^{-}_{uu} =0\,,\quad T^+_{vv} + T^-_{vv} = 0   \, ,\label{Virasoroconstraint}
\eea
that implement the equations of motion of variation with respect to the background worldsheet metric. The Virasoro conditions enforce that the boundary conditions are generally covariant and imply that physical phase space of classical solutions of the complex Liouville gravity theory does not factorize into the product of a separate classical phase spaces of the $\varphi_+$ and $\varphi_-$ fields. Instead the space of classical solutions of $\varphi_+$ and $\varphi_-$ are linked together via an (almost) one-to-one mapping.

Working on-shell, as is usual for a lightcone gauge treatment, we can express the Virasoro conditions \eqref{Virasoroconstraint} in terms of the uniformizing coordinate $U_\pm$ and $V_\pm$. The constraints \eqref{Virasoroconstraint} then take the form $\left\{e^{\i \smpc U_+}, u \right\}  = \left\{e^{\i \smpc U_-},u\right\}$ and $\LL\{e^{-\i \smpc V_+}, v \RR\}  = \LL\{e^{-\i \smpc V_-}, v \RR\}$, or equivalently
\footnote{One checks that \begin{equation}
    \LL\{e^{\i U_+}, u \RR\}  - \LL\{e^{\i U_-}, u \RR\}  =  e^{2\i  U_-} \LL(\de_u U_- \RR)^2  \LL\{e^{\i U_+}, e^{\i U_-} \RR\}\,.
\end{equation}}
\bea
   \LL\{e^{\i \smpc U_+}, e^{\i \smpc U_-} \RR\} = 0 \,,\ & & \ \LL\{e^{-\i\smpc V_+}, e^{-\i \smpc V_-} \RR\} =  0 \,.
\eea
We thus learn that the $\varphi_+$ and $\varphi_-$ uniformizing coordinates are related by a M\"obius transformation
\begin{equation} \label{Virasoroconstraint2}
    e^{\i \smpc U_+} = L\cdot e^{\i\smpc U-}\,,\quad  e^{-\i\smpc V_+}    = R\cdot e^{-\i\smpc V-}\,,\quad L, R \in SU(1,1)\,.
\end{equation}
To determine the explicit form of the transformations $L$, $R$, we first note that the relations \eqref{Virasoroconstraint2} must be compatible with the monodromy conditions \eqref{hmonodromy1}. This means that
\bea
\hh_+ L = L\spc \hh_-\,,\quad \hh_+ R = R\spc \hh_-\,.
\eea
From this, we learn that $\hh_+$ and $\hh_-$ are not independent, but must be elements of the same conjugacy class. In terms of the explicit form \eqref{hmonodromy2} of $\hh_+$ and $\hh_-$, this implies that $b_+=b_-$. We will call this condition the global WDW constraint.

Assuming that the SU(1,1) transformations \eqref{Virasoroconstraint2} are again elliptic, and using the residual gauge freedom to shift $V_+$ by an arbitrary constant, we find that \eqref{Virasoroconstraint2} can be fixed to the form
\begin{equation}
 e^{\i \smpc U_+} =  e^{2 \i (\theta_+ -\smpc \theta_-)} e^{\i\smpc U_-}  \, , \quad e^{- \i\smpc V_+} = e^{-i\smpc V_-}  \, .\label{3.15uplus}
\end{equation} 
Combining \eqref{3.15uplus} with \eqref{uniformizing coordinate conformal transformationminus} we hence find that the dynamical uniformizing coordinates $U_+$ and $V_+$ are also equal to the classical lightcone coordinates $\U$ and $\V$ up to a global conformal transformation labeled by a conjugate pair of dynamical zero modes $b_+$ and $\theta_+$, conform \eqref{uniformizing coordinate conformal transformationminus}
\begin{equation} \label{virasoro final constraint}
    U_+ = b_+ \U + 2  \theta_+ \, , \quad V_+ =b_+ \V  \, , \quad\ b_+ = b_-\,.
\end{equation}

Putting everything together, we learn that the solutions \eqref{liouville metric} of the classical $\varphi_+$ and $\varphi_-$ equations of motion, written in terms of $(U,V)$ and zero modes, take the by now familiar form 
\bea
 \label{liouville gsigma metric final}
    e^{2\varphi_+}= \frac{b_+^2}{\sin(\theta_+ - b_+ (\U-\V)/2  )^2}\,, \quad e^{2\varphi_-}=  \frac{b_-^2}{\sin(\theta_- + b_-(\U-\V)/2 )^2}\,
    ,\\[2.5mm]
    \pi_+ =\,  b_+ \cot\bigr(\theta_+\! - b_+ (\U\nspc -\nspc\V)/2  \bigr)\,,\quad \pi_-  = \,  \,  b_- \cot\bigl(\theta_- \!+ b_- (\U\nspc -\nspc\V)/2  \bigr)\, .
\eea
with $\pi_\pm = \partial_T \varphi_\pm$ the canonical momentum. The stress energy tensors for these solutions take the form
\bea
 T^+_{\U\U} = T_{\V\V}^+ = - \i \spc {b_+^2}/4\sfb^2 , \quad & & \quad T^-_{\U\U} = T^-_{\V\V} = \spc \i {b_-^2}/4\sfb^2.
\eea
Hence $b_+$ and $b_-$ are indeed required to be equal by virtue of the global Virasoro constraint \eqref{Virasoroconstraint}. This concludes our general description of the classical lightcone gauge fixed CLS worlsheet theory.

So far our discussion was classical. Transitioning to a quantum theory, the zero modes are promoted to quantum mechanical operators. One easily verifies that $(b_+,\theta_+)$ and $(b_-,\theta_-)$ form canonically dual pairs $[b_+,\theta_+] = -[b_-,\theta_-] = {\hbar}$, $\hbar = 2\pi \sfb^2$.
The WDW condition $b_+-b_- =0$ is a first class constraint that, via the usual symplectic reduction, enforces a gauge symmetry under simultaneous shifts of 
\bea
(\theta_+,\theta_-) \to (\theta_+ + \alpha,  \theta_-+ \alpha)\,.
\eea
We are thus free to  put $\theta_-=0$, leaving a two dimensional physical phase space parametrized by 
\bea
b\equiv b_+=b_- \,,\quad \theta\equiv \theta_+-\theta_- \,.
\eea

In conclusion, we have shown that the conformal symmetry and Virasoro constraints eliminate all local dynamics: unlike the standard critical string, CLS supports no local transversal oscillations and the  only physical sector that survives are the zero modes. This is equivalent to the statement that, for the purpose of computing the partition function, the minisuperspace quantization of 2D dilaton gravity is exact. The minisuperspace treatment is discussed in more detail in section~\ref{sec: hamiltonian analysis}.

\def\mX{{Y}}

\begin{figure}[t]
\begin{center}
\quad \begin{tikzpicture}[scale=1.8]

% Define cube vertices

\coordinate (A) at (0,0,0);
\coordinate (B) at (2,0,0);
\coordinate (C) at (2,2,0);
\coordinate (D) at (0,2,0);
\coordinate (E) at (0,0,2);
\coordinate (F) at (2,0,2);
\coordinate (G) at (2,2,2);
\coordinate (H) at (0,2,2);

\coordinate (A1) at (0.32,0,0);
\coordinate (B1) at (1.68,0,0);
\coordinate (C1) at (1.68,2,0);
\coordinate (D1) at (0.32,2,0);
\coordinate (E1) at (0.32,0,2);
\coordinate (F1) at (1.68,0,2);
\coordinate (G1) at (1.68,2,2);
\coordinate (H1) at (0.32,2,2);

\coordinate (A2) at (0,0.22,0);
\coordinate (B2) at (2,0.22,0);
\coordinate (C2) at (2,1.78,0);
\coordinate (D2) at (0,1.78,0);
\coordinate (E2) at (0,0.22,2);
\coordinate (F2) at (2,0.22,2);
\coordinate (G2) at (2,1.78,2);
\coordinate (H2) at (0,1.78,2);

\coordinate (A3) at (0,0,0.42);
\coordinate (B3) at (2,0,0.42);
\coordinate (C3) at (2,2,0.42);
\coordinate (D3) at (0,2,0.42);
\coordinate (E3) at (0,0,1.58);
\coordinate (F3) at (2,0,1.58);
\coordinate (G3) at (2,2,1.58);
\coordinate (H3) at (0,2,1.58);

\coordinate (A4) at (1.15,0,0);
\coordinate (B4) at (1.3,0,0);
\coordinate (A5) at (0,1.15,0);
\coordinate (D5) at (0,1.3,0);
% Draw vertical edges
\draw[thick,purple,<-] (A1) -- (B1) node [midway,below]{\small $\ll_-\quad$};
\draw[thick,white,<->] (A4) -- (B4);
\draw[thick,gray,->] (B2) -- (C2)node [midway,left]{\small $R{\strut}^{\strut}$};
\draw[thick,purple,->] (C1) -- (D1) node [midway,below]{\small $\ll_+$};
\draw[thick,gray,<-] (D2) -- (A2) node [midway,left]{\small $L{\strut}^{\strut}$};
\draw[thick,white,<->] (A5) -- (D5);
\draw[thick,purple,->] (F1) -- (E1) node [midway,below]{\small $\ll_-$};
\draw[thick,gray,<-] (G2) -- (F2)node [midway,left]{\small $R{\strut}_{\strut}$};
\draw[thick,purple,<-] (H1) -- (G1) node [midway,below]{$\qquad\ll_+$};
\draw[thick,gray,->] (E2) -- (H2)node [midway,left]{\small $L{\strut}^{\strut}$};
\draw[thick,blue,->] (E3) -- (A3) node [midway,right]{\small $\hh_-{\strut}^{\strut}$};
\draw[thick,blue,<-] (B3) -- (F3)node [midway, right]{\small $\hh_-{\strut}^{\strut}$};
\draw[thick,blue,<-] (C3) -- (G3)node [midway, right]{\small $\!\!\hh_+$ \Large ${}^{\strut}$};
\draw[thick,blue,<-] (D3) -- (H3)node [midway, right]{\small $\!\!\hh_+$ \Large ${}^{\strut}$};
% Label vertices
\node[darkgray,] at (A) {\small $U_-(\mbox{\footnotesize {\U+1}})\qquad$};
\node[darkgray,] at (B) {\small $\qquad V_-(\mbox{\footnotesize {\V+1}})$};
\node[darkgray,] at (C) {\small $\qquad V_+(\mbox{\footnotesize {\V+1}})$};
\node[darkgray,] at (D) {\small $U_+(\mbox{\footnotesize {\U+1}})\qquad$};
\node[darkgray,] at (E) {\small $U_-(\mbox{\footnotesize {\U}})$};
\node[darkgray,] at (F) {\small $V_-(\mbox{\footnotesize {\V}})$};
\node[darkgray,] at (G) {\small $V_+(\mbox{\footnotesize {\V}})$};
\node[darkgray,] at (H) {\small $U_+(\mbox{\footnotesize {\U}})$};
\end{tikzpicture}~~~~~~~~~~~~~\raisebox{+3mm}{\begin{tikzpicture}[scale=1.8]

% Define cube vertices

\coordinate (A) at (0,0,0);
\coordinate (B) at (2,0,0);
\coordinate (C) at (2,2,0);
\coordinate (D) at (0,2,0);
\coordinate (E) at (0,0,2);
\coordinate (F) at (2,0,2);
\coordinate (G) at (2,2,2);
\coordinate (H) at (0,2,2);

\coordinate (A1) at (0.32,0,0);
\coordinate (B1) at (1.68,0,0);
\coordinate (C1) at (1.68,2,0);
\coordinate (D1) at (0.32,2,0);
\coordinate (E1) at (0.32,0,2);
\coordinate (F1) at (1.68,0,2);
\coordinate (G1) at (1.68,2,2);
\coordinate (H1) at (0.32,2,2);

\coordinate (A2) at (0,0.22,0);
\coordinate (B2) at (2,0.22,0);
\coordinate (C2) at (2,1.78,0);
\coordinate (D2) at (0,1.78,0);
\coordinate (E2) at (0,0.22,2);
\coordinate (F2) at (2,0.22,2);
\coordinate (G2) at (2,1.78,2);
\coordinate (H2) at (0,1.78,2);

\coordinate (A3) at (0,0,0.42);
\coordinate (B3) at (2,0,0.42);
\coordinate (C3) at (2,2,0.42);
\coordinate (D3) at (0,2,0.42);
\coordinate (E3) at (0,0,1.58);
\coordinate (F3) at (2,0,1.58);
\coordinate (G3) at (2,2,1.58);
\coordinate (H3) at (0,2,1.58);

\coordinate (A4) at (1.15,0,0);
\coordinate (B4) at (1.3,0,0);
\coordinate (A5) at (0,1.15,0);
\coordinate (D5) at (0,1.3,0);
% Draw vertical edges
\draw[\lgray,] (A1) -- (B1) node [midway,below]{$\quad$};
\draw[thick,white,<->] (A4) -- (B4);
\draw[\lgray,] (B2) -- (C2)node [midway,left]{${\strut}^{\strut}$};
\draw[thick,purple,->] (C1) -- (D1) node [midway,below]{\small $\ll$};
\draw[\lgray,<-] (D2) -- (A2) node [midway,left]{\small $\ll{\strut}^{\strut}$};
\draw[thick,white,<->] (A5) -- (D5);
\draw[\lgray,] (F1) -- (E1);
\draw[\lgray,] (G2) -- (F2);
\draw[thick,purple,<-] (H1) -- (G1) node [midway,below]{$\qquad\ll$};
\draw[\lgray,->] (E2) -- (H2)node [midway,left]{\small $\ll{\strut}^{\strut}$};
\draw[\lgray,->] (E3) -- (A3) node [midway,right]{\small $\hh{\strut}^{\strut}$};
\draw[\lgray,<-] (B3) -- (F3)node [midway, right]{\small $\hh{\strut}^{\strut}$};
\draw[thick,blue,<-] (C3) -- (G3)node [midway, right]{\small $\hh{\strut}^{\strut}$};
\draw[thick,blue,<-] (D3) -- (H3)node [midway, right]{\small $\hh{\strut}^{\strut}$};
% Label vertices
\node[\lgray] at (A) {\small $U_-(\mbox{\footnotesize {\U+1}})\qquad$};
\node[\lgray] at (B) {\small $\qquad V_-(\mbox{\footnotesize {\V+1}})$};
\node[black] at (C) {\small $\qquad V_+(\mbox{\footnotesize {\V+1}})$};
\node[black] at (D) {\small $U_+(\mbox{\footnotesize {\U+1}})\qquad$};
\node[\lgray] at (E) {\small $U_-(\mbox{\footnotesize {\U}})$};
\node[\lgray] at (F) {\small $V_-(\mbox{\footnotesize {\V}})$};
\node[black] at (G) {\small $V_+(\mbox{\footnotesize {\V}})$};
\node[black] at (H) {\small $U_+(\mbox{\footnotesize {\U}})$};
\end{tikzpicture}}
\vspace{-5mm}
\end{center}
\caption{Overview of the SU(1,1) M\"obius transformations and holonomies relating the uniformizing coordinates and their transformed values after going once around the cylinder. Using the residual gauge invariance under the global  conformal symmetry group, one can trivialize all transition functions except the holonomy $\hh=\hh_+$ and $\ll=\ll_+$ relating the $(U_+,V_+)$ coordinates. }\label{fig:cube}
\end{figure}
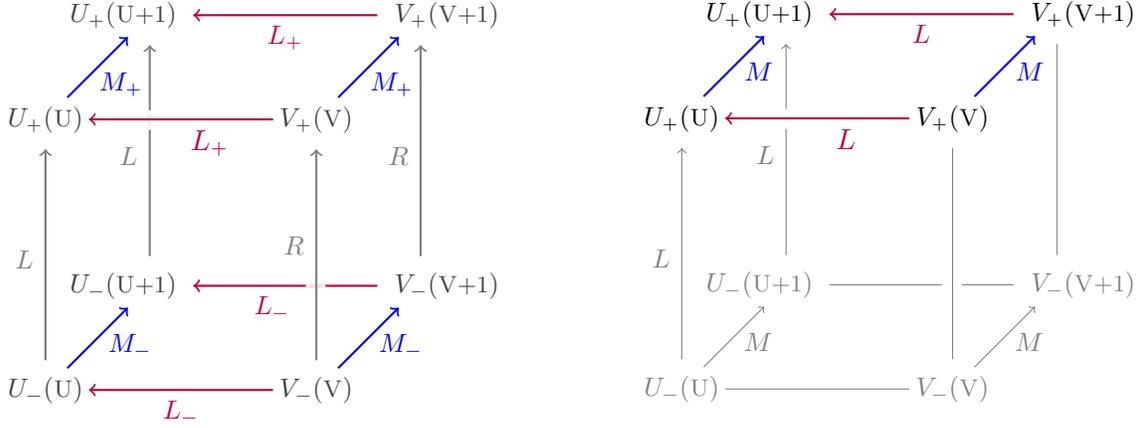
\subsection{Zero mode geometry}\label{sect:holonomies}
\def\cA{\mathcal{A}}
\vspace{-1mm}

To prepare for the transition to the quantum theory, it is useful to gain more insight into the global geometry of the physical phase space.
We have shown that the physical CLS phase space is captured by the M\"obius transformations $L$ \eqref{lmonodromy} relating the $U_+$ and $V_+$ coordinates and by their holonomies $M$ \eqref{hmonodromy2} around the cylinder.  Consider the dynamical lightcone coordinates $(U_\pm(\U),V_\pm(\V))$ and their image $(U_\pm(\U+1),V_\pm(\V+1))$ after going around the cylinder once.\footnote{Recall that
 $(U_\pm(\U), V_\pm(\V))$ depend only on the corresponding lightcone-gauge coordinate and that 
going around the cylinder once amounts to shifting both coordinates by one unit $(\U, \V) \to (\U\pm 1,\V\pm 1)$.} Figure \ref{fig:cube} shows these 8 coordinates placed on the corners of a cube connected by edges labeled by the corresponding M\"obius transformations. The transformations are subject to the homotopy requirement that going around any face should be trivial. As indicated on the right, the light-cone gauge trivializes all transition functions except the two transformations 
that act on the $\varphi_+$ uniformizing coordinates $(U_+,V_+)$ 
\begin{equation}
\label{lhmonodromy}
%\boxed{\ \ 
\ll = \bigg(\! \begin{array}{cc} e^{\i \theta} \! &\!  0\\[0mm] 0 \! &\! e^{-\i \theta} \end{array}\!\bigg)\,,\quad
\hh  = \bigg(\! \begin{array}{cc} e^{\i b/2} \! &\!  0\\[0mm] 0 \! &\! e^{-\i b/2} \end{array}\!\bigg)\,.
\end{equation}
This motivates considering the following cross-ratio combination \cite{Faddeev:2008xy}
\bea
\label{crossratio}
Y(\T)^2 \is \frac{\bigl(e^{\i U_+(\U+1)} \!- e^{\i V_+(\V+1)}\bigr)\bigl(e^{\i U_+(\U)}\! - e^{\i V_+(\V)}\bigr)}{\bigl(e^{\i U_+(\U+1)} \! - e^{\i U_+(\U)}\bigr)\bigl(e^{\i V_+(\V+1)} \! - e^{\i V_+(\V)}\bigr)}\,,\quad \T = \mbox{\Large $\frac{\U\,-\,\V}{2}$}\,.
\eea
The right-hand side is invariant under simultaneous M\"obius transformations of all four coordinates and depends only on the time coordinate $\T$. It provides a natural coordinatization of the physical phase space. Plugging in \eqref{virasoro final constraint} gives 
\bea
\mX(\T) = \frac{\sin(\theta - b\T)}{\sin(b/2)}\,.
\eea 
Comparing with \eqref{liouville gsigma metric final} shows that $Y(\T)$ relates to the constant mode of the Liouville exponentials via:
\bea
%e^{g(\T)/2} \is 
\frac{e^{-\spc\varphi_{{\!}_+}}\!(\T)}{e{}^{-\varphi_{{\!}_-}\!}{(0)}\smpc} = \frac{\mX(\T)}{\mX(0)} = \,\frac{\sin(\theta - b \spc \T)}{\sin(\theta)}\,.
\eea
Indeed, the expression \eqref{crossratio} can be read as a discretized version of the classical solution \eqref{liouville gsigma metric final} of $\varphi_+$.

Alternatively, we can view the cross-ratio \eqref{crossratio} as a SU(1,1) holonomy variable associated to the plaquette spanned by the four top corners of the cube in figure \ref{fig:cube}.  Following Faddeev and Volkov\cite{Faddeev:2008xy}, it is natural to introduce an integer spaced coordinate grid $(n,m)$ on the $(\U,\V)$-plane and represent the uniformizing coordinates evaluated on the grid points as
\bea
e^{\i U_+(n)} = \ll\cdot \hh^n  \cdot e^{-\i V_+(0)} , \quad & & \quad e^{-\i V_+(m)} = \hh^m \cdot e^{-\i V_+(0)}
\eea
Evaluating the cross ratio variables $Y(\T)$ at the corresponding time instances $\T_n = -\frac{n+1}2$, one defines
\bea
\mX_n \equiv \mX(\T_n)= \frac{\sin\bigl((n+1) b/2   +\theta\bigr)}{\sin(b/2)}\,.
\eea
This gives a natural sequence of phase space variables satisfying the recursive relation \cite{Faddeev:2008xy, Verlinde:2024zrh,Gaiotto_verlinde:2024kze}
\bea
\label{ysystem}
\mX_{n+1} \mX_{n-1} = 1_{} + \mX_n^2\,.
\eea
This relation is recognized as a discretized version of the Liouville equation of motion on the cylinder, in natural time units equal to the propagation time around the cylinder \cite{Faddeev:2008xy}. 

An equally natural sequence of phase space coordinates are\footnote{
The variables $Z_k$ can be geometrically represented as the cross ratio \eqref{crossratio} associated to a dual set of lattice points
\bea
e^{\i \tilde{U}_+(n)} = \hh\cdot \ll^n  \cdot e^{-\i V_+(0)}\,,\quad e^{-\i \tilde{V}_+(m)} = \ll^m \cdot e^{-\i V_+(0)}\,.
\eea
The interpretation of $Y_n$ and $Z_k$ as open Wilson lines becomes clear by expressing them as \cite{Gaiotto_verlinde:2024kze}
\begin{equation}
    Y_n = \frac{s\! \wedge\! \ll \hh^{n+1}  s}{s\!\wedge\! \hh s}\,,\quad  Z_k = \frac{s\!\wedge\! \hh \ll^{k+1} s}{s\!\wedge\! \ll\spc s}\,.
\end{equation}
Here $s$ denotes any two component spinor, and $\wedge$ denotes the  inner product defined with an insertion of a 2D $\epsilon$-symbol.
In a follow-up paper we will give the $Y_n$ and $Z_n$ a natural 3D interpretation in terms of line operators in $SL(2,\mathbb{C})$ CS theory, or equivalently, geodesic length operators in 3D de Sitter gravity. The quantum version of the relations \eqref{ysystem}-\eqref{zsystem} are the familiar skein relations, which in turn can be directly linked to the SYK chord rules \cite{Verlinde:2024znh}}
\bea
Z_k = \frac{\sin\bigl(b/2 + (k+1)\theta\bigr)}{\sin(\theta)}\,.
\eea
These variables satisfy a similar recursive formula
\bea
\label{zsystem}
Z_{k+1}Z_{k-1} = 1 + Z_k^2\,.
\eea
The variables $Y_n$ and $Z_k$ and relations \eqref{ysystem} and \eqref{zsystem} will be useful for the quantization and computation of the CLS partition function and for uncovering the connection with 3D gravity \cite{Verlinde:2024znh, Gaiotto_verlinde:2024kze}.

\subsection{Future boundary  condition}\label{sect:3.3fzzt}
\vspace{-1mm}

%%%%%%%%%%%%%%%%%%

To complete the embedding of the DSSYK partition function in CLS, we should impose  the boundary condition \eqref{2.8gbc} and the crosscap identification \eqref{g sigma g field boundary conditions}. As indicated in figure 2, we place the reflecting bc at the future boundary $T=\frac{\U-\V}{2} = 0$ and view the crosscap identification as an initial condition.
%at $T=-1/4$. We first take a closer look the future boundary. 
%As explained above and in section 2, the future boundary conditions at $\U = \V$ are summarized by $\theta_- = 0$ and $\oint e^{\varphi_+}\vert_{U = V} = \beta \leftrightarrow  b_+ = {\beta}\,{\sin\theta_+}$. The first equation is a gauge choice, the second defines the inverse temperature~$\beta$. Both should be read as classical conditions on the zero modes. How can we uplift these into conformal boundary conditions for the $\varphi_-$ and $\varphi_+$ Liouville CFTs?

\begin{figure}[t]
\begin{center}
\begin{tikzpicture}[thick,scale=0.7, every node/.style={scale=0.9}]
\draw[black,fill=blue!10, thick] (-4,0)--(4,0)--(4,-2)--(-4,-2) --cycle ;
\draw[black, thin,lightgray] (-4,0)--(0,4)--(4,0)--(0,-4)--cycle ;
\draw[darkgreen,very thick] (-4,0)--(4,0);
\draw[blue,thick] (-4,-2)--(4,-2);
\draw[thin,gray,->] (-2.1,1.9)--(-2.05,1.95);
\draw[thin,gray,->] (-2,2)--(-1.95,2.05);
\draw[thin,gray,<-] (2.1,1.9)--(2.05,1.95);
\draw[thin,gray,<-] (2,2)--(1.95,2.05);
\draw[thick,<<-] (-1.95,-2)--(-2,-2);
\draw[thick,<<-] (2.05,-2)--(2,-2);
\draw[thick,<-] (4,-.9)--(4,-1);
\draw[thick,<-] (-4,-.9)--(-4,-1);
\draw (-0.2,-0.4) node[color=blue] {$T=0$};
\draw (-0.2,0.4) node[color=darkgray] {$e^{\varphi_+}=\beta$};
\draw (-0.2,-2.4) node[color=blue] {$T=-1/4$};
\draw (-5,0) node[color=blue] {\small $X=0$};
\draw (5,0) node[color=blue] {\small $X=1$};
\draw[->] (-5,-4)--(-3.9,-4);
\draw (-3.65,-4) node {$X$};% = \frac{\tau_1+\tau_2}{2\beta}$};
\draw[->] (-5,-4)--(-5,-3);
\draw (-5,-2.7) node {$T$};% = \frac{\tau_1-\tau_2}{2\beta}$};
\end{tikzpicture}~~~~~~~~~~~~
\raisebox{2mm}{\begin{tikzpicture}[thick,xscale=0.72,yscale=0.77, every node/.style={scale=0.9}]
\draw[very thick,fill=blue!05!white] (3,-2)--(3,2)--(-3,2)--(-3,-2)--cycle;
\draw[very thick,darkgreen,fill=white] (0,2)  ellipse (3cm and 1cm);
\draw[very thick,blue,fill=blue!10!white] (0,-2)  ellipse (3cm and 1cm);
\draw[very thick, <<->>] (-1.1,-2.45)--(1.1,-1.55);
\draw[very thick, <<->>] (1.1,-2.45)--(-1.1,-1.55);
\draw (0,0.68) node[color=blue] {$T=0$};
\draw (0,-3.35) node[color=blue] {$T=-1/4$};
\draw (0,1.4) node[color=darkgray] {$e^{\varphi_+}=\beta$};
\end{tikzpicture}\qquad}
\end{center}
  \caption{\small\textbf{Left:} After lightcone gauge fixing, the worldsheet theory reduces to a $\varphi_+$ Liouville CFT  living in the blue shaded region (equivalent to a M\"obius strip) in the $(X,T)$ plane. \textbf{Right:} The same  domain represented as a worldsheet cosmology with a crosscap initial condition at the physical time $T=-1/4$ and a $e^\varphi_+ = \beta$ boundary condition at $T=0$. } 
  \label{FIGURE crosscap}
\end{figure}
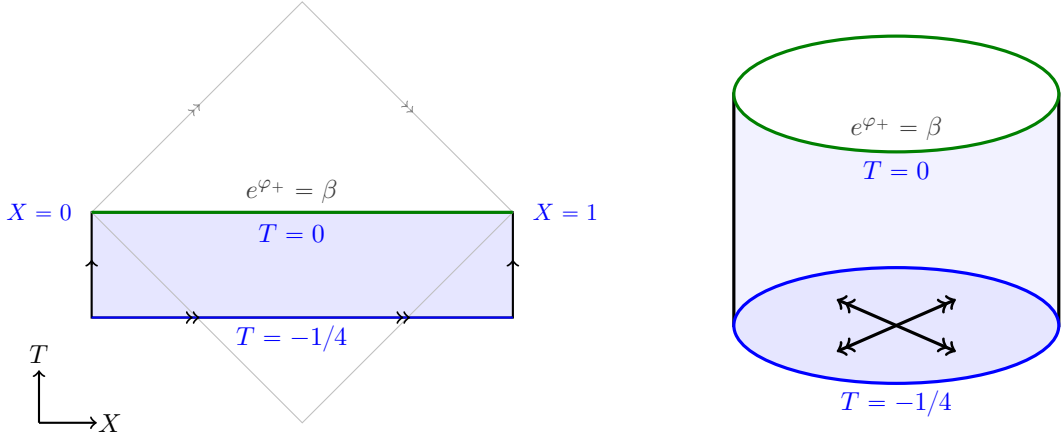

A standard set of conformally invariant boundary conditions are the FZZT boundary conditions \cite{Fateev:2000ik,Teschner:2000md}. They are labeled by a choice of boundary cosmological constant $\mu_{B_-}$ and  the boundary action 
\bea
\label{mubaction}
I_{\text{B}_-} = \frac{1}{2\pi \sfb^2}\int\d x \, \mu_{\text{B}_-} e^{\varphi_-} 
\eea 
Via the equations of motion of the combined bulk and boundary action, the boundary term \eqref{mubaction} fixes the boundary value of the ``extrinsic curvature'' to \cite{blommaert_wormholes_2025} $
%\label{kamu}
K_-\bigr|_{u=v} \equiv e^{-\varphi_-}\pi_-\bigr|_{u=v} = \mu_{\text{B}-}\,.$
Plugging in \eqref{liouville gsigma metric final}, we read off that  $\mu_{\text{B}_-}$ specifies the zero mode variable $\theta_-$ via 
\bea
\label{mubtheta}
\mu_{\text{B}_-} = \cos(\theta_-).
\eea
Notice that this parametrization implies that $\mu_{B_-}$ takes values over a finite interval between -1 and 1. The presence of this class of boundary conditions labeled by an angle $\theta_-$ is a novel feature of CLS in comparison with ordinary minimal string theory. We will elaborate on the relation \eqref{mubtheta}  in section \ref{sec: liouville crosscap amplitude}.
We will denote the boundary state \eqref{mubaction} by $|{\rm FZZT}(\mu_{\text{B}_-})\rangle$, or in case the more specific relation \eqref{mubtheta} holds, by $|{\rm FZZT}(\theta_-)\rangle$. As discussed in the previous section (see also section 4.2), the canonical dual $b_-$ to the FZZT parameter $\theta_-$ appears in the WDW constraint $b_+-b_- =0$. Hence, when restricted to the physical subspace of the combined  CLS worldsheet theory, we have the freedom to set $\theta_-=0$. This amounts to setting $\mu_{\text{B}_-} = 1$. We'll denote the corresponding boundary state by $|{\rm FZZT}(0)\rangle$.

The zero mode boundary condition of the $\varphi_+$ field can be summarized as
\bea
\oint  \d \U\, e^{\varphi_+(\U,\V)} \bigr|_{\U = \V} = 
\oint \d u\, e^{\varphi_+} \bigr|_{u = v}= \beta\, = \frac{b_+}{\sin(\theta_+)}\,.
\label{plusbc}
\eea
Note that this condition  is coordinate invariant. 
We will denote the conformal boundary state that implements the relation \eqref{plusbc} by $|e^{\varphi_+}\!=\beta\rangle$. Formally, this conformal boundary state is a Legendre transform of the FZZT boundary state
\bea
|e^{\varphi_+}\!=\beta\rangle = \int\! \d \mu_{\textB} \, e^{\beta \mu_{\textB}/\hbar }\, |{\rm FZZT}(\mu_{\textB})\rangle\,,
\eea
where the integral over $\mu_{\textB}$ runs over a suitably defined contour. We will describe this contour in more detail in section 5. 
The last equality  between the inverse temperature $\beta$ and the zero mode variables $b_+$ and $\theta_+$ in \eqref{plusbc} follows from plugging in the explicit classical solution \eqref{liouville gsigma metric final}.

With these two definitions in place, our proposed final state specifying the future boundary condition on the combined $\varphi_+$ and $\varphi_-$ theory is
\bea
|\psi_{\rm final}\rangle\is |e^{\varphi_+}\!=\beta\rangle \otimes |{\rm FZZT}(0)\rangle\,.
\eea
Both states in this tensor product are conformal boundary states that satisfy the standard Ishibashi conditions. From this one easily derives that the combined state lies in the physical CLS Hilbert space selected by the non-zero mode Virasoro conditions. %\footnote{Adding the Ishibashi conditions of the $\varphi_+$ and $\varphi_-$ theory gives $(L^+_{n}+L^-_{n})|\psi_{\rm final}\rangle = 
% (\tilde{L}^+_{-n}+\tilde{L}^-_{-n})|\psi_{\rm final}\rangle$. This
% shows that the total left-moving Virasoro conditions $L^+_{n}+L^-_{n} = 0$ are satisfied, modulo a spurious right-moving state of the form $(\tilde{L}^+_{-n}+\tilde{L}^-_{-n})|{\rm anything}\rangle$, and vice versa.} 
However, it does not solve the zero-mode Virasoro condition, or equivalently, the minisuperspace WDW constraint
\bea
\label{wdwconstrt}
  H_++ H_- = 0\,,\quad & &\ \ H_\pm = 2\pi \hbar \Bigr(L_0^\pm+\tilde{L}^\pm_0 - \frac{c_\pm\!-\nspc 1}{12}\,\Bigl)\,.
\label{hpmdef}
\eea
Instead, this condition will need to be imposed  inside the computation of the overlap, by means of an explicit integral over a lapse parameter $N$
\bea
       Z(\beta)\is \int_0^{\infty}\!\!\mathrm{d}N \,
        \bra{\psi_{\rm final}} e^{-N (H_++H_-)}\spc \ket{\psi_{\rm initial}}\,
        \label{3.00Zproposal}
\eea
For our choice of $|\psi_{\rm initial}\rangle$ (introduced below), the lapse variable $N$ parametrizes the complex structure of a M\"obius strip geometry, or equivalently, of the disk with a crosscap.

\subsection{Crosscap initial condition}\label{sect:3.3cross}
Next, we impose the crosscap identification  for $\varphi_+(\X,\T)$ as deduced from the $G\Sigma$ theory.  Due to the reflection of the (Euclidean) time direction, smoothness of the solution implies that the time derivative of $\varphi_+$ at the location of the crosscap identification vanishes. Looking at the explicit classical solutions  \eqref{liouville gsigma metric final} we deduce that the crosscap is located at the initial time instant $\T = -1/4$ (see figure \ref{FIGURE crosscap}). The crosscap boundary condition is thus:
\begin{equation} \label{crosscap condition}
    \frac{\de \varphi_+}{\de \T} \! \LL(\X, \T=-1/4 \RR) = 0 \, ,\quad \varphi_+\!\LL(\X, \T=-1/4 \RR) = \varphi_+\!\LL(\X+ 1/2 ,\T=-1/4 \RR) \,.
\end{equation}
 The antipodal crosscap identification $\X \sim \X + 1/2$ is automatically satisfied for the classical solution, since the lightcone gauge completely removed any $x$ dependence from \eqref{liouville gsigma metric final}. Solving the above constraint yields the following identity for the zero-mode variables in \eqref{liouville gsigma metric final}
\begin{equation} \label{crosscap condition solved}
    \cos(\theta_+\! +b_+/4) = 0
\end{equation}
This relation, and its utility for the computation of the partition function were  described in section 2.1. In particular, combined with \eqref{plusbc}, equation \eqref{crosscap condition solved} yields the familiar relation
\begin{equation}
    \beta(\theta_+) =  \frac{2\pi - 4\theta_+}{\sin(\theta_+)} 
\end{equation}
between the SYK inverse temperature and the spectral parameter $\theta_+$ labeling the DSSYK energy \cite{Maldacena:2016hyu}. 

We would again like to associate a conformal boundary state to the above classical initial condition. We will denote this initial state by
\bea
|\psi_{\rm initial}\rangle\; = \;|\text{C}\rangle \otimes |\T \! =\! - 1/4\rangle
\eea
Here $|\text{C}\rangle$ denotes the standard crosscap boundary state for the $\varphi_+$ theory - defined as the unique state that solves both the crosscap Ishibashi and Cardy conditions \cite{blumenhagen2009boundary}. We will describe this state in more detail in section 5.1. 
The role of the initial state $|\T\! =\! - 1/4\rangle$ of the $\varphi_-$ theory is that it sets the ``initial time instant'' as measured via the physical time variable $T_-$.  As we will motivate in section \ref{sec: hamiltonian analysis} and \ref{sec: liouville crosscap amplitude}, the natural candidate for such an initial state is the time evolved ZZ-boundary state\footnote{A heuristic motivation is that $\tau_-=0$ corresponds with a region where the Liouville field diverges, which we associated with $\ket{\text{FZZT}(0)}$. In the spirit of equation \eqref{2.28zz}, this is replaced by $\ket{\text{ZZ}}$ because we are considering an initial rather than a final state. This motivation is more clear after reading section \ref{sec: hamiltonian analysis}. Ultimately, the proof is our computation in section \ref{sec: liouville crosscap amplitude}.}
\bea
|\T\! =\! - 1/4\rangle = e^{\i H_-/4\hbar}\nspc\ket{\text{ZZ}}\,,
\eea
with $H_-$ defined in \eqref{hpmdef}.
This solves the same crosscap Ishibashi conditions as the standard crosscap state $|C\rangle$, but it does not solve the crosscap Cardy condition.

Let us return to our discussion of the lightcone gauge fixing and the mapping between CLS and the SYK collective field theory. We would like to claim that, upon choosing the Liouville lightcone gauge described above, the state $|\psi_{\rm final}\rangle$ effectively implements the $e^{g(u,v)}|_{u=v} = \beta$ boundary condition of the $G\Sigma$ theory. This looks surprising, since the $e^{g(u,v)}|_{u=v} = \beta$  boundary condition breaks the apparent conformal invariance of the bulk $G\Sigma$-theory, whereas the $\varphi_+$ boundary state $|e^{\varphi_+} = \beta\rangle$ preserves conformal invariance. How can both boundary conditions be physically the same? 

The full resolution of this apparent contradiction goes beyond the discussion in this paper. Instead, let us list some facts and considerations that we believe support our claim.

\vspace{-1mm}

\begin{enumerate}
\item{Our proposed identification $g = 2\varphi_+$ is not a local identification between the collective field $g$ and the Liouville field $\varphi_+$, but comes with the extra prescription that one should first pick the CLS lightcone gauge. So, $g(\U,\V)$ is in fact a highly non-local physical operator of the covariant CLS worldsheet CFT, which can be made to look local in the lightcone gauge.}
\item{The CLS worldsheet is a topological theory. Indeed, its physical Hilbert space (and phase space) are spanned by zero modes. One can introduce boundary condition changing operators (BCCOs) that interpolate between different FZZT boundary conditions. These BCCOs are given by diffeomorphism invariant integrals over the boundary. They naturally appear in pairs, corresponding to bi-local operators $e^{\alpha g(\U,\V)}$ in the collective field theory,\footnote{We are using intuitively the match between computing expectation values of such bulk correlators in $G\Sigma$ and boundary correlators in Liouville gravity \cite{Blommaert:2023wad,Blommaert:2024ydx}. Those are associated with open Verlinde lines i.e. BCCOs. We leave making such correspondence more precise to future work.} where the bulk location ($\U,\V$) is determined by the location of the BCCOs in the CLS lightcone gauge. The SYK energy is identified with the boundary cosmological constant, and the SYK time $\T=\frac{\U-\V}{2}$ is measured by means of a coordinate invariant integral along the CLS boundary \cite{Verlinde:2024zrh}.}
\item{$G\Sigma$ collective theory is also a topological theory with, for any given correlator, a finite dimensional phase space of classical solutions. One can insert local operators $e^{\alpha g(\U,\V)}$ but their correlation functions can be computed by means of topological chord rules that only depend on the ordering of operators \cite{Berkooz:2018jqr,Lin:2022rbf,Lin:2023trc}. The DSSYK rules that govern how to re-order operators match with the re-ordering of the BCCOs in the CLS: both are given by quantum 6j-symbols of SU(1,1)$_{\sfq}$}
\item{In section 5 we will perform a direct calculation of the CLS partition function associated with the geometry depicted in figure 2, with the covariant future boundary condition and crosscap initial conditions. As we have argued above, and will substantiate in more detail in sections \ref{sect4.3xcapstatefromsd} and \ref{sec: liouville crosscap amplitude}, the corresponding crosscap amplitude in CLS takes the form
\begin{equation}
        \boxed{\ \ Z(\beta)=\int_0^{\infty}\!\!\mathrm{d}N\, 
    \bra{e^{\varphi_+}\!\nspc=\beta\spc}\, e^{-N H_+} \ket{\text{C}}      \,
        \bra{\text{FZZT}(0)}e^{-N H_-}\spc e^{\i  H_-/4\hbar}\nspc\ket{\text{ZZ}}\,{}^{\strut}_{\strut}\ }\label{3.00Zproposal2}
\end{equation}
We will show that the result exactly matches with the DSSYK partition function $Z(\beta)$ in \eqref{2.13zsyk}. This calculation will be performed at the full quantum level, and the match holds at all orders in the coupling $\hbar = \lambda = 2\pi \sfb^2$.}
\end{enumerate}

 %%%%%%%%%%%
% SECTION %
%%%%%%%%%%%
\section{CLS quantum mechanics and Liouville time} \label{sec: hamiltonian analysis}
One of the main points of this work is that the collective field theory of DSSYK is a gauge-fixed version of the CLS. Namely, $g=2\varphi_+$ in the gauge where $\varphi_-$ is used to define a diff invariant time coordinate. So far we provided evidence for this at the classical level. In this section, we make this relation precise at the level of minisuperspace quantum mechanics. In section \ref{sect4.1minisuper} we review aspects of the minisuperspace quantization of sine dilaton gravity \cite{blommaert_wormholes_2025} in the language of the CLS. We also show how to translate the collective field boundary conditions \eqref{2.8gbc} and \eqref{2.10gbccrosscap} into minisuperspace Liouville variables. In section \ref{subsect:obs} we construct gauge-invariant observables by ``dressing to the time of $\varphi_-$'' and show that this results in ``physical'' $G\Sigma$ quantum mechanics. In section \ref{sect4.3xcapstatefromsd}, we relate the minisuperspace Liouville boundary conditions at the crosscap with our proposal \eqref{3.00Zproposal} for the appropriate crosscap boundary state in the full CLS string theory. To complete the proof of this equivalence, in section \ref{sec: liouville crosscap amplitude}, we will finally compute the crosscap amplitude with that boundary state, and exactly reproduce the DSSYK partition function \eqref{2.13zsyk}, as computed using the $G\Sigma$ path integral.

\subsection{Phase space recap}

We start with a brief summary of the minisuperspace description of sine dilaton gravity \cite{blommaert_wormholes_2025}, modified to CLS language. As a starting point, we recall the $x$-independent classical solutions \eqref{2.20liousol} 
%of Liouville gravity
\bea \label{lsolrep1}
    e^{\varphi_\pm}=\frac{b_\pm}{\sin(\theta_\pm \mp \i\smpc b_\pm \tau_\pm)}\, %,\quad\e^{\varphi_-}=\frac{b_-}{\sin(\theta_-+\i\smpc b_-\tau_-)}\,,\\[2.5mm]
\ \ & & \ \ \pi_\pm = b_\pm \cot(\theta_\pm\! \mp \i \smpc\sfbb_+\spc \tau_\pm)\,.
%\quad     \pi_- = b_-\cot(\theta_-\! + \i \smpc \sfbb_-\spc \tau_-)\,.
\label{lsolrep2}
\eea
Here we introduced two independent time variables $\tau_+$ and $\tau_-$, one for each sector. Each time flow is generated by the respective Hamiltonian
\bea
    H_+=\pi_+^2+e^{2\varphi_+}\,,\quad H_-=\pi_-^2+e^{2\varphi_-}\,.\label{4.6h}
\eea
On the classical solutions %\eqref{lsolrep1}-
\eqref{lsolrep2} they evaluate to constants\footnote{In Liouville CFT language, the $b_\pm$ variables label the Liouville momenta in the closed string channel, as one appreciates by comparing equations \eqref{4.6h} and \eqref{5.6}.
}
\bea
    H_+ = -b_+^2\,,\ & & \ H_-=-b_-^2\,.\label{4.6hnew}
\eea

Phase space is the space of all classical solutions evaluated at one given time instant. In total, for each Liouville sector, we may identify three natural canonically conjugate pairs of phase space variables: (i) the Liouville zero mode $\varphi_\pm$ and conjugate momentum $\pi_\pm$; (ii) the integration constants $\theta_\pm$ and their dual momenta $b_\pm$ which appear in the general classical solutions \eqref{lsolrep1}; and (iii) the Hamiltonian $H_\pm$ and its conjugate ``time'' variable
\begin{equation}
\tau_{\pm} =  \i\smpc {\theta_\pm}/{b_\pm}
\end{equation}
The symplectic form on the total minisuper phase space reads
\bea
    \omega \is \left\{
    \begin{array}{c}{\i\spc (\d \pi_+\!\wedge \d \varphi_+- \d \pi_- \wedge \d \varphi_-)\,},\\[1.5mm]
    {\i\spc (\d b_+\!\wedge \d \theta_+- \d b_-\!\wedge \d \theta_-)\,, }\\[1.5mm]
    { \d \ptau_+ \! \wedge dH_+ - \d \ptau_-\!\wedge dH_-\, ,}\end{array}\right.\qquad \hbar=2\pi \sfb^2\,.\label{omega}
\eea
Note that $(b_\pm,\theta_\pm)$ and $(H_\pm,\ptau_\pm)$ define proper action angle variables. Table below gives the set of canonical transformations that relates all these phase space variables. 
\def\darkblue{blue!85!black}
\def\ppp{\mbox{\normalsize $\tau$}}
\def\psinh{\mbox{\normalsize $\sinh$}}
\def\ptanh{\mbox{\normalsize $\tanh$}}
\def\sqrtH{\mbox{\small $\sqrt{H_\pm\!\!}$}}
\bigskip
\begin{table}[hbtp]
\qquad\qquad\quad
   \begin{minipage}[c]{0.35\textwidth}
   \textcolor{black}{\boxed{\parbox{3cm}{\vspace{-.5mm}
\begin{align}
\nonumber\ \ 
& \spc [\pi_\pm, \varphi_\pm]\spc =  {\hbar}\ \ \\[4.75mm]
& \, [\, b_\pm, \theta_\pm\spc ] \, = {\hbar} \ \ \nonumber\\[4.75mm]
&  [\spc H_\pm, \ptau_\pm]\spc =  \i \spc {\hbar}\ \ \nonumber
\end{align}\vspace{-3mm}}}}
   \end{minipage}\hspace{-12mm}\begin{minipage}[c]{0.65\textwidth}
\textcolor{black}{\boxed{
\parbox{8.5cm}{\vspace{-2.5mm}
\bea
-\, b_\pm^2 \ \, =& \pi_\pm^2 \spc + \spc e^{2\varphi_\pm} & = \ \, H_\pm \nonumber \\[2.75mm]
 \cos(\theta_\pm) \,  = & \  e^{-\varphi_\pm}\pi_\pm \   & = \ \cosh( \spc \sqrtH \;\spc \ptau_\pm )  
\nonumber \\[1.5mm]
e^{\varphi_\pm} \; = &\frac{\raisebox{2.5pt}{${b_\pm}$}}{\raisebox{-3pt}{$\sin(\theta_\pm)$}} \ &=\, \frac{\i\spc \sqrtH}{\mbox{\small $\psinh(\sqrt{H_\pm\!} \, \ppp_\pm)$}}\nonumber \\[-.5mm]
\pi_\pm \ \, = & b_\pm \cot(\theta_\pm) \ &= \, \frac{\i\spc \sqrtH}{\mbox{\small $\ptanh(\sqrt{H_\pm\!} \, \ppp_\pm)$}}
\nonumber\eea
\vspace{-4mm}}}}
   \end{minipage}\qquad\qquad
   \caption{The three conjugate pairs of phase space variables for CLS quantum mechanics and their relations.} 
\end{table}
\noindent
As already alluded to in section \ref{sect:actor2}, the physical phase space of CLS quantum mechanics is subject to a WDW Hamiltonian constraint
\begin{equation}
    H_\text{WDW}=H_+-H_-=0\,.\label{H_WDW4.5}
\end{equation}
In the sine dilaton language, this WDW constraint comes from variation with respect to the lapse $N$ \cite{Iliesiu:2020zld,Held:2024rmg,blommaert_wormholes_2025}; in CLS language it is the zero mode of the Virasoro constraint. 

\def\mX{Y}

%%%%%%%%%%%%%%%%%%%%%%%%%%
\subsection{Liouville time}\label{sect4.1minisuper}
In order to construct gauge-invariant observables, we may notice an analogy between the Hamiltonian constraint \eqref{H_WDW4.5} and the construction of CLPW \cite{Chandrasekaran:2022cip} in a different context. They consider a perturbative matter QFT in the dS static patch, in the presence of an ``observer'' point particle with quantized mass. That theory has a constraint $H_\text{WDW}=H_\text{mat}+q=0\,,$ 
with $q$ the observer's energy. Introducing a quantum mechanical time variable $[\ptau,q]=\frac \hbar \i$ this becomes
\begin{equation}
    -\i \hbar \frac{\d}{\d \tau}=H_\text{mat}\,.
\end{equation}
This is a Schrodinger equation that described evolution of the matter QFT with $-\ptau$ playing the role of the physical time coordinate. In other words, adding the observer is a way to get rid of the gravitational constraint and have ordinary physical time evolution ``with respect to the observer's time''. As we clarify further in section \ref{subsect:obs}, physical statements such as "the matter configuration is $\psi$ when the observer's clock reads $\T$" are implemented by considering the state $\ket{\psi}\otimes \ket{\ptau=-\T}$ or the associated projector.

One can construct gauge-invariant observables in minisuperspace CLS by exploiting the analogy of \eqref{H_WDW4.5} under the replacement
\begin{equation}
    H_\text{mat}\to H_+\,,\quad   -\i \hbar \frac{\d}{\d \tau}\to -H_-\,.
\end{equation}
This means that $\ptau_-$ becomes a physical time coordinate with respect to which the $\varphi_+$ system experiences ordinary quantum mechanical evolution. Indeed, minisuperspace CLS is subject by the WDW constraint \eqref{H_WDW4.5}.
\begin{equation}
    \i\hbar \frac{\d\ }{\d \ptau_-\!}\,=\, \pi_+^2+e^{2\varphi_+}\,.
\end{equation}
Introducing $g=2\varphi_+\,,$
we recover exactly the Schrodinger equation for the collective field theory description of DSSYK \eqref{2.4gaction}. Crucially, $\ptau_-$ has become the physical Lorentzian time coordinate
\begin{equation}
    \boxed{\ \i \hbar \spc \frac{\d\  }{\d \ptau_-\!\!\!}=\frac{\pi_g^2}{4}+e^{g}\ {}^{\strut}_{\strut}}\label{4.12physicaltime}
\end{equation}
Here, $\pi_g= \d g/\d \ptau_-$ is the canonical momentum dual to $g$. 

Comparing with the $G\Sigma$ solutions \eqref{2.29solbis} we see that one may interpret indeed $\ptau_-= \tau$. Recall that in the $G\Sigma$ theory $g$ satisfies the boundary condition $e^{g(0)/2}=\beta$ \eqref{2.8gbc}. This translates into the physical statement that $e^{2\varphi_+}$ equals $\beta$ when the clock $\ptau_-$ of the $\varphi_-$ field reads zero. Hence, the final state which implements this future boundary condition and the crosscap state boundary condition \eqref{2.10gbccrosscap} 
in CLS quantum mechanics take the form
\bea
\ket{\psi_{\mathrm{final}}} \is \ket{\ptau_-=0}\otimes \ket{e^{2\varphi_+}=\beta}\,.\label{4.14}\\[2.5mm]
\ket{\psi_{\rm initial}} \is  \ket{\ptau_-=\i /4}\otimes \ket{\pi_+=0}\,.\label{4.14b}
\eea
The condition $\pi_+=0$ dictates that the derivative of $g$ ($= 2\varphi_+$) vanishes, and the state $\ptau_-=\i/4$ says at which ``Liouville time'' this occurs. Applying this dictionary identifies the DSSYK partition function with the following transition amplitude in  CLS quantum mechanics
\bea
        Z(\beta)\is  \bra{\ptau_-\!=0}\otimes \bra{e^{\varphi_+}=\beta}
        \int
        \mathrm{d} N\,e^{-\mathrm{i}N H_\text{WDW}}\ket{\ptau_-=\i /4}\otimes\ket{\pi_+=0}
      \,.\label{4.15zbeta}
\eea
%The integral over $N$ is what remains of the full Lorentzian gravitational path integral in minisuperspace \cite{Marolf:1996gb,DiazDorronsoro:2017hti,Blommaert:2025bgd}, and the bra-and ket simply represent initial and final boundary conditions in the path integral. 
We will argue in section \ref{sect4.3xcapstatefromsd} that this equation is correct up to a slight refinement of the definition of the boundary state which is classically invisible. As mentioned around \eqref{2.31sdpathint}, the DSSYK partition function $Z(\beta)$ was recovered from the canonical quantization of sine dilaton gravity \cite{blommaert_wormholes_2025} using a peculiar initial state $\ket{\psi_\text{initial}}$. We will argue in section \ref{sect4.3xcapstatefromsd} that $\ket{\psi_\text{initial}}$ is gauge-equivalent to $\ket{\ptau_-=\i /4}\otimes \ket{\pi_+=0}$. 
%But first, let us make the relation between \eqref{4.15zbeta} and the $G\Sigma$ minisuperspace path integral more precise.

%%%%%%%%%%%%%%
\subsection{Relational observables}\label{subsect:obs}

The purpose of this section is to substantiate the discussion around \eqref{4.12physicaltime} and map amplitudes in CLS minisuperspace quantum mechanics to $G\Sigma$ quantum mechanics.

\def\fff{{\!\!\!\!\spc}^f\,}
\def\ff{{\!\!\!}^f\,}
Consider the following rewriting of a transition amplitude in CLS minisuperspace quantum gravity
\begin{align}
    &\bra{\ptau^{f}_{-}
    }\otimes \bra{\varphi^{}_{+}\fff}\int\mathrm{d}N\, e^{\mathrm{i}N (H_-- H_+)}\ket{\ptau^i_{-}}\otimes \ket{\varphi^i_{+}}\nonumber\\&\qquad\qquad\qquad\qquad=\int
    \mathrm{d}N\,\bra{\ptau_-\!=0}e^{-\mathrm{i} H_-(\ptau^{f}_{-}-\ptau_{-}^i)/\hbar-N H_-}\ket{\ptau_-\!=0}\bra{\varphi^f_{+}}e^{-\mathrm{i}NH_+}\ket{\varphi_{+}^i}\nonumber\\ &\qquad\qquad\qquad\qquad=\int
    \mathrm{d}N\int
    \mathrm{d} E_-\,e^{-\mathrm{i} E_-((\ptau^{f}_{-}-\ptau_{-}^i)/\hbar-N)}\bra{\varphi^{}_{+}\fff}e^{-\i N H_+}\ket{\varphi^i_{+}}\nonumber\\[1.5mm]
&\qquad\qquad\qquad\qquad=\bra{\varphi^{}_{+}\fff}e^{-\mathrm{i} H_+(\ptau^{f}_{-}-\ptau^i_{-})/\hbar}\lket{\varphi^i_{+}}\,.\label{33gaugefixed}
\end{align}
In the third line we introduced a completeness relation in the eigenbasis of $H_-$. Performing the integral over $E_-$ produces a delta function resulting in the expression on the last line. 

What has happened here? We started with a calculation in quantum gravity, where time evolution is redundant, as shown by the integral over the lapse. By evaluating the inner product of the $\varphi_-$ theory, we arrived at an ordinary QM transition amplitude in the $\varphi_+$ system with Hamiltonian $H_+$. The physical time evolution $(\ptau^{}_{-}\ff-\ptau_{-}^i)$ in the $\varphi_+$ system is fixed by the boundary states $\ket{\ptau^{}_{-}\ff}$ and $\ket{\ptau_{-}^{\spc i}}$ in the original quantum gravity transition amplitude. This proves the statement that considering in quantum gravity the state $\ket{\ptau_-}\otimes \ket{\varphi_+}$ physically corresponds with saying that ``the matter configuration is $\varphi_+$, when the clock reads $\ptau_-$''. As we already mentioned several times, this is how the $G\Sigma$ theory is embedded in CLS (or sine dilaton), with the identification $g=2\varphi_+$. These constructions are  sometimes also referred to as using ``quantum reference frames'' with $\ptau_+$ being the reference frame in question \cite{dewitt1967quantum,page1983evolution,DeVuyst:2024pop}.

As a particularly important example of \eqref{33gaugefixed}, consider our proposal for the CLS QM description of the $G\Sigma$ path integral \eqref{4.15zbeta}. Applying the above reduction procedure, and using $\hbar = 2\pi \sfb^2$, we find that the full expression \eqref{4.15zbeta} for the partition function $Z(\beta)$ collapses to 
\bea
   Z(\beta)\is \bra{\pi_g=0}e^{H_g/8\pi b^2} \ket{g=0}\,.\label{4.17relational}
\eea
We recognize on the right-hand side the minisuperspace Hamiltonian description of the $G\Sigma$ path integral \eqref{2.4gaction}. Notice in particular that the correct prefactor of $H_g$ in the exponential arises. In the next section \ref{sect4.3xcapstatefromsd}, we will directly evaluate the left-hand side of \eqref{4.17relational} using the sine dilaton path integral, and confirm its identification with $Z(\beta)$ as defined in \eqref{2.13zsyk}. In section \ref{sec: liouville crosscap amplitude} we will perform the full calculation in the covariant Liouville CFT formulation of CLS. A calculation of \eqref{4.17relational} was also presented in section \ref{sect2.1gsigma}.\footnote{As a short aside, let us explain how to construct gauge-invariant operators by dressing to $\ptau_-$. This construction mimics that of \cite{Chandrasekaran:2022cip}, to which we refer for details. Gauge invariant observables commute with $H_\text{WDW}=H_+-H_-$. One class of such invariants are functions $f(H_+)$, thus functions of the $G\Sigma$ Hamiltonian $H_g$. For the other invariant operators we recall the ``time evolved'' operators $
    e^{\varphi_+}(s)=e^{-\mathrm{i}H_+ s}e^{\varphi_+}e^{\mathrm{i}H_+s}$.
    %$={\i\spc \sqrt{H_+\!}}/{\text{sinh}(\sqrt{H_+\!}\,(\ptau_++s))}\,.$
The dressed version of this operator is simply \cite{Chandrasekaran:2022cip}
\begin{equation}
    e^{\varphi_+}(s+\ptau_-) =  \frac{\i \spc \sqrt{H_+\!}}{\text{sinh}(\sqrt{H_+\!}\,(\ptau_++s+\ptau_-))}
\end{equation}
This indeed commutes with $H_+-H_-$. One also checks that inserting such operators in quantum gravity amplitudes like the left-hand side of \eqref{33gaugefixed} corresponds with ``usual'' time evolved operator expectation values in the gauge-fixed $G\Sigma$ quantum mechanics (the right-hand side of \eqref{33gaugefixed}).}

Finally, a point that will be relevant in section \ref{sect4.3xcapstatefromsd} is that quantum gravity amplitudes such as \eqref{33gaugefixed} have the following gauge equivalence
\begin{equation}
    \bra{\ptau_-}\otimes \bra{\ptau_+}\int_{-\infty}^{+\infty}\!\!\!\mathrm{d} N\,e^{-\mathrm{i}N H_\text{WDW}}=\bra{\ptau_-+s} \otimes \bra{\ptau_+-s}\int_{-\infty}^{+\infty}\!\!\!\mathrm{d} N\,e^{-\mathrm{i}N H_\text{WDW}}\,.
\end{equation}
This follows from a simple shift of the integration variable $N$. What this means is that the combination $\ptau_++\ptau_-$ is physical, whereas the difference is a redundant (gauged) variable. This is because evolution with $H_\text{WDW}$ is gauged in gravity. In string theory language, using the relation between $p$ variables and FZZT momenta $s$ in equation  \eqref{2.22fzzt}, this becomes the statement that only the sum of FZZT momenta $s_++s_-$ physically effects amplitudes:
\begin{equation}
    \ket{\text{FZZT}(s_+)} \otimes \ket{\text{FZZT}(s_-)}\sim \ket{\text{FZZT}(s_++s)} \otimes \ket{\text{FZZT}(s_--s)}\,.\label{4.21gauge}
\end{equation}
In \cite{Collier:2024mlg_zz}, this redundancy was called ``Seiberg-Shih equivalence''.\footnote{This equation holds when interpreting the FZZT momenta as living on the full real axis, such that the $S_{p q}$ matrices are just single exponentials rather than $\cosh(4\pi p q)$.}

\def\sfq{\mathsf{q}}

%%%%%%%%%%%%%%%%
\subsection{Comparing boundary states}\label{sect4.3xcapstatefromsd}

 In the following two subsections, we describe the steps needed to lift our semi-classical minisuperspace formulation of CLS into an exact non-perturbative quantum treatment. To this end, we will connect our set-up with two existing approaches to the same problem. First we explain how our set up is related to the calculation in sine dilaton gravity \cite{blommaert_wormholes_2025} which exactly reproduces the DSSYK partition function. In the next subsection, we describe how our set-up relates an old beautiful and exact treatment of the Liouville zero modes by Faddeev and Volkov \cite{Faddeev:2008xy}. This approach clarifies the relation with skein relations of Verlinde lines and with the DSSYK chord rules. Our description will be somewhat schematic but will serve as a useful warm up for the exact CLS calculation in Section 5.

We wish to compute to partition function $Z(\beta)$ given in \eqref{4.15zbeta}. As explained above, and in section 2, at the classical level,  the initial crosscap state $\ket{\ptau_-\!=\i /4}\ket{\pi_+\!=0}$ imposes the boundary condition 
\begin{equation}
\label{initialcondition}
   \cos\bigl(\theta + b/4\bigr) = 0,
\end{equation}
with $\theta=\theta_+-\theta_-$. This equation implements the physical imprint of the crosscap state in the zero mode sector by setting $\partial_{\textsc{T}}\varphi_+=0$ at the ``initial time'' $\T=-1/4$. The idea of our computation is to turn this equation into an exact difference equation for the initial state wavefunction $\psi_{\rm initial}$, from which we can then obtain the partition function $Z(\beta)$, upon taking the overlap with the final state.

The gauge-redundancy \eqref{4.21gauge} allows us to consider equivalently the state 
$\ket{\ptau_-=0}\otimes\ket{\psi_\text{initial}}$, with $\ptau_-= \theta_-=0$ and where
\bea
\ket{\psi_\text{initial}} =
    \sum_{b/\hbar=-\infty}^{+\infty}\psi_\text{initial}(b_+) \ket{\spc b\spc }\, = \int_0^\pi \!d\theta \,   \psi_{\rm initial}(\theta) \ket{\theta} ,
\eea
is a judiciously constructed state which satisfies \eqref{initialcondition} as a difference equation - recalling that $b_+$ and $\theta$ are canonical conjugates. The discretization of $b_+$ arises because $\theta_+$ is periodic, as one can appreciate from \eqref{2.22fzzt}. Specifically, we have $\bra{\theta}b_+\rangle = \cos(b_+\theta/\hbar)$ and
\begin{equation}
    \psi_{\rm initial}(\theta)=\sum_{b/\hbar=-\infty}^{+\infty}\psi_\text{initial}(b)\cos(b\spc\theta/\hbar)
\end{equation}

At this point we can compare with the known calculation \cite{blommaert_wormholes_2025} in sine dilaton gravity minisuperspace quantum mechanics that reproduces the DSSYK partition function. 
The exact form of the wavefunctions $\psi_\text{initial}(\theta)$ and $\psi_\text{initial}(b_+)$ are given by \cite{berkooz2018chord,blommaert_wormholes_2025}
\bea
\psi_{\rm initial}(\theta)\!\is\! \big(e^{\pm 2\i\theta}; e^{-\hbar}\big)_\infty\, ,\\[2mm]
\psi_\text{initial}(b)\!\is \!\cos(\pi b/2\hbar)\cosh(b/4)e^{- b^2/8\hbar}\,.\label{psin}
\eea
In \cite{blommaert_wormholes_2025}, this initial state was viewed as providing a suitably constructed beginning-of-the-world brane. 
From both wave-function representations, one directly verifies that the state $\ket{\psi_\text{initial}}$ is the solution to the difference equation %\AB{An $i$ has kreeped in because clearly the correspondence requires $b\to \i b$.}
\begin{equation}
e^{\i b/2}\,\ket{\psi_{\text{initial}}}=e^{\i \pi-2\i \theta +\hbar}\,\ket{\psi_\text{initial}}\,.\label{4.29diffeq}
\end{equation}
This is classically equivalent to \eqref{initialcondition}! This  match between the minisuperspace QM of CLS and the DSSYK model supports our general approach and our proposed interpretation of the initial state as a crosscap boundary state. We will find more supporting evidence in the next sections.

\subsection{Skein algebra and q-harmonic oscillator}\label{sect:skein}

To lift the minisuperspace quantization of CLS into a robust rigorous framework, it is a valuable step to establish precise crosslinks with other approaches to quantizing Liouville gravity. Here we describe the correspondence between CLS quantum mechanics and the algebraic approach of \cite{Faddeev:2008xy} to quantizing the Liouville zero modes, and to some other well developed methods for quantizing the space of constant curvature metrics. 

In section \ref{sect:holonomies}, we introduced two useful sequences of CLS phase space variables
\bea
\label{yzrep}
Y_n= \frac{\sin\bigl((n+1)\smpc b/2 + \theta\bigr)}{\sin(b/2)}\,,\quad 
Z_k= \frac{\sin\bigl( b/2 +( k+1)\theta\bigr)}{\sin(\theta)}\,.
\eea
As explained there, these variables represent open line operators associated with the phase space ${\cal M}$ of classical solutions to the Liouville equations of motion. This space ${\cal M}$ is isomorphic to the space of 2D metrics with constant curvature, which in turn can be recast into the form of the moduli space of flat SU(1,1) gauge fields, modulo gauge transformations. In our setting, this moduli space is parametrized by two holonomies given in \eqref{lhmonodromy},  the holonomy $M$ around the cylinder; and a dual holonomy $L$ that relates the two light-cone coordinates $U_+$ and $V_+$. Geometrically, the dual holonomy $L$ connects some marked point of the future boundary\footnote{As we explain in the next section, the introduction of a marked point is needed for gauge fixing the residual translation invariance, or equivalently, for eliminating a corresponding zero mode in the CLS functional integral.} with its anti-podal point on the doubled geometry obtained by unfolding the cross-cap boundary condition.\footnote{To understand morally why this happens, notice that at the crosscap an identification between $U_+$ and $V_+$ is made, such that traveling from the asymptotic boundary through the crosscap, and back to the boundary, indeed results in picking up a holonomy related to $L$ as defined in \eqref{lrelation}.}

The relation between $Y_n$ and $Z_k$ and the holonomies $L$ and $M$ can be made explicit as follows. The holonomies $L$ and $M$ are $2\times 2$-matrices that naturally act on a two-component spinor. 
Pick any such a two-component spinor $s$. Then the following combinations are independent of the choice of $s$\footnote{We can think of $s$ as the value of a constant section $s$ of the flat SU(1,1) bundle at the marked point. The only restriction on $s$ is  that $s\!\wedge\! Ls$ and $s\! \wedge \! M s$ are non-zero. For simplicity, one could pick $s= \left(\begin{array}{c} \! 1\! \\[-.35mm] \! 1 \! \end{array} \right).$} \bea
\label{yzholo}
Y_n = \frac{s\! \wedge\! \ll \hh^{n+1}  s}{s\!\wedge\! \hh s} \,,\quad   Z_k = \frac{s\!\wedge\! \hh \ll^{k+1} s}{s\!\wedge\! \ll\spc s}\, .
\eea
Here $\wedge$ indicates the inner product defined with the insertion of a 2D $\epsilon$-symbol. Thus $s\! \wedge\! s=0$.

The classical variables $Y_n$ and $Z_k$ each satisfy an set of recursion relations \eqref{ysystem} and \eqref{zsystem}. These relations are readily verified from the classical expressions \eqref{yzrep}. In terms of the holonomy representation \eqref{yzholo} the classical recursion relations are known as Ptolemy relations. For our purpose, their main utility is that they have a known exact quantum manifestation in terms of the so-called skein relations familiar from knot theory. Introducing the notation  $
\sfq = e^{-\hbar/2}= e^{-\pi \sfb^2}$, the skein relations read
\bea
\label{yskein}
\mX_{n-1}\mX_{n+1}\! =  1 + \sfq\, \mX_n^2\,, \quad  
\mX_{n+1}\mX_{n-1}\! = 1 + \sfq^{-1}\smpc \mX_n^2\,,\quad\ \ 
\mX_{n+1}\mX_{n}\! = \sfq \spc \mX_{n}\mX_{n+1}\\[2mm]
Z_{k+1}Z_{k-1}\! =  1 + \sfq\, Z_k^2\,, \quad  
Z_{k-1}Z_{k+1}\! = 1 + \sfq^{-1}\smpc Z_k^2\,,\quad\ \ 
Z_k Z_{k+1}\! = \sfq \spc Z_{k+1}Z_k
\label{zskein}
\eea
For small $\hbar$, these relations are easily derived from the expressions \eqref{yzrep} and the canonical commutation relations $[b,\theta]=\hbar$. The non-perturbative relations \eqref{yskein}-\eqref{zskein} imply a normal ordering prescription that extends to all orders in $\hbar$.
Hence, the task of defining CLS quantum mechanics is now reduced to the task of finding an appropriate unitary Hilbert space representation of this algebra satisfying all the necessary physical requirements and positivity properties. 

For our discussion, the $Z_k$ algebra and in particular the elements $Z_{1}, Z_0$ and $Z_{-1}$ are most relevant. The algebra of these three elements 
%\bea 
%Z_{1}Z_{-1}\!\! \is \! 1 + \sfq \, Z_0^2\,,\quad\ \ \ Z_0 Z_1 = \sfq Z_1Z_0\nonumber\\[-2.75mm]\\[-2.75mm]\nonumber
%Z_{-1}Z_{1} \!\is\! 1 + \sfq^{-1} Z_0^2\,,\quad Z_0 Z_{-1} = \sfq^{-1} Z_{-1}Z_0 
%\eea
can be recast in the form of a $\sfq$-deformed harmonic oscillator algebra by considering the combination 
\bea
A= \frac{Z_{1}/Z_0}{\sqrt{1-\sfq^2}}\,, \quad\ A^\dag = \frac{Z_{-1}/Z_0}{\sqrt{1-\sfq^2}}\,, \quad \ [A,A^\dag]_{\sfq^2}= 1\,,
\eea
where $[A,B]_{\sfq^2} = AB - \sfq^2 BA$. This $\sfq$-oscillator algebra has a natural Hilbert space representation in terms of the number basis $\ket{n}$ of the $\sfq$-harmonic oscillator, on which $A^\dag, A$ and $Z_0$ act via
\bea
A^\dag \ket{n} = \ket{n+1}\,, \quad A\ket{n}  \! \is \! \spc[n]_{\sfq^2}\ket{n-1}\,, \quad Z_0 \ket{n} = - \sfq^{-n}\ket{n}\,.
\label{zact}
\eea
The $\sfq$-oscillator algebra and the $\sfq$-harmonic oscillator Hilbert space capture the chord rules and chord Hilbert space of the double scaled SYK model.
Our goal is to show that it also captures the quantum theory of the CLS on the disc with a crosscap. The evidence in support of this identification is three-fold (i) the one-to-one map between the operator algebras, (ii) the existence of a natural vacuum state annihilated by the lowering operator $A$, (iii) direct computation of the partition functions on both sides.

The key element of our proposed Liouville/SYK dictionary is the identification of the initial crosscap state $\ket{\psi_{\rm initial}}$ with the vacuum state $\ket{0}$ of the $\sfq$-harmonic oscillator. The vacuum condition $A\ket{0}=0$ is equivalent to the identity 
\bea
\label{zone}
Z_1 \ket{\psi_{\rm initial}} \!\is \! \frac{\sin\bigl( b/2 +2\theta\bigr)}{\sin(\theta)}\ket{\psi_{\rm initial}}  = 0\\[2mm]
Z_0 \ket{\psi_{\rm initial}} \!\is \! \frac{\sin\bigl( b/2 + \theta\bigr)}{\sin(\theta)}\ket{\psi_{\rm initial}}  = -\ket{\psi_{\rm initial}} 
\label{zzero}
\eea
The first identity is classically equivalent to the relation \eqref{initialcondition} derived from the condition that $\partial_T\varphi =0$ at the location of the crosscap. Indeed, classically, we can satisfy both relations by setting $b= 2\pi -4\theta.$ However, we can do better!

The crosscap boundary condition is a topological statement: it makes the worldsheet non-orientable and changes the homology class of one-cycles. The topology of the crosscap has non-trivial impact on the correlators of physical operators. Let us see if the property \eqref{zone} can be understood in this way. 

\begin{figure}[t]\centering
\quad a)\quad \begin{minipage}[c]{0.3\textwidth} \centering \begin{tikzpicture}[thick,xscale=0.7,yscale=.6, every node/.style={scale=0.9}]
\draw[very thick,fill=blue!05!white] (3,-2.5)--(3,2.5)--(-3,2.5)--(-3,-2.5)--cycle;
\draw[very thick,darkgreen,fill=white] (0,2.5)  ellipse (3cm and 1cm);
\draw[very thick,blue,fill=blue!10!white] (0,-2.5)  ellipse (3cm and 1cm);;
\draw[very thick,red] (-3,0)  arc (180:360:
3cm and 1cm);
\draw[very thick,red,dashed] (3,0)  arc (0:180:
3cm and 1cm);
\draw[very thick, <<->>] (-1.1,-2.95)--(1.1,-1.95);
\draw[very thick, <<->>] (1.1,-2.95)--(-1.1,-1.95);
\end{tikzpicture}\end{minipage}~~~~~~~~~~~~b)\begin{minipage}[c]{0.42\textwidth}\centering
    \includegraphics[height = 3cm]{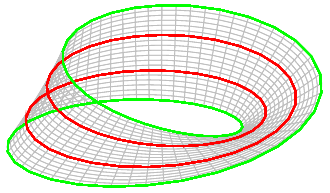}
  \end{minipage}\hfill\\[6mm]
 $\mbox{\quad c)}$ \begin{minipage}[c]{0.35\textwidth}\centering
    \includegraphics[height = 2.5cm]{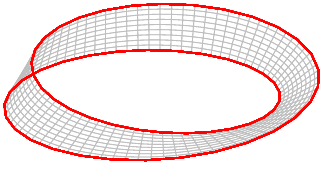}
  \end{minipage}~~~~~~~~d)\begin{minipage}[c]{0.4\textwidth}
  \centering
    \includegraphics[height = 2.9cm]{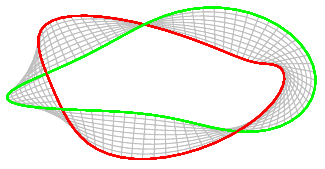}
  \end{minipage}
  \hfill
  \caption{The worldsheet a) has a 3D embedding as a M\"obius strip b). Cutting such a strip along the red one-cycle produces two linked strips: a smaller M\"obius strip c) and a doubly twisted strip d). This shows us that in the 3D embedding, the boundary (green) and the one cycle (red) both have self-linking number 2. }
  \label{FIGURE mobius-figure}
\end{figure}
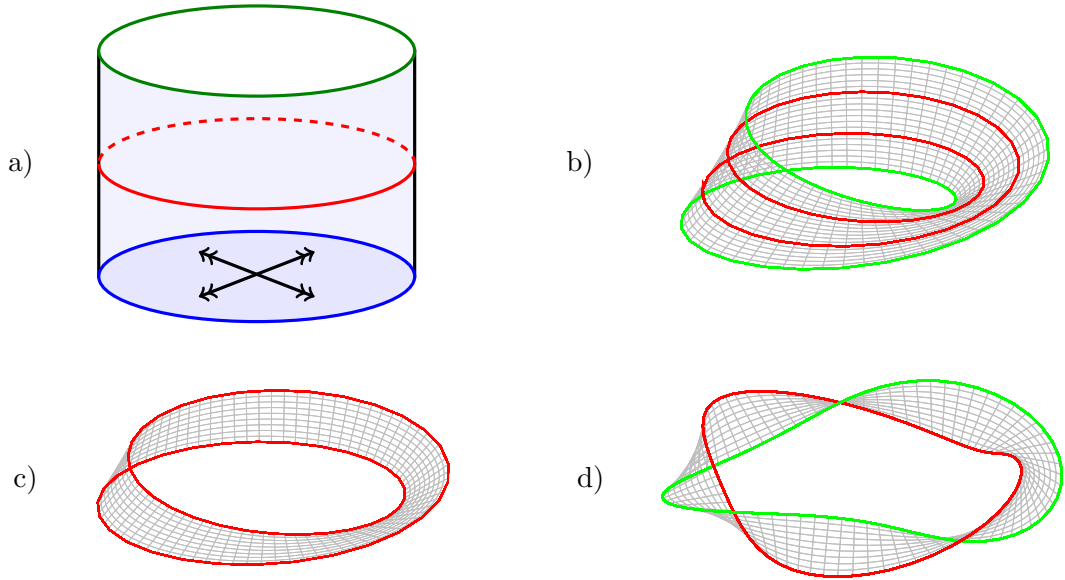

Consider the geometric representation \eqref{yzholo} of $Z_1, Z_0, Z_{-1}$, in terms of holonomy matrices $M$ and $L$. This geometric representation of the $Z$ operators suggests that the vacuum condition \eqref{zone} amounts to the topological statement that the holonomy operator associated to $Z_{1}$ is trivial
\bea
\label{trivial-holonomy}
\boxed{\ Z_1=0 \  \leftrightarrow \  M L^2 = \mathbb{1}\,.\Large\strut\,  }
\eea
This equation has the following topological explanation. Imagine cutting the worldsheet along the red one-cycle shown in figure \ref{FIGURE mobius-figure}a). In the 3D embedding of the worldsheet as a M\"obius strip, the result is two linked strips: a smaller M\"obius strip  and a doubly twisted strip shown in figure  \ref{FIGURE mobius-figure}c) and \ref{FIGURE mobius-figure}d). The doubly twisted strip makes clear that the red and green cycle have linking number 2. Since the red and green cycles are homologous, this tells us that each have self-linking number 2.

The matrix $M$ represents the holonomy around the cylinder. So, it is associated with the red one-cycle with self-linking number 2. As we will explain in section \ref{sect:threedee}, the second matrix $L$ can be thought of as the holonomy of a one-cycle in 3D that links the green boundary curve once. With this insight, the geometric meaning of the relation \eqref{trivial-holonomy} becomes clear: composing $M$ with $L^2$ amounts to undoing the self-linking by inverting the double twist, resulting in a trivial one-cycle. This explains the identification between $Z_1$ and the lowering operator $A$.  This topological characterization of the boundary condition $Z_1 = \sin(b/2+\theta) = 0$ will be useful in the discussion of the 3D lift of the CLS amplitude in section 6. There we will show that the condition \eqref{trivial-holonomy} implies that the relevant 3D geometry takes the form of 3D projective space $\mathbb{RP}^3$, also known as elliptic euclidean de Sitter spacetime.

Our other proposed identification \eqref{zact} between $Z_0$ and the number operator $\sfq^{-n}$ also finds direct confirmation via the following property of the $\sfq^2$-Hermite polynomials (see e.g. \cite{atakishiyeva_integral_2009}) 
\bea
\Bigl(e^{\i \theta}e^{\i \frac{\hbar}{2} \frac{\d\ }{\d\theta}}- 
e^{-\i \theta}e^{-\i \frac{\hbar}{2} \frac{\d\ }{\d \theta}}\Bigr) H_n(\cos\theta|\sfq^2) \is 2 \i \sfq^{-n}  \sin(\theta)H_n(\cos\theta|\sfq^2)
\eea 
The left-hand side is the quantum implementation of the operator $\sin(b/2+\theta)$. It would be interesting to understand the relation with the identification of DSSYK chord number $n$ as bulk length \cite{Blommaert:2024ydx}.

%%%%%%%%%%%%%%%%
% SECTION %
%%%%%%%%%%%
\section{The calculation} \label{sec: liouville crosscap amplitude}

\vspace{-1mm}

In this section, we demonstrate the following equivalence between a particular CLS crosscap amplitude and the partition function $Z(\beta)$ of DSSYK \eqref{2.13zsyk}\footnote{We leave ghosts implicit in these equations, see section \ref{subsec: string amplitude} for details.}
\bea
    Z(\beta)\is\int_0^{\i \infty}\!\! \mathrm{d} \tau \, \bra{e^{\varphi_+}\!=\beta}e^{2  \pi \i \tau L_0^+}  \ket{\text{C}}
    \, \bra{\text{FZZT(0)}}e^{2 \pi \i (\tau + \frac 1 2) L_0^-} 
 \ket{\text{ZZ}}
    \,.\label{5.1zbeta}
\eea
The object on the right is a closed string propagator between specific boundary states, one of which is a crosscap boundary state \cite{Hikida:2002bt,blumenhagen2009boundary}. Our goal is to compute this using string theory techniques \cite{Collier:2024kmo_base, Collier:2024kwt,Collier:2024mlg_zz}. In section \ref{sect5.1clsbackground}, we recall the relevant boundary states in Liouville CFT and the associated modular $S, T$ and $P$ matrices. In section \ref{subsec: string amplitude} we compute the string amplitude \eqref{5.1zbeta}. As usually \cite{mertens_liouville_2021,Kostov:2002uq}, the practical order to perform this calculation is to first consider an amplitude with an FZZT state
\bea
    Z(s)\is\int_0^{\i \infty}\!\! \mathrm{d} \tau \, \bra{\text{FZZT}(s)}e^{2 \pi \i \tau L_0^+}  \ket{\text{C}}
    \, \bra{\text{FZZT(0)}}e^{2 \pi \i  (\tau + \frac 1 2)  L_0^-} \ket{\text{ZZ}}
    \,.
    \label{5.2zfzzt}
\eea
From this, one then obtains the fixed $\varphi_+$ (or fixed length) amplitude \eqref{5.1zbeta} by a suitable inverse Laplace transform with respect to the boundary cosmological constant \cite{mertens_liouville_2021,Kostov:2002uq,Collier:2024mlg_zz}. An intermediate step in this calculation gives the ``spectral amplitude''
\begin{equation}
    {\mathcal{Z}}_\text{spec}(\theta)=\big( e^{\pm 2 i \theta}; e^{-2\pi \sfb^2}\big)_\infty=\frac{\d}{\d s}Z(s(\theta)+\i \sfb/2)-\frac{\d}{\d s}Z(s(\theta)-\i \sfb/2)\,,\quad 2\pi \sfb\spc s =\theta\,.\label{5.3zpec}
\end{equation}
This amplitude computes what should be interpreted as the spectral density in the dual matrix integral \cite{Collier:2024mlg_zz,Collier:2024lys_matrix}, but for a worldsheet that has a (particular) crosscap asides from the ``spectral'' boundary. Our main point is that this spectrum \eqref{5.3zpec} matches exactly with the spectral density of DSSYK in \eqref{2.13zsyk}. The relation is quite nontrivial, we believe.

Finally, in section \ref{sect5.3zsyk} we detail the relation with the DSSYK partition function at fixed $\beta$. It turns out that this is slightly different form the usual fixed length amplitude.
Before proceeding with the calculation of  \eqref{5.2zfzzt}, we mention a subtle point, which will be important in section \ref{sect5.3zsyk}. At the level of the action, FZZT boundary conditions corresponding to a Cardy state of $s$ can be implemented in two dual ways. The usual way \cite{Fateev:2000ik, Teschner:2000md} is to add a boundary cosmological constant to the action as in \eqref{2.17liouac}
\begin{equation}
    I=-\frac{\i}{2\pi \sfb^2}\mu_{\textB}\int\d x\,e^{\varphi_+}\,,\quad \mu_{\textB}=\cos(2\pi\sfb s)\,.\label{5.15mubaction}
\end{equation}
The second option is to exploit the $\sfb\to -1/\sfb$ symmetry of Liouville CFT \cite{Kostov:2002uq,Fateev:2000ik}, and instead add a boundary action with a ``dual'' cosmological constant to the action
\begin{equation} \label{action boundary dual cc}
    I = \frac{\i \sfb^2}{2\pi } \, \tilde{\mu}_{\textB} \int \d x \, e^{- \i \varphi_\pm/\sfb^2}   \, ,\quad \tilde{\mu}_{\textB}=\cosh(2\pi s/\sfb)\,.
\end{equation} 
These actions look different, but they project on the same momentum $s$ in the open string channel \cite{Seiberg:2003nm}.

%%%%%%%%%%%%%%%%
\subsection{CLS boundaries and branes}\label{sect5.1clsbackground}

We start by recalling some relevant aspects of the complex Liouville string \cite{Collier:2024kmo_base, Collier:2024kwt}. We will use variables that are naturally real in CLS. For instance, the central charges $c_+,c_-$ of the two Liouville fields given in \eqref{2.14cc} involve the real parameter $b$. Natural primary operators \cite{Collier:2024kwt} have conformal dimensions
\begin{equation}
    \alpha_+=\frac{Q_+}{2}+\i P_+\,,\quad h_+=\alpha_+(Q_+-\alpha_+)\,,\quad Q_+=\i \sfb +\frac{1}{\i \sfb}\,,
\end{equation}
where $P_{\pm}=e^{\pm \i \pi/4}p_{\pm}$ with real variables $p_\pm$. See also footnote \ref{fn6}. The weights are thus largely complex
\begin{equation}
    h_{\pm}= \frac{c_\pm-1}{24}\pm \i p_{\pm}^2\,.
\end{equation}

In the computation of closed string cylinder and crosscap amplitudes, such as the one that we are interested in \eqref{5.2zfzzt}, an important role is played by the modular $S, T$ and $P$ matrices - which encode the transformation of the Virasoro characters under modular transformations \cite{blumenhagen2009boundary}\cite{Brunner:2002em}
\bea
\chi_q(-1/\tau)\!\is\!\int \d q\,S_{pq}\,\chi_p(\tau)\,,\quad \chi_p(\tau+1)=T_p\,\chi_p(\tau)\,,\notag \\[-2.5mm]
\label{chimod}
\\[-2.5mm] \tilde{\chi}_q\bigl(-1/(4\tau)\bigr)\!\is \! \int \d p\,P_{q p}\,\tilde{\chi}_p(\tau)\,,\quad \tilde{\chi}_p(\tau)=T_p^{-1/2}\chi_p(\tau+1/2)\,.\notag
\eea
Here ,the characters of the Virasoro algebra are  defined as
\bea
    \chi^{\pm}_{p}(\tau)\! \is \! \Tr_{{\cal H}_{p_\pm}} \Bigl(e^{\i \tau H_{\pm}}\Bigr) = \frac{e^{\mp 2\pi \tau p^2 }}{\eta(\tau)} \,,\quad\  H_\pm=2\pi \Bigl(L_{0}^\pm -\frac{c_\pm}{24}\Bigr) \,.\notag \\[-2mm]\label{5.6}\\[-2mm]\notag
     \tilde\chi^{\pm}_{p}(\tau) \!\is\! \Tr_{{\cal H}_{p_\pm}} \Bigl(T_{p_\pm}^{-1/2} e^{\i H_{\pm}}\Bigr) = \frac{e^{\mp 2\pi \tau p^2 }}{\eta(\tau+1/2)}\,.
\eea
where ${\cal H}_{p_\pm}$ denotes the Virasoro module labeled by the Liouville momentum $p_\pm$ and
$\tilde{\chi}^\pm_p$ is the twisted Virasoro character of the M\"obius strip.\footnote{Equations for the Virasoro character associated with the identity $\mathbb{1}$ are obtained using the following property
$$
    \chi^+_\mathbb{1}(\tau)=\chi^+_{p=1/2\sfb+\i \sfb/2}(\tau)-\chi^+_{p=1/2\sfb-\i \sfb/2}(\tau)\,.
$$}
The relevant modular matrices for our purposes are
\bea
S_{p q} =    \cosh{(4 \pi p q)}  \,, & \quad & S_{\mathbb{1} p} =\sin(2\pi\sfb p) \sinh(2\pi p/\sfb) \, ,\, \notag \\[-2mm]\label{sptmatrix}\\[-2mm]\notag
T_{\pm p} = e^{\mp  2 \pi p^2} \, ,\qquad \,& \quad & P_{\mathbb{1}p}=  \cos(\pi\sfb p) \cosh(\pi p/b) \,.
\eea
For later reference, we note that the characters ${\chi}^\pm_p$ span dual linear spaces with inner product %(with $\tau = \i t$) 
\bea
\label{chinner}
\langle \chi_p\ket{\chi_q} & \equiv &  
\int_0^{\i \infty}\!\! \d\tau\,  
\chi^+_p(\tau){\chi}^-_q(\tau){\spc \eta(\tau)^2} =
\int_0^{\i \infty}\!\! \d \tau \,e^{-2\pi \tau(p^2-q^2)} = \frac{1}{p^2-q^2-\i \epsilon}
\eea
The $\eta(\tau)^2$ in the integral represents the ghost determinant. An analogous formula holds for $   \tilde\chi^{\pm}_{p}$. 

We are interested in boundary states of the CLS. The CFT boundary states with well-defined local boundary conditions are the Cardy states. In general, Cardy states are labeled by the primary fields of the CFT. They can be decomposed into a basis of Ishibashi boundary states which we denote by $\ishiket{p_\pm}$. Ishibashi states have the convenient property of diagonalizing the Hamiltonian evolution in the closed channel (where Cauchy slices are circles). In particular, following the normalization of \cite{Brunner:2002em}
\bea \label{CFT amplitude definitions}
    {\ishibra{p_{\pm}}} e^{\i \tau H_\pm} {\ishiket{q_\pm}}\is  \chi^\pm_p(\tau) \, \delta(p-q) \,,\nonumber\\[-2mm]
    \\[-2mm]\nonumber
    {\ishibra{p_\pm}} e^{\i\tau H_\pm} e^{\i \pi L^\pm_{0}} {\ishiket{q_\pm}}  \is  \chi^\pm_p(\tau+1/2) \, \delta(p-q)  \, . 
\eea
Notice in the second equality the presence of $e^{\i \pi L_0^{\pm}} $, which leads to the extra $+1/2$ in the character; its role is to turn the Ishibashi state ${\ishiket{p_\pm}}$ into a ``crosscap Ishibashi state'' ${\ishiket{p_\pm}}_{{}_\text{C}}= e^{\i \pi L_0^{\pm}} {\ishiket{p_\pm}}$\spc , which span a basis for Cardy states with crosscap boundary conditions \cite{blumenhagen2009boundary}
\bea
\label{ishicond}
\bigl(L^\pm_n-\tilde{L}^\pm_{-n}\bigr){\ishiket{p_\pm}}\,=\!\! & 0, &\, \notag \\[-2mm]\\[-2mm]\notag
\bigl(L^\pm_n-(-1)^n\tilde{L}^\pm_{-n}\bigr){\ishiket{p_\pm}}_{{\spc}_\text{C}}\!\! \is\! 0\,.
\eea

An important Cardy state is the ZZ state. This is defined by the property that a strip sandwiched between two ZZ boundary states has only the identity sector $\mathbb{1}$ propagating in the open channel \cite{Zamolodchikov:2001ah,Martinec:2003ka}. The corresponding boundary state is therefore
\begin{equation}
\label{zzstate}
    \ket{\text{ZZ}}=\int \d p\,\sqrt{S_{\mathbb{1}p}}\ishiket{p}\,.
\end{equation}
The generic Cardy states in CLS are FZZT states \cite{Fateev:2000ik, Teschner:2000md}. They decompose into Ishibashi states as
\begin{equation} \label{fzzt state}
    {\lket{\TT{FZZT}(s)} }  = \int \d p \, \frac{S_{s p}}{ \sqrt{S_{\mathbb{1} p} }} \, {\ishiket{p}} = \int \d p \, \frac{\cosh(4\pi p s)}{ \sqrt{\sin(2\pi\sfb p) \sinh(2 \pi p/b)}} \, {\ishiket{p}} \, .
\end{equation}
The FZZT states are the Cardy states associated with a Liouville primary with a momentum $s$ propagating in the open channel \cite{Fateev:2000ik, Teschner:2000md}.\footnote{More precisely, a strip sandwiched between an ZZ state and an FZZT(s) state supports the sector with Liouville momentum $s$ in the open string channel.} Hence ZZ and FZZT branes diagonalize the evolution on Cauchy slices stretching between the two boundaries of a cylinder. 

Another type of boundary state that features in our amplitude \eqref{5.2zfzzt} is the crosscap state $\ket{\text{C}_+}$. The decomposition of this state into crosscap Ishibashi states is determined by  comparing with the M\"obius strip amplitude computed in the open channel \cite{Hikida:2002bt,blumenhagen2009boundary}
\begin{equation}  \label{crosscap liouville}
    \ket{\text{C}_+}  =  \int \d p \, \frac{P_{\mathbb{1} p} T_{p}^{-1/2} }{ \sqrt{S_{\mathbb{1} p} }} \, e^{\i \pi L^+_{0}}{\ishiket{p}} = \int \d p \, \frac{\cos(\pi \sfb p)\cosh(\pi p/\sfb)}{\sqrt{\sin(2\pi \sfb p) \sinh(2\pi p/\sfb)}} e^{\pi p^2} \, e^{i \pi L^+_{0}}{\ishiket{p}}  \, .
\end{equation}
This boundary state $\ket{\text{C}_+}$ solves the Cardy conditions and thus describes a true physical crosscap.

The CLS amplitude \eqref{5.2zfzzt} we wish to evaluate is the overlap between the initial and final state
\bea \label{psi initial cft}
\ket{\psi_\text{initial}} \is \ket{\text{C}_+}\otimes e^{\i \pi L^-_{0}}\ket{\text{ZZ}} \\[3mm]
\ket{\psi_\text{final}(s)} \is \ket{\text{FZZT}(s)}\otimes \ket{\text{FZZT(0)}}
\eea
The initial state is the tensor product of the regular crosscap state $\ket{\text{C}_+}$ with a ``crosscap ZZ boundary state" $\ket{\text{ZZ}}_{{}_\text{C}} = e^{\i \pi L^-_{0}}\ket{\text{ZZ}}$. The latter state solves the crosscap reflecting boundary conditions \eqref{ishicond}.
However, it does not solve the crosscap Cardy conditions, so it is more appropriate to interpret it as a time-evolved ZZ boundary state. This has the key property that its overlap with the $\ket{\text{FZZT}(0)}$ state is essentially trivial. This property will become useful shortly.

%%%%%%%%%%%%%%
\subsection{Spectral crosscap amplitude} \label{subsec: string amplitude}
The goal of this section is to compute the spectral amplitude ${\mathcal{Z}}_\text{spec}(\theta)$ and match this with the DSSYK spectrum \eqref{5.3zpec}. We start by considering a CLS M\"obius strip-like amplitude \eqref{5.2zfzzt} computed as a cylinder diagram with an FZZT state on one side of the cylinder and the SYK crosscap state on the other side 
\bea
    Z(s)=\int_0^{\i \infty}\!\!\mathrm{d} \tau \,Z_\text{ghost}\bra{\text{FZZT(0)}}e^{2 \pi \i (\tau +1/2) L_0^-} \ket{\text{ZZ}}\bra{\text{FZZT}(s)}e^{2\pi \i \tau L_0^+}\ket{\text{C}_+}\,.\label{5.17}
\eea
The $\tau$ integral represents the modular integral of the closed string. We included the ghost contribution, which on the crosscap differs from the usual cylinder contribution by replacing $\tau\to \tau+1/2$:
\begin{equation}
    Z_\text{ghost}(\tau) = \eta(\tau+1/2)^2 \, .
\end{equation}

The contribution to the integrand in \eqref{5.17} due to the $\varphi_+$ Liouville theory evaluates to
\bea
   \bra{\text{FZZT}(s)}e^{\i \tau H_+}\ket{\text{C}_+}\is \int \d p \, \frac{S_{s p} P_{\mathbb{1}p}}{S_{\mathbb{1}p}} \,  T_p^{-1/2} \, \chi^+_p(\tau+1/2)\\[2mm]
  \is\int \d p \, \frac{\cosh(4 \pi p s)e^{\pi p^2}}{\sin(\pi\sfb p)\sinh(\pi p/\sfb)}  \, \chi^+_p(\tau+1/2)\, .
\label{zplusb}
\eea
Here we expanded $\ket{\text{C}_+}$ and $\ket{\text{FZZT}(s)}$ into (crosscap) Ishibashi states by using \eqref{crosscap liouville} and \eqref{fzzt state}. The inner product of Ishibashi states with the Hamiltonian propagator was then evaluated using \eqref{CFT amplitude definitions}. 

Similarly, the contribution due to the $\varphi_-$ Liouville theory evaluates to
\begin{equation}
\label{zminus}
    \bra{\text{FZZT(0)}}e^{\i (\tau+1/2) H_-}\ket{\text{ZZ}}=\int\d q\,\chi_q^-(\tau+1/2)\,.
\end{equation}
Notice that this contribution from the $\varphi_-$ sector is essentially trivial. This is of course built into our construction and perfectly compatible with our physical interpretation of the $\varphi_-$ sector as providing a clock, without any internal dynamics; other than that keeps track of the time-difference between the initial crosscap state and the final FZZT boundary condition. 

The contributions from the descendants encoded in the characters cancel with those from the ghosts, consistent with the fact that only primaries are physical states in the theory,
thus leaving
\begin{align}
    Z(s)&=\int\! \d q\int\! \d p \, \frac{\cosh(4 \pi p s)\, e^{\pi q^2}}{\sin(\pi \sfb p)\sinh(\pi p/\sfb)} \int_0^{\i \infty}\!\!\! \d \tau \,e^{-2\pi \tau (p^2-q^2)}\\[2.5mm]
    &  = \int \frac{\d p}{p} \,\frac{\cosh(4 \pi p s)\, e^{\pi p^2}}{\sin(\pi \sfb p)\sinh(\pi p/\sfb)}  \,.\label{5.21zfzzt}
\end{align}
The second equality uses the steps between equations (2.16) and (2.18) of \cite{Collier:2024mlg_zz}. One first performs the $\tau$ integral and then evaluates the $q$ integral by picking up a resulting pole at $p=q$. This simply projects on physical states propagating in the closed string channel.\footnote{The same result is obtained by integrating $\tau$ over the full real axis, corresponding with Lorentzian background metrics. It has been argued that this is indeed the correct definition of the lapse contour \cite{Marolf:1996gb,DiazDorronsoro:2017hti,Blommaert:2025bgd}, with $\tau$ replacing $N$ in \eqref{4.15zbeta}.}
 
Thus far, we have not detailed the precise $p$ and $q$ contours. The details can be found in \cite{Collier:2024mlg_zz}. For the purpose of computing the spectral amplitude ${\mathcal{Z}}_\text{spec}(\theta)$ we will not need to perform the final integral in \eqref{5.21zfzzt} but a simpler $p$ integral will suffice.

Consider now the ``fixed length'' amplitude \cite{mertens_liouville_2021,Kostov:2002uq,Collier:2024mlg_zz}. This is computed using the following standard procedure. One starts by viewing the FZZT amplitude as a function of the dual boundary cosmological constant by setting $s=\sfb\,\text{arccosh}(\tilde{\mu}_{\textB})/2\pi$ in equation \eqref{5.21zfzzt}. Secondly, we introduce a marked point on the dual FZZT brane to eliminate the zero mode due to unbroken translation invariance along the FZZT brane. Introducing the marking simply corresponds with taking a derivative of $Z(s(\tilde{\mu}_{\textB}))$ with respect to $\tilde{\mu}_{\textB}$, bringing down the marking operator $e^{-\i \varphi_+/\sfb^2}$ in \eqref{action boundary dual cc}. Hence, the marked FZZT amplitude is 
\begin{equation}
    Z_{\bullet}(s(\tilde{\mu}_{\textB}))\, \equiv \, \frac{\d\ \ }{\d \tilde{\mu}_{\textB\!\!\!}}\;Z(s(\tilde{\mu}_{\textB}))\,.
\end{equation}
The fixed length amplitude is defined as the inverse Laplace transform of the marked FZZT amplitude:
\begin{equation}
     Z(\spc\tilde{\ell}\,)=\int_{-\i \infty}^{+\i \infty}\!\!\d \tilde{\mu}_{\textB}\,e^{\tilde{\mu}_{\textB}\tilde{\ell}}\, Z_{\bullet}(s(\tilde{\mu}_{\textB}))\label{5.23zltilde}
\end{equation}
This inverse Laplace transform fixes the zero mode of $e^{-\i \varphi_+/\sfb^2}$ as a boundary condition, as we see from the boundary action \eqref{action boundary dual cc}. In the CLS it appears to be most natural to consider the above Legendre transform with respect to $\tilde{\mu}_{\textB}$.\footnote{Earlier work \cite{mertens_liouville_2021,Kostov:2002uq} considered a Legendre transform with respect to the normal boundary cosmological constant $\mu_{\textB}$.}  Indeed, as shown in \cite{Collier:2024mlg_zz}, this dual boundary cosmological constant $-\tilde{\mu}_{\textB}$ acquires an interpretation as an energy in the dual two-matrix model and $\tilde{\ell}$\, has the interpretation of a dual temperature. (See equations (2.47) and (2.61) in  \cite{Collier:2024mlg_zz}.) We proceed along these lines.

\begin{figure}[t]
    \centering
   \begin{tikzpicture}[thick,scale=0.6, every node/.style={scale=0.8}]
\draw[red, thick,->] (4.25,.25)--(.25,.25)--(.25,4.5) node[below right] {\Large $\gamma_+$};
\draw[darkgreen, thick,->] (-4.5,-.25)--(-.25,-.25)--(-.25,-4.25) node[above left] {\Large $\gamma_-$};
\draw[black, thick,->] (-4.75,0)--(4.75,0) ;
\draw[black, thick,->] (0,-4.5)--(0,4.75) ;
\draw[blue] (1,-1)  node {\large $\mathsf{x}$};
\draw[blue] (2,-2)  node {\large $\mathsf{x}$};
%\draw[blue] (4,-4)  node {\large $\mathsf{x}$};
\draw[blue] (3,-3)  node {\large $\mathsf{x}$};
\draw[blue] (-1,1)  node {\large $\mathsf{x}$};
\draw[blue] (-2,2)  node {\large $\mathsf{x}$};
%\draw[blue] (-4,4)  node {\large $\mathsf{x}$};
\draw[blue] (-3,3)  node {\large $\mathsf{x}$};
%\draw[blue] (-4.5,-4.5)  node {\large $\mathsf{x}$};
\draw[blue] (0,0)  node {\large $\mathsf{x}$};
\draw (-5.25,4)--(-4.5,4);
\draw (-5,4.5) node {\Large $p$};;
\draw (-4.5,4)--(-4.5,4.75);
\end{tikzpicture}~~~~~~~~~
   \begin{tikzpicture}[thick,scale=0.6, every node/.style={scale=0.8}]
\draw[red, thick,->] (-2.8,2.8) arc (-45:135:
.3cm and .3cm) ;
%\draw[red, thick,->] (-3.8,3.8) arc (-45:135:.3cm and .3cm) ;
\draw[red, thick,->] (-1.8,1.8) arc (-45:135:
.3cm and .3cm) ;
\draw[red, thick,->] (-.8,.8) arc (-45:135:
.3cm and .3cm) ;
\draw[red, thick,->] (.2,-.2) arc (-45:135:
.3cm and .3cm) ;
\draw[red, thick,->] (1.2,-1.2) arc (-45:135:
.3cm and .3cm) ;
\draw[red, thick,->] (2.2,-2.2) arc (-45:135:
.3cm and .3cm) ;
\draw[red, thick,->] (3.2,-3.2) arc (-45:135:
.3cm and .3cm) ;
%\draw[red, thick,->] (4.2,-4.2) arc (-45:135:.3cm and .3cm) ;
\draw[darkgreen, thick,->] (-3.2,3.2) arc (135:315:
.3cm and .3cm) ;
\draw[red, thick,->] (-2.8,2.8) arc (-45:135:
.3cm and .3cm) ;
\draw[darkgreen, thick,->] (-2.2,2.2) arc (135:315:
.3cm and .3cm) ;\draw[red, thick,->] (-2.8,2.8) arc (-45:135:
.3cm and .3cm) ;
%\draw[darkgreen, thick,->] (-4.2,4.2) arc (135:315:.3cm and .3cm) ;
\draw[red, thick,->] (-2.8,2.8) arc (-45:135:
.3cm and .3cm) ;
\draw[darkgreen, thick,->] (-1.2,1.2) arc (135:315:
.3cm and .3cm) ;\draw[red, thick,->] (-2.8,2.8) arc (-45:135:
.3cm and .3cm) ;
\draw[darkgreen, thick,->] (-.2,.2) arc (135:315:
.3cm and .3cm) ;\draw[red, thick,->] (-2.8,2.8) arc (-45:135:
.3cm and .3cm) ;
\draw[darkgreen, thick,->] (.8,-.8) arc (135:315:
.3cm and .3cm) ;\draw[red, thick,->] (-2.8,2.8) arc (-45:135:
.3cm and .3cm) ;
\draw[darkgreen, thick,->] (1.8,-1.8) arc (135:315:
.3cm and .3cm) ;
\draw[darkgreen, thick,->] (2.8,-2.8) arc (135:315:
.3cm and .3cm) ;\draw[red, thick,->] (-2.8,2.8) arc (-45:135:
.3cm and .3cm) ;
%\draw[darkgreen, thick,->] (3.8,-3.8) arc (135:315: .3cm and .3cm) ;
\draw[black, thick,->] (-4.5,0)--(4.25,0) ;
\draw[black, thick,->] (0,-4.25)--(0,4.5) ;
\draw[blue] (1,-1)  node {\large $\mathsf{x}$};
\draw[blue] (2,-2)  node {\large $\mathsf{x}$};
%\draw[blue] (4,-4)  node {\large $\mathsf{x}$};
\draw[blue] (3,-3)  node {\large $\mathsf{x}$};
\draw[blue] (-1,1)  node {\large $\mathsf{x}$};
\draw[blue] (-2,2)  node {\large $\mathsf{x}$};
%\draw[blue] (-4,4)  node {\large $\mathsf{x}$};
\draw[blue] (-3,3)  node {\large $\mathsf{x}$};
\draw[blue] (0,0)  node {\large $\mathsf{x}$};
\draw (-5.25,4)--(-4.5,4);
\draw (-5,4.5) node {\Large $p$};;
\draw (-4.5,4)--(-4.5,4.75);
\end{tikzpicture}
  \caption{The $p$ contour consists of the union $\gamma_+ \cup \gamma_-$ of the red and green contour drawn in the left panel. The blue \textcolor{blue}{$\mathsf{x}$}-marks indicate the location of the poles $p= \i \sfb n$. After a contour deformation the integral reduces to a discrete sum over residues. (We rotated the figure by 45$^\text{o}$ to match the convention of \cite{Collier:2024kwt_worldsheet}).}
  \label{FIGURE poles}
\end{figure}
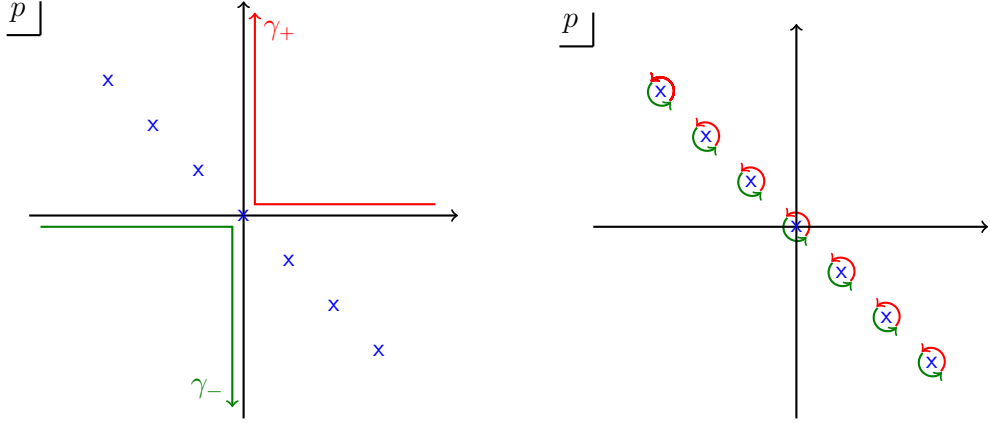

Following \cite{mertens_liouville_2021,Kostov:2002uq} and \cite{Collier:2024mlg_zz}, the integral in \eqref{5.23zltilde} can be rewritten by noticing that $Z_{\bullet}(s(\tilde{\mu}_{\textB}))$ is some meromorphic function of $\tilde{\mu}_{\textB}$ with a branch-cut between $-\infty$ and $-1$, which one can parameterize with a real parameter $\theta$:
\begin{equation}
    \tilde{\mu}_{\textB}=-\cosh(\theta/\sfb^2)\,.
\end{equation}
Deforming the $\dmub$ contour to wrap around this branchcut one obtains
\bea
    Z(\spc\tilde{\ell}\,) \! \is \! \int_0^\infty \d \theta\,e^{-\tilde{\ell}\cosh(\theta/\sfb^2)}\,{\mathcal{Z}}_\text{spec}(\theta)\,.\label{5.25zltilde}
\eea
Here ${\mathcal{Z}}_\text{spec}(\theta)$ has the interpretation as a spectral density in the two-matrix model dual of CLS \cite{Collier:2024lys_matrix}. It equals the discontinuity of $Z_{\bullet}(s(\tilde{\mu}_{\textB}))$ across the aforementioned branchcut, including a measure factor %$\d \tilde{\mu}_{\textB}/\d\theta$. %\textcolor{red}{My pitch for the name: something like $\rho_C$ (or $\rho_\TT{crosscap}$), to contrast with $\rho_\TT{disk} = \sin(\theta) \sinh(\theta/\sfb^2)$ 
%(though I know it is not a true crosscap so not sure) -DT}
\bea
\label{zdisc}
    {\mathcal{Z}}_\text{spec}(\theta)\! \is \! \frac{\d\tilde{\mu}_{\textB}}{\d \theta}\; \text{Disc}_{} Z_{\bullet}\bigl(s(\tilde{\mu}_{\textB})\bigr) = \frac{\d\tilde{\mu}_{\textB}}{\d \theta}\; \bigl(Z_{\bullet}\bigl(s(\tilde{\mu}_{\textB} +i\epsilon)\bigr)-Z_{\bullet}\bigl(s(\tilde{\mu}_{\textB} -i\epsilon\bigr)\bigr)
\eea
Plugging in the relation $\tilde{\mu}_{\textB}(s) = -\cosh(2\pi s/\sfb^2)$ yields
\bea
\label{5.26}
  {\mathcal{Z}}_\text{spec}(\theta)\!
  \is \! 
    \sinh(\theta/\sfb^2) \Bigl( Z_{\bullet}\bigl(s(\theta)\nspc+\nspc \i \sfb/2\bigr)-Z_{\bullet}\bigl(s(\theta)\nspc -\nspc \i \sfb/2\bigr)\Bigr)\,,\quad 2\pi \sfb s(\theta)=\theta\,.
\eea
Inserting into this equation the integral expression \eqref{5.21zfzzt} for the FZZT amplitude, we obtain the final result for the spectral amplitude
\begin{equation}
    %\boxed{\ 
    {\mathcal{Z}}_\text{spec}(\theta)=\int \d p\, \frac{\cosh( 2 p \spc\theta /\sfb )\cos(\pi \sfb p)}{\sinh(\pi p/\sfb)}e^{\pi p^2}=\sum_{n=-\infty}^{+\infty} (-1)^n e^{- \pi \sfb^2 n^2} \text{cosh}(\pi \sfb^2 n)  \cos(2 n \theta )\,.\; %}
    \label{5.27}
\end{equation}
Notice that the $p$-integral only has poles at $p=\i \sfb n$. These originate from the zeros of $S_{\mathbb{1}p}$ at $p_n=\i n \sfb/2$. The zeros at half-integers were canceled by zeros of $P_{\mathbb{1}p}$ in \eqref{crosscap liouville}. The zeros at $p=\i n /2\sfb$ were canceled by zeros in $P_{\mathbb{1}p}$, and by zeros arising from taking the discontinuity. In the second equality, we evaluate the $p$-integral by picking up the poles at $p=\i \sfb n$. This requires a choice of $p$ contour. The relevant $p$ contour indicated in figure \ref{FIGURE poles} is identical to the one used in equation (2.40) of \cite{Collier:2024kwt} in the CLS computation of the three holed sphere amplitude.\footnote{The integrand is odd under $p\to -p$. Then with an appropriate way of defining the contour close to $p=0$ we can double the contour in equation (2.40) of \cite{Collier:2024kwt} to one that goes around all the poles at $p=\i\sfb n$. Notice that the three holed sphere amplitude $A_{0,3}(p_1,p_2,p_3)$ is also closely related with some DSSYK quantity, the ``doubled two-point function'' in \cite{Narovlansky:2023lfz,Verlinde:2024zrh}.} 

Equation \eqref{5.27} can further be rewritten as
\begin{equation}
    {\mathcal{Z}}_\text{spec}(\theta)=\sum_{m=-\infty}^{+\infty}\psi_\text{initial}(m)\cos(m \theta)\,,\quad \psi_\text{initial}(m)=\cos(\pi m/2)\cosh(\pi \sfb^2 m/2)e^{-\pi \sfb^2 m^2/4}\,.\label{5.28}
\end{equation}
This wavefunction $\psi_\text{initial}(m)$, which we found by computing a CLS crosscap amplitude, exactly matches the wavefunction $\psi_\text{initial}(b_+=\hbar m)$ that we discussed in the context of the minisuperspace quantization of sine dilaton gravity in equation \eqref{4.29diffeq} (see equation (2.30) of \cite{blommaert_wormholes_2025}). The discretization of the closed channel momenta at $p=\i \sfb m/2$ is the CLS version of the fact that physical states in sine dilaton gravity have a discrete spectrum $b_+=\hbar m$ because of the periodicity of the dilaton $\Phi \sim \Phi+2\pi$ \cite{Blommaert:2023opb,Blommaert:2024whf,Blommaert:2024ydx, Blommaert_spheres:2025rgw, blommaert_wormholes_2025}.\footnote{As noted in the introduction, this shift invariance is unobstructed in our geometric situation: a disk $D_{\otimes}$ with a crosscap has zero Euler character. Hence the $\Phi R$ coupling between the $\Phi$ zero mode and $\chi(D_{\otimes}) = \int_{D_{\otimes}}\!\! R$ vanishes.}

One further summation identity \cite{Okuyama:2024eyf} leads to
\begin{equation}
    \boxed{\ {\mathcal{Z}}_\text{spec}(\theta)=\big(e^{\pm 2\i\theta};e^{-2\pi \sfb^2}\big)_\infty\,{\Large {\strut}}}\label{zspec}
\end{equation} 
With the identification \eqref{1.1}, we see that the spectral amplitude ${\mathcal{Z}}_\text{spec}(\theta)$ matches  the spectrum \eqref{2.13zsyk} of DSSYK. This match between the spectrum in CLS on a cylinder ending on the crosscap boundary state $e^{\i \pi L_{0-}}\ket{\text{ZZ}}\otimes \ket{\text{C}_+}$ and the DSSYK spectrum is our main result. It strongly supports our physical considerations in sections \ref{sec: lightcone quantization} and section \ref{sec: hamiltonian analysis}.

%%%%%%%%%%%%%%%%%%%%%%%
\subsection{SYK partition function from CLS}\label{sect5.3zsyk}

\vspace{-1mm}

Inserting the spectrum \eqref{zspec} into equation \eqref{5.25zltilde} we obtain the ``fixed length'' CLS crosscap amplitude
\bea
    Z(\tilde{\ell}\spc)\!\is\!\int_0^\infty \d \theta\,\big(e^{\pm 2\i\theta};e^{-2\pi \sfb^2}\big)_\infty\, e^{-\tilde{\ell}\cosh(\theta/\sfb^2)}\,.\label{5.30}
\eea
The reason we place ``fixed length'' between quotes is that the quantity held fixed is the dual length $\tilde{\ell}$, the zero mode of $e^{-\i \varphi_+/\sfb^2}$. We compare this amplitude with the DSSYK partition function \eqref{2.13zsyk}
\bea
     Z(\beta) \! \is \! \int_0^\pi \d\theta  \, \big( e^{\pm 2 i \theta}; e^{-2\pi \sfb^2}\big)_\infty \,e^{\frac{\beta}{2\pi \sfb^2}\cos(\theta)}\,.\label{5.31zsyk}
\eea
We see that the DSSYK partition function and the CLS amplitude \eqref{5.30} feature the exact same spectral density but differ by their choice of Hamiltonian function $E(\theta)$. 

The difference in the choice of Hamiltonians has a natural interpretation in terms of the matrix integral formulation of CLS of \cite{Collier:2024lys_matrix}. 
The SYK partition function $Z(\beta)$ corresponds to inserting, instead of the Boltzman factor $\Tr( e^{-\beta M_1}) = \Tr\bigl( e^{-\beta {H_{\rm CLS}}}\bigr)$ defined in terms of the first matrix $M_1$ with eigenvalues $E_{\rm CLS}$,
the Boltzman factor $\Tr( e^{-\beta M_2}) = \Tr\bigl(e^{-\beta {H_{\rm SYK}}}\bigr)$ of the other matrix $M_2$ with eigenvalues $E_{\rm SYK} = \text{cos}(\sfb^2\arccosh(E_{\rm CLS}))$ inside the matrix model expectation value. 

This point should not come as a surprise. Indeed, the very same feature arises when comparing the disk amplitude in sine dilaton gravity \cite{Blommaert:2024whf} with the disk in CLS \cite{Collier:2024mlg_zz}, as we summarize in appendix \ref{sect:appendix}. The two calculations agree on the spectrum, but use a different definition of the Boltzmann factor. In sine dilaton gravity, the ADM energy is $\cos(\theta)$, whereas in CLS the natural energy function associated with the theta parameter is $\cosh(\theta/\sfb^2)$. Both prescriptions provide natural disk amplitudes. Therefore, we also accept that \eqref{5.30} and \eqref{5.31zsyk} are natural crosscap amplitudes.\footnote{A natural interpretation of the DSSYK partition function is that it represents the CLS partition function with fixed actual length $\ell$ (conjugate to $\mu_\text{B}$, as one sees from \eqref{5.15mubaction}). This is almost correct. The FZZT boundaries $s=\theta/2\pi \sfb\pm \i \sfb/2$ which contribute to the spectral amplitude in \eqref{5.26} correspond with $\mu_\text{B}=\cos(\theta\pm \pi \sfb^2)$. The action can be trusted in the classical limit $\sfb\to 0$, in which case these boundary cosmological constants indeed reduce to $\cos(\theta)$. Thus, classically the DSSYK partition function \eqref{5.31zsyk} indeed computes a fixed length CLS crosscap amplitude. However, quantum mechanically (for finite $\sfb$) this is not exactly true.}

Asides from the distinction in Hamiltonians, there is a difference between the CLS amplitude \eqref{5.30} and the DSSYK partition function \eqref{5.31zsyk} in the integration range of $\theta$. Comparing the expressions at zero length
%\bea 
%\label{shift symmetry infinity}
   % \!\is\!\int_0^\infty \d \theta\,\big(e^{\pm 2\i\theta};e^{-2\pi \sfb^2}\big)_\infty=\infty \int_0^\pi \d \theta\,\big(e^{\pm 2\i\theta};e^{-2\pi \sfb^2}\big)_\infty\,.
%\eea
we observe that $Z(\tilde{\ell}=0)= \infty \; Z(\tilde{\ell}=0)$. So the naive CLS amplitude equals the DSSYK answer but overcounts it by a factor of $\infty$.  Physically, this divergence arises because we are considering the theory on a M\"obius strip, which has Euler character zero. One can check that this means that the sine dilaton action \eqref{2.19sdaction} (and hence also the CLS action) is invariant under shifts of the dilaton field $\Phi\to \Phi+2\pi$. This shift symmetry leads to a divergence, which can be eliminated by gauging the shift symmetry  \cite{Blommaert:2024whf}.

A possibly related fun fact is that in the study of the CLS partition function in \cite{Collier:2024mlg_zz} it was independently argued that (for the disk amplitude) the $\theta$ integral over the matrix model spectrum should effectively be cut off at $\theta=\pi$, because the spectral density (of the dual matrix whose eigenvalues are equal to $\tilde{\mu}_{B}$) turns negative at this point. The authors of \cite{Collier:2024mlg_zz} argue this truncation is justified by taking into account non-perturbative effects due to ZZ D-instantons. 

%%%%%%%%%%%%%%%%%%%

\subsection{Open string channel}
\label{SECTION open channel}

\vspace{-1mm}

\par In the previous sections, we have seen how the DSSYK spectral density can be computed from a closed string perspective as the CLS amplitude on a disk with FZZT and crosscap boundary conditions. To gain more insight, it is instructive to look at the calculation in the open string channel. In the open channel %provided that the Cardy condition is satisfied for both Liouville theories 
the amplitude turns into a trace over a physical Hilbert space of CLS states propagating along an open strip. This representation of the partition function is therefore expected to give direct insight into how the DSSYK Hilbert space can be cast in the form of an open CLS Hilbert space. 

We start by rewriting the $c_+$ partition function \eqref{zplus} in the open string channel. Applying the modular $P$ transformation \eqref{chimod}-\eqref{sptmatrix} to the twisted character (here we use $\tau = \i t$ for convenience)
\bea
    Z_+(s; t) \is \bra{\text{FZZT}(s)}e^{- \pi t H_+}\ket{\text{C}_+}   = \int \d p \,  \frac{ S_{sp}P_{\mathbb{1}p}}{S_{\mathbb{1}p}} \, \tilde{\chi}^+_p(\i t)  \notag \\[-2mm]\\[-2mm]\notag
   \is  \int\! \d q \int\! \d p \,   \frac{  S_{sp} P_{\mathbb{1}p} P_{pq}}{S_{\mathbb{1}p}} \; \tilde{\chi}^+_{q} \!\LL(\i/4 t \RR)  \, . 
\eea
Next, we  perform the procedure explained in the previous section of deforming the $p$ integral so that it picks up the discontinuity.\footnote{Here we ignore the marking and assume that the inner product \eqref{chinner} defined by the integral over the modulus can be replaced by $\langle \tilde{\chi}_p\ket{\tilde{\chi}_q} = \delta(p-q)$. 
This simplification can be justified a posteriori by the calculations in the previous section.} 
In practice the two steps amount to making the following replacement  $S_{sp}/S_{\mathbb{1}p} \to \TT{Disc}[\de_{s} S_{sp}/S_{\mathbb{1}p}] =
p \sin(2 \pi \sfb p) S_{sp}/S_{\mathbb{1}p}$ in the above formula. 
To evaluate the resulting $p$ integral, we proceed as before and pick up the residues of the poles at the discrete momenta $p = \i \sfb n$. Finally, we apply the Poisson resummation formula to the sum over $n$ converting the sum over discrete momenta in the closed string channel into a sum over a discrete set of Liouville momenta $q_n$ in the open string channel
\bea
\label{zplusopen}
Z_+(s; t) \, = \,  \de_\theta \sum_{{n\in \mathbb{Z},\pm} }\tilde{\chi}^+_{q_{n}\pm \frac{\i\sfb} {2} }\!\LL(\i/4t\RR), \ \ & & \ \ q_{n} = \frac{2\theta -\pi + 2\pi n}{2\pi \sfb}, \ \theta = 2\pi \sfb\spc s
\eea

We observe that --- setting aside the derivative coming from the marking --- the crosscap state indeed satisfies the Cardy property that each Virasoro module in the sum occurs with unit coefficient. Note, however, that the open string momentum $q$ is not simply given by the FZZT quantum number $s = \theta/2\pi \sfb$. Instead, we see that  when paired up with the crosscap state, a given FZZT(s) brane produces a partition function spanned by an infinite set of momenta $q_n\pm \frac{i\sfb}{2}$.
The $c_+$ partition function can thus be written as the trace
\bea \label{open trace c_+}
     Z_+(s; t) \!\is \! \de_\theta \Tr_{\ca{H}_+(\theta)} 
     \LL[ \Omega \, 
     e^{- 2 \pi t'\spc (L_0-\frac{c_+}{24})}  \RR]    \, 
     , \quad\quad  t' = \frac{1}{4 t}  \, 
\eea
over a Hilbert space ${\ca{H}_+(\theta)}$ spanned by the conformal families of primaries with momenta $q_n\pm \frac{\i\sfb}{2}$
\bea
    \ca{H}_+(\theta) \!\is \! \bigoplus_{{n\in \mathbb{Z}, \pm}} \, \ca{H}^{\text{Vir}}_+ \bigl(q_n \pm \mbox{\large $\frac{i \sfb}{2}$} \bigr)
    % I am writing with the 2\pi b factor because that's the actual Liouville momentum, or is it better to just drop the 2\pi b? (note that there is also an additional factor of 2 that you might not expect, as 2\pi b p' \propto 2\theta, not \theta
    % \bigl({\theta\!-\! \frac \pi 2\! +\! \pi n\! +\!  \mbox{\large $\frac{\i \hbar}{2}$}\sigma}\bigr)
    %{\bigoplus_{\sigma = \pm} \bigoplus_\TT{desc.} \LL\vert \frac{1}{\pi \sfb}  \LL(\theta - \frac{\pi}{2} + \pi n + \sigma \frac{\i \pi \sfb^2}{2} \RR) \RR\rangle_+
\eea
In \eqref{open trace c_+}, the modulus $t'$ is the length of the M\"obius strip in the open channel and $\Omega$ is the Dehn-twist operator that implements the twisted periodicity condition of the M\"obius strip. Specifically, $\Omega$
exchanges the left and right moving excitations while leaving the zero modes invariant \cite{Brunner:2002em}.\footnote{Formally, one may write $\Omega = T^{-1/2} e^{\i \pi L^+_0} =  T^{1/2} e^{-\i \pi L^+_0}$. The operator $e^{\i \pi L^+_0}$ exchanges all left and right movers but also acts on the zero mode. The latter action is canceled by the $T^{-1/2}$.} 

\par We can perform a similar closed-to-open channel transformation on the $\varphi_-$ theory given in  \eqref{zminus}). Performing the modular transformation \eqref{chimod}, we can write $Z_-(t)$ in the open channel as follows:
\bea
    Z_-(t) \is  \bra{\text{FZZT}(0)}e^{\i \pi (\i t + 1/2) L_0^-}\ket{\text{ZZ}}=\int\d p\, T_p^{-1/2} \tilde{\chi}_p^-(\i t) \notag \\[-2mm]\label{zminusopen}\\[-2mm]\notag
\is     \int\!\d q \int \! \d p \,  T^{-1/2}_p \spc P_{pq} \, \tilde{\chi}_{q}^-(\i/4t) 
%=  \! \int\! \d q \, T_{q}^{1/2} \tilde{\chi}_{q}^-(\i t') 
= \int\! \d q \, \chi_{q}^-(\i t'\! -\!  1/2)
\,.
\eea
In the last step we performed the simple gaussian integral $\int \d p \,   T^{-1/2}_p P_{pq}  = T^{1/2}_{q}$ and made the replacement $T_{q}^{1/2} \tilde{\chi}_{q}^-(\i t') = \chi_{q}^-(\i t'\! -\!  1/2)$. We can thus formally write the $c_-$ partition function as the trace 
over the Hilbert space spanned by a uniform continuous spectrum of  Virasoro modules
\bea
    Z_-(\spc t'\spc ) \!\is \!  \Tr_{\ca{H}_-}\!\LL[e^{-\i \pi L_0} e^{-2\pi t' (L_0 - \frac{c_-}{24})} \RR] \,, \qquad \quad 
    \ca{H}_- = \bigoplus_{q \in \mathbb{R}} \; \ca{H}^{\text{Vir}}_-(q)  \, .
\eea
Comparing with the $\varphi_+$ partition function, we  note several differences. First, the $\Omega$ operator in \eqref{open trace c_+} is replaced by the operator $e^{-\i \pi L^-_0}$. The operator $e^{-\i \pi L^-_0}$ also implements a M\"obius strip twist, but unlike $\Omega$, it also acts on the zero modes and thus
differs from $\Omega$ by a factor of $ T^{1/2}$.

\begin{figure}[t]
\begin{center}
\begin{tikzpicture}[thick,xscale=0.7,yscale=.7, every node/.style={scale=0.9}]
\draw[very thick,fill=blue!05!white] (3,-2)--(3,2)--(-3,2)--(-3,-2)--cycle;
\draw[very thick,darkgreen,fill=white] (0,2)  ellipse (3cm and 1cm);
\draw[very thick,blue,fill=blue!10!white] (0,-2)  ellipse (3cm and 1cm);
\draw[very thick, <<->>] (-1.1,-2.45)--(1.1,-1.55);
\draw[very thick, <<->>] (1.1,-2.45)--(-1.1,-1.55);
\draw (0,2) node[color=black] {$\ket{\mu_{\textB}(s)}
=\ket{E=\frac{\cos(\theta)}{\hbar}}$};
\draw[thick,->] (-1,-.25)  arc (240:300:
2cm and .85cm) node[midway,above]{$s = \frac{\spc \theta}{2\pi \sfb}$};
\draw (0,-4) node[color=blue] {\large $\bra{\spc s\!=\!\frac{ \smpc\theta}{2\pi \sfb}\spc}\psi_{\rm initial}\rangle$};
\end{tikzpicture}~~~~~~~~~~~~~~~~~~~~~~\begin{tikzpicture}[thick,xscale=0.7,yscale=.7, every node/.style={scale=0.9}]
\draw[very thick,fill=blue!05!white] (3,-2)--(3,2)--(-3,2)--(-3,-2)--cycle;
\draw[very thick,darkgreen,fill=white] (0,2)  ellipse (3cm and 1cm);
\draw[very thick,blue,fill=blue!10!white] (0,-2)  ellipse (3cm and 1cm);
\draw[thick, ->] (-.25,-.6)--(-.25,.6) node[midway,right] {$p_n = \frac{\i\spc\sfb n}{2}$};
\draw[very thick, <<->>] (-1.1,-2.45)--(1.1,-1.55);
\draw[very thick, <<->>] (1.1,-2.45)--(-1.1,-1.55);
\draw (0,2) node[color=black] {$\ishiket{p_n=\frac{\i \sfb n}{2}}\!
=\ket{b_+\!= \hbar n}$};
\draw (0,-4) node[color=blue] {\large $\langle\!\nspc\bra{\spc p_n\!=\! \mbox{\large $\frac{\i \sfb n}{2}$}\,} \psi_{\rm initial}\rangle$};
%\draw (0,-3.5) node[color=black] {\large cross-cap};
\end{tikzpicture}
\vspace{-4mm}
\end{center}
\caption{The FZZT brane (left) projects on a  Liouville momentum $s=\frac{\theta}{2\pi \sfb}$ in the open string channel while the Ishibashi state (right) projects on a Liouville momentum $p=\frac{i\sfb n}{2}$ in the closed string channel. The novel property of CLS is that $p$ runs only over discrete values, while $\theta$ runs over a finite interval $[0,\pi]$.  }
\end{figure}
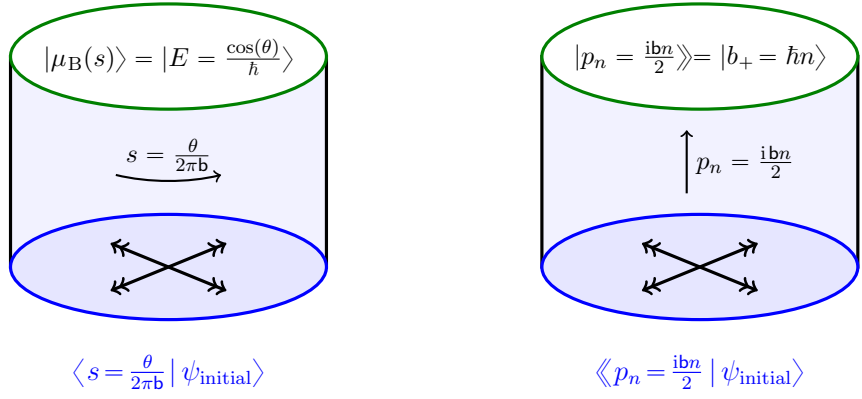

To obtain the spectral partition function $\ca{Z}_{\rm spec}(\theta)$ of the total CLS worldsheet theory, we can proceed in two equivalent ways. First, we can directly multiply the two partition functions 
\eqref{zplusopen} and \eqref{zminusopen}, include the ghost partition function and perform the integral over the modulus $t'$. This yields\footnote{Following the stepwise treatment of the previous subsection, including the marking, one encounters integrals over the open channel modulus $t'$ of the form $\int_0^{\infty} \frac{\d {t}}{{t}} \, 
\eta^2(\i {t} + \frac 12) \de_p 
\hat{\chi}^+_p({t}) \chi^-_q(\i {t} - \frac 12) = 2p e^{- \pi q^2}\int_0^\infty\! dt \, e^{-t(p^2-q^2) } = \frac{2p e^{- \pi q^2}}{p^2-q^2-\i \epsilon} = e^{- \pi q^2}\delta_{pq}
$. Here the delta function anticipates the subsequent contour prescription for the integrals over $p$ and $q$.}
\bea
\ca{Z}_{\rm spec}(\theta) \is \int\!\frac{dt'}{t'} \; {Z_-(\spc {t'}\spc )Z_+(\spc s; t' \spc)} \eta\bigl(\i t' + \mbox{\large $\frac 12$}\bigr)^2\notag  \\[-2mm]
\label{ztot}
\\[-2mm]\notag \is  \sin(\theta) \sum_{n = -\infty}^{+\infty} (-1)^n e^{- \frac{1}{\pi \sfb^2} \LL(\theta - \frac{\pi}{2} + \pi n \RR)^2} = \bigl(e^{\pm 2 \i \theta}; e^{-2\pi \sfb^2}\bigr)_\infty  
\eea
Again, we obtain a match with the DSSYK partition function. However, we can gain more physical insight by viewing  the total partition sum \eqref{ztot} as the trace of the physical state space of the CLS 
\bea
\ca{H}_{\rm phys}(\theta) = \bigl(\ca{H}_+(\theta) \otimes \ca{H}_- \otimes \ca{H}_{bc}\bigr)_{\rm BRST} 
\eea
Here the subscript BRST means implementing the BRST cohomology, i.e.  performing the projecting onto the space spanned by states annihilated by the BRST charge, modulo BRST exacts states. In each Virasoro module all descendant states are projected out or spurious and only the ground states with Liouville momente $q_n \pm \frac{\i \sfb}2$ survive in the cohomology. 

After imposing the on-shell condition that the total conformal dimension must add up to $h_+\!+h_-\! = \! 1$, we are  left with a physical Hilbert space $\ca{H}_\TT{phys}(\theta)$ spanned by all diagonal primary states of the form
\bea
    % \!\is\! 
    %\bigoplus_{{n\in \mathbb{Z},\pm}} \,
   \ca{H}_\text{phys}(\theta) \is \TT{Span}\biggl\{\; \ket{q_n \pm \mbox{\large $\frac{\i \sfb}2$}\spc}_{\!+} \!\otimes \ket{ q_n \pm \mbox{\large $\frac{ \i \sfb}2$}\spc}_{\nspc -} \, ,\ \ q_{n} = \frac{2\theta -\pi + 2\pi n}{2\pi \sfb},\ \  n\in \Z ,\sigma=\pm \,\biggr\}
\eea
The total CLS crosscap partition function \eqref{ztot} can then be represented twisted trace defined on the physical open string Hilbert space 
\bea
    \ca{Z}_{\rm spec}(\theta) = \Tr_{\ca{H}_\TT{phys}(\theta)} \LL[\Omega \otimes e^{- \i \pi L^-_0} \RR] \is \sum_{n\in \mathbb{Z},\pm} e^{- \pi (q_n \pm \frac{\i\sfb}{2})^2}.
    %= \sum_{n\in \mathbb{Z},\pm}e^{- \frac{1}{\pi \sfb^2} \LL(\theta - \frac{\pi}{2} + \pi n + \sigma \frac{\i \pi \sfb^2}{2} \RR)^2} \\
   %sin(\theta) \sum_{n = -\infty}^{+\infty} (-1)^n e^{- \frac{1}{\pi \sfb^2} \LL(\theta - \frac{\pi}{2} + \pi n \RR)^2} 
   %= \bigl(e^{\pm 2 \i \theta}; q\bigr)_\infty  \, .
\eea
We observe that the physical Hilbert space and total partition function $Z(\theta)$ are both invariant under the discrete shift symmetry
\bea
    \ca{H}_\TT{phys}(\theta+\pi) = \ca{H}_\TT{phys}(\theta)\quad & & \quad \ca{Z}_{\rm spec}(\theta+\pi) = \ca{Z}(\theta).
\eea
This shift symmetry is already a property of the $\varphi_+$ partition function $Z_+(\theta,t)$, as seen from \eqref{zplusopen}, and it is preserved in the total partition function thanks to the fact that the $\varphi_-$ partition function \eqref{zminusopen} is invariant under arbitrary shifts in $q$.
Finally, collecting all sectors into a single Hilbert space
\begin{equation}
    \ca{H}_{\rm phys} = \bigoplus_{\theta \in [0,\pi]} \ca{H}_\TT{phys}(\theta) \, ,
\end{equation}
we can rewrite the zero-temperature DSSYK partition function as the trace
%\footnote{This representation of the SYK partition function as the twisted trace over an open CLS Hilbert space is somewhat reminiscent of the Schur-SYK duality of \cite{Gaiotto_Teschner:2024osr, Gaiotto_verlinde:2024kze} matching the SYK partition function with the supersymmetric Schur half-index $Z(\beta = 0) = \Tr_{\TT{Op}^\de} (-1)^{2R} q^{j_3 + R}$ of 4D $\CMcal{N}=2$ pure SU(2) super-Yang--Mills on the half-space $\mathbb{R}_+ \times \mathbb{R}^3$ \cite{Gaiotto_verlinde:2024kze}.}
\bea
    Z(\beta = 0)\!\is\!\int_0^\pi \! \d \theta \,  
    %\Tr_{\ca{H}_\TT{BRST}(\theta)} \LL[\Omega_+ \otimes e^{- \i \pi L^-_0} \RR] = 
    \ca{Z}_{\rm spec}(\theta) \, =\, \Tr_{\ca{H}_{\rm phys}} \LL[\Omega \otimes e^{- \i \pi L^-_0} \RR] 
\eea
%\begin{equation}   Z(\beta = 0) = \Tr_{\TT{Op}^\de} (-1)^{2R} q^{j_3 + R} \, . \end{equation}
%\textcolor{red}{Some kind of 4D/2D duality? This whole last paragraph on Schur-SYK might increase confusion so it can be removed}

\smallskip

% %%%%%%%%%%%
% % SECTION %
% %%%%%%%%%%%
\section{Towards the SYK hologram of dS$_3$}\label{sect:threedee}

\vspace{-1mm}

We have shown that the DSSYK spectral density ${\mathcal{Z}}_\text{spec}(\theta)$ can be represented as the partition function of a covariant complex Liouville gravity theory on a disk with FZZT and crosscap boundary conditions. The crosscap boundary state looks asymmetric between the two Liouville theories
\bea
\label{xzcap}
    \ket{\spc \psi_\text{initial}} \is \ket{\text{C}_+}\otimes e^{\i \frac{\pi}{2} L_0^-} \ket{\text{ZZ}}\,.
\eea
It would be desirable to have a better geometric understanding of this state. One thing which is clear, is that the state is \emph{not} implementing the ordinary crosscap boundary condition on the CLS worldsheet theory. The latter would be described by the tensor product $\ket{\text{C}_+}\spc \otimes\spc \ket{\text{C}_-}$. In appendix \ref{app:B}, we compare the two types of states and show that they are indeed different.

This raises the question: is there some covariant geometric requirement (other than  reproducing the SYK partition function) that naturally selects the  crosscap boundary condition \eqref{xzcap}? We will  argue that the answer is ``Yes". Our argument  relies on an identification of the spectral amplitude with some line observable in dS$_3$ quantum gravity. This involves the identification between the space of Virasoro conformal blocks with the Hilbert space of a 3D gravity theory  and the interpretation of the 2D  gravity path integral as providing the 3D inner product in 3D de Sitter gravity, as made manifest by its first order formulation in terms of SL(2,$\mathbb{C}$) theory \cite{Verlinde:1989ua,Collier:2023fwi, Collier:2025lux}. The contents of this section are meant to be suggestive and correct, but not fully self-contained. For the reader interested in more detailed accounts of the general ideas outlined below, we refer to original literature on the subject \cite{Witten:1988hc, witten1989quantum, Witten:1989ip, Verlinde:1989ua, Dimofte:2011gm, Dimofte:2011py, Gaiotto_Teschner:2024osr, Cotler:2019nbi, Collier:2023fwi, Gaiotto_verlinde:2024kze,Collier:2025lux}.  

\def\II{\text{I\!\!\spc\smpc I}}

At a more concrete level, the following (empirical) quantitative correspondence will be helpful for uncovering the 3D lift of our geometrical set-up. The initial wavefunction and spectral partition function\footnote{Here $\ishibra{m}$ and $\bra{\text{FZZT}(\theta)}$ are short-hand for $\ishibra{m}\otimes \bra{\text{ZZ}}$ and $\bra{\text{FZZT}(\theta)}\otimes \bra{\text{ZZ}}$. } $\psi_{\rm initial}(m) = \ishibra{m}\psi_\text{initial}\rangle$ and $\ca{Z}_\text{spec}(\theta) = \bra{\text{FZZT}(\theta)}\psi_\text{initial}\rangle$ can be written as a sum of matrix elements of the modular matrices $P$, $T$ and $S$ as follows
\bea
\label{psi_initial_3d}
\psi_\text{initial}(m) \is \,P_{\mathbb{1}m} T_{m}^{-1/2} = \int \! \d p\, S_{\mathbb{1}p} T^2_p S_{pm} = (ST^2S)_{\mathbb{1}m},\\[2mm]
\ca{Z}_\text{spec}(\theta)\, =\hspace{-2mm} & & \hspace{-5mm} \sum_{m\in 2\mathbb{Z}} \psi_\text{initial}(m)\cos(m\theta)  = 
\sum_{m\in 2\mathbb{Z}}  P_{\mathbb{1}m} T_{m}^{-1/2} S_{m \theta}.
\label{zspec_3d}
\eea
Both these expressions can be recognized as expectation values of a line operator in a 3D SL(2,$\mathbb{C}$) CS theory defined on a (suitable) 3D geometry \cite{Verlinde:2024znh, Verlinde:2024zrh, Gaiotto_verlinde:2024kze}. Below, we will summarize this 3D interpretation and its relation to the 2D set up developed in the previous sections, and, along the way, we will try to give a natural 3D  motivation for our special choice of boundary state \eqref{xzcap}.

\begin{figure}[t]
\centering
\begin{tikzpicture}[rotate=90,xscale=.7,yscale=-.7] 
\tikzset{
    partial ellipse/.style args={#1:#2:#3}{
        insert path={+ (#1:#3) arc (#1:#2:#3)}
    }
}   
\draw[black,fill=darkgreen!12!cyan!10!white] (0,0) [partial ellipse=-180:180:2.5cm and 1.9cm];
\draw[black] (0,0) [partial ellipse=-180:180:2.5cm and 1.9cm];
\draw[white,fill=white] (0,0) [partial ellipse=-180:180:1cm and .5cm];
\draw[black] (0,0) [partial ellipse=-0:-180:1cm and .5cm];
\draw[black] (0,-.1) [partial ellipse=-0:180:1.2cm and .6cm];
 \draw[red, very thick] [partial ellipse=0:360:1.75cm and -1.39cm];
%\draw[\darkred,thick,decorate, decoration={snake, segment length=1.5mm, amplitude=.1mm}](0.15,-1.25) [partial ellipse=13:353:.3cm and .7cm];
\draw[blue] (-2,-2.5) node {\mbox{Torus I}};
%\draw[blue,very thick] (-1.95,-1.5) -- (-1.95,-.4);
 \draw[red] (-.5,-.8) node {\mbox{\footnotesize $\hat{W}_m$}};
\draw[white,fill=white] (0,2.5) [partial ellipse=-180:180:.1cm and .05cm];
%\draw[thick] (-1.65,0) circle (.35mm);
%\draw[thick] (1.65,0) circle (.35mm);
\end{tikzpicture}\hspace{-3mm}
 \begin{tikzpicture}[scale=.4,baseline={([yshift=-2.3cm]current bounding box.center)}]
        \draw[<-, thick,darkgray] (60:1) arc (60:120:4) node[midway, below] {paste} node[midway, above] {\mbox{\small  ${T}_\text{B}^{2} = \nspc {S}\spc T_A^2 S$\Large$\strut$}} ;
\end{tikzpicture}\hspace{-7mm}\raisebox{1mm}{\begin{tikzpicture}[rotate=90,xscale=.7,yscale=-.7] 
\tikzset{
    partial ellipse/.style args={#1:#2:#3}{
        insert path={+ (#1:#3) arc (#1:#2:#3)}
    }
}   
\draw[black,fill=darkgreen!12!cyan!10!white] (0,0) [partial ellipse=-180:180:2.5cm and 1.9cm];
\draw[black] (0,0) [partial ellipse=-180:180:2.5cm and 1.9cm];
\draw[white,fill=white] (0,0) [partial ellipse=-180:180:1cm and .5cm];
\draw[black] (0,0) [partial ellipse=-0:-180:1cm and .5cm];
\draw[black] (0,-.1) [partial ellipse=-0:180:1.2cm and .6cm];
%\draw [thick,dotted,partial ellipse=-180:180:2.5cm and -2.5cm];
\draw[\darkgreen,thick,dotted](0.15,-1.2) [partial ellipse=13:353:.3cm and .7cm];
\draw[blue] (-2,-2.5) node {\mbox{Torus II}};
%\draw[blue,very thick] (-1.95,-1.5) -- (-1.95,-.4);
%\draw[thick] (-1.65,0) circle (.35mm);
%\draw[thick] (1.65,0) circle (.35mm);
\end{tikzpicture}}~~~~\raisebox{1.7cm}{\Large $=$}~~~~
\raisebox{-.3cm}{\begin{tikzpicture}[rotate=0,xscale=.75,yscale=-.75] 
\tikzset{
    partial ellipse/.style args={#1:#2:#3}{
        insert path={+ (#1:#3) arc (#1:#2:#3)}
    }
}
 \draw[blue!07!white] [fill=darkgreen!12!cyan!10!white] [partial ellipse=0:360:3.2cm and 3.2cm];
% \draw[blue!07!white] [fill=darkgreen!12!cyan!10!white] (-3,-2.5) -- (-3,2.5) --(3,2.5) -- (3,-2.5) -- cycle; 
\draw[red, very thick] [partial ellipse=0:360:1.3cm and -1.8cm];
\draw[\darkblue] (-2,-1.6) node {\mbox{\large $\mathbb{RP}^3$}};
 \draw[red] (-.55,-.4) node {\mbox{\footnotesize $\hat{W}_m$}};
\end{tikzpicture}}
\bigskip
\caption{The wave function $\psi_{\rm initial}(m)$ can be identified with the expectation value of a circular Wilson line in the spin $m$ represention of an SU(1,1) CS theory on 3D projective space $\mathbb{RP}^3$. This 3D geometry can be obtained by gluing two tori via the identification $T_B^2 = S \spc T_A^2 S$.} 
\label{figure-glue-one}
\end{figure}
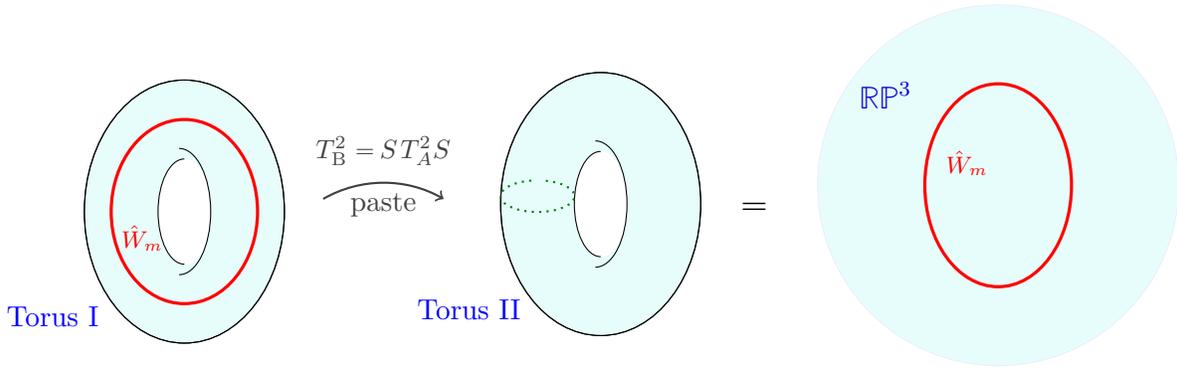
\subsection{Conformal blocks as Hilbert states}

A closer look at the expression \eqref{psi_initial_3d} for the initial wavefunction $\psi_\text{inital}(m)$ reveals that it equals the expectation value of the circular Wilson line $W_m$ in the spin $m/2$ representation in SU(1,1) Chern-Simons theory with imaginary level $k = \i /\sfb^2$ defined on 3D real projective space 
\bea
\label{rp3exp}
\boxed{\ \; \psi_\text{initial}(m) \,=\, \bigl\langle \, W_m  \bigr\rangle_{\mathbb{RP}^3}\,,\quad W_m = \Tr_{m} \text{P}\exp \oint_{\strut}^{\strut} A_+.\ }
\eea
A short pictorial proof of this equality is depicted in figure \ref{figure-glue-one}. $\mathbb{RP}_3$ can be obtained by gluing two filled tori (donuts) via a modular SL(2,$\mathbb{Z})$ transformation $T_B^2 = ST_A^2 S$, where $T_A$ and $T_B$ denote the Dehn twist around the (contractible) A-cycle and (non-contractible) $B$-cycle of the torus (see e.g. \cite{Castro:2011xb}). Inserting the Wilson line around the B-cycle of one of the tori produces the expectation value \eqref{rp3exp}. The identity \eqref{psi_initial_3d} then follows from the standard surgery rule \cite{witten1989quantum} of 3D CS theory  (see section \ref{sect:6.2surgery} for a related discussion).
In the remainder of this subsection, we sketch how this Wilson line amplitude relates directly with the 2D CLS crosscap amplitude discussed in the previous sections, recalling some elements of the 2D-3D dictionary between complex Liouville CFT and SL(2,$\mathbb{C}$) CS theory \cite{Verlinde:2024znh,Collier:2025lux}.\footnote{The mathematical relation \eqref{rp3exp} between the 3D CS expectation value and the SYK partition function is directly linked to the Schur-SYK correspondence, see e.g. figure 4 in \cite{Gaiotto_verlinde:2024kze}.}

CFT partition functions are not numbers, but functionals on the space of 2D metrics. Let us make this dependence explicit for the $\varphi_+$ partition function with Ishibashi and crosscap boundary conditions
\bea
\label{zplus}
\psi^+_m\!\spc(\spc g) \is
\ishibra{m} {\text{C}_+}\rangle(g) \, =\, P_{\mathbb{1}m} T_m^{-1/2} \chi^+_m\!\spc(\spc g) .
\eea
Here $m$ labels the discrete Liouville momentum $p_m = \i \sfb m/2$ of the primary sector in the closed string channel.\footnote{The primary state in Liouville with discrete momentum $p_m = \i\sfb m/2$ satisfies a $m$-th order null state equation that can be interpreted as the statement that this state transforms under a spin $m$ representation of SU(1,1)$_q$. Hence, one should anticipate that the 3D lift of $\psi_m(g)$ involves a spin $m$ Wilson line $W_m$ piercing the 2D geometry.}
The idea behind equation \eqref{zplus} is that we let the $\varphi_-$ field emerge via the Weyl factor $g = e^{2\varphi_-}\eta$ of the dynamical 2D metric in the conformal gauge.
The dependence of the conformal factor of the 2D metric is prescribed by the conformal anomaly
\bea
\psi^+_m(g)\! \is \!
 \psi^+_m(\tau\nspc+\nspc 1/2) e^{c_+ S_L(\varphi_-)}
, \qquad \ \ Z_{\rm ghost}(g) = {e^{-26 S_L(\varphi_-)}}{\eta(\tau\nspc+\nspc1/2)^2}.
\eea
Performing the functional integral over all 2D metrics  produces the complex Liouvile gravity partition function with crosscap and Ishibashi boundary conditions for the $\varphi_+$-field
\bea
\label{gintegral}
\int\![\d g]\,\psi^+_{m}(g) \!\is\!
\int_0^\infty\!\! d\tau \; {\psi^+_m(\tau\nspc+\nspc 1/2)\,  Z_-(\tau)}{\eta(\tau\nspc+\nspc 1/2)^2}, \quad \ %\notag\\[-2mm]\\[-2mm]\notag
Z_-(\tau) = \int [d\varphi_-] \, 
e^{-c_-S_L(\varphi_-)} .
\label{zminus3d}
\eea

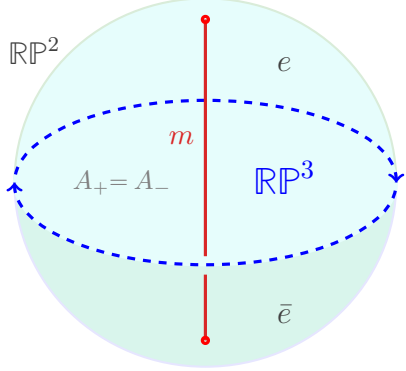
\begin{figure}[t]
\begin{center}
\begin{tikzpicture}[thick,xscale=.35,yscale=.35,rotate=90, every node/.style={scale=1}]
\tikzset{
    partial ellipse/.style args={#1:#2:#3}{
        insert path={+ (#1:#3) arc (#1:#2:#3)}
    }
}   
% \draw[blue!07!white] [fill=darkgreen!12!blue!07!white] (-6,-14) -- (-6,10) --(18,10) -- (18,-14) -- cycle; 
\draw[blue!10!white,fill=darkgreen!50!cyan!15!white]  
(6,-2)  [partial ellipse=90:270:7cm and 7.25cm];
\draw[darkgreen!15!white,fill=darkgreen!12!cyan!10!white] (6,-2) [partial ellipse=-90:90:7cm and 7.25cm];
\draw[very thick,darkgreen!12!cyan!10!white,fill=darkgreen!12!cyan!10!white] (6,-2) [partial ellipse=0:360:3.125cm and 7.25cm];);
\draw[very thick,blue,<-,fill=darkgreen!12!cyan!10!white,dashed]  (6,-2) [partial ellipse=90:270:3.125cm and 7.25cm];
\draw[very thick,blue,<-,fill=darkgreen!12!cyan!10!white,dashed]  (6,-2) [partial ellipse=-90:90:3.125cm and 7.25cm];
\draw [darkgray] (11,4.5) node {\large $\mathbb{RP}^2$};
\draw [blue] (6.2,-5) node {\Large $\mathbb{RP}^3$};
\draw [darkgray] (1,-5) node {\large $\bar{e}$};
\draw[gray] (6,1) node {\small $A_+\!\! =\nspc A_-\ $};
\draw[very thick, gray!30!red]  (0,-2)  -- (2.5,-2);
\draw[very thick, gray!30!red]  (3.2,-2) -- (12.2,-2) node[midway,left] {$m$} ;
\draw [darkgray] (10.5,-5) node {\large $e$};
\draw[very thick,red] (0,-2) circle (1.15mm);
\draw[very thick,red] (12.2,-2) circle (1.15mm);
%\draw[thick, darkgray, <<->>,opacity=.7] (11.5,-1)--(12.5,-3);
%\draw[thick, darkgray, <<->>,opacity=.7] (11.5,-3)--(12.5,-1);
%\draw [darkgray] (10,1) node {${A_+}$};
\end{tikzpicture}
\vspace{-2mm}
\end{center}
\caption{
The SL(2,$\mathbb{C}$) CS path-integral over $A_\pm$ on $\mathbb{RP}^3$ with a topological boundary condition $A_+\!=\!A_-$ at the equatorial plane  unfolds into a path integral over $A_+$ on  $\mathbb{RP}^3$, i.e.~a 3-ball $B_3$ with antipodal points on its $S^2$ boundary identified. 
The $A_+$ path integral on $B_3$ with a Wilson line $W_m$ specifies a density matrix $\rho%(e,\bar{e}) = \langle e 
= \ket{\psi_m}\bra{\psi_m}%\bar{e}\rangle
$ with bra and ket states given by wavefunctionals  $\langle e\ket{\psi_m}$  and $\bra{\psi_m}\bar{e}\rangle$ 
of the respective components of  $A_+\!=(e,\bar{e})$ evaluated on the top and bottom~$\mathbb{RP}^2$. This density matrix equals the partition function $\ishibra{m}\text{C}_+\rangle(e,\bar{e})$  
of the $\varphi_+$ theory with crosscap and Ishibashi boundary conditions. Performing the path integral over the background metric $g=(e,\bar{e})$ amounts to gluing the top and bottom half-spheres, producing the expectation value $\langle W_m\rangle$ on $\mathbb{RP}^3$. The $\varphi_-$ field represents the conformal mode of $g$. }
\vspace{-2mm}
\end{figure}

The conformal blocks $\chi^+_m(g)$ span a linear space of solutions to the conformal Ward identities. They can be viewed as states of a 3D gravity Hilbert space \cite{Witten:1989ip,Verlinde:1989ua,Verlinde:2024zrh, Collier:2025lux}, which (formally) looks like (a subspace of) the tensor product of two SU(1,1) CS Hilbert spaces with opposite imaginary level $k_\pm = \pm \i/\sfb^2$. Let $A_+$ and $A_-$ denote the two CS fields. 
Splitting the metric $g$ into a zweibein $(e,\bar{e})$, we can identify the partition function $Z^+_m\spc(e,\bar{e})$ with an element of the tensor product Hilbert space 
\bea
\label{psiop}
\psi^+_m(e,\bar{e}) \ & \longleftrightarrow & \ 
\ket{\spc \hat{\psi}_m^+\spc }  \, \in \, {\cal{H}}_{\text{CS}_+}\otimes {\cal{H}}^*_{\text{CS}_-}
\eea
of the two CS theories. The geometric motivation for this identification is illustrated in figure 10. 

Figure 10 depicts the Wilson line insertion inside an $\mathbb{RP}^3$ geometry, represented as a 3-ball $B_3$ with antipodal points on its $S^2$ boundary identified. The $\mathbb{Z}_2$ quotient of the boundary $S^2$ is the real projective plane $\mathbb{RP}^2$.\footnote{Note, however, that this $\mathbb{RP}^2$ is not a boundary component of $\mathbb{RP}^3$, but is embedded inside $\mathbb{RP}^3$ as a one-sided surface. For a recent study of AdS and dS holography applied to 2D CFTs on the $\mathbb{RP}^2$ crosscap geometry, see \cite{Wei:2024zez}.} The Wilson line connects two antipodal points on the $S^2$ but intersects the $\mathbb{RP}^2$ in one point.
The SL(2,$\mathbb{C}$) CS path-integral over the complex gauge field $(A_+,A_-)$ on $\mathbb{RP}^3$ is subject to a topological boundary condition $A_+\!=\!A_-$ along the equatorial plane; it can be unfolded into a path integral over $A_+$ on all of $\mathbb{RP}^3$. To obtain the Wilson line expectation value, we first perform the path-integral with fixed boundary conditions on the $S^2$ and then we glue the two half-spheres together by integrating over the specified boundary values of $A_+$, subject to the anti-podal identification.

 Let $A_+\!=(e,\bar{e})$ denote the two components of $A_+$ on the $S^2$ boundary. The CS path integral on $B_3$ with a Wilson line $W_m$ inserted and with $e$ fixed on the top half-sphere and $\bar{e}$ on the bottom half-sphere specifies a density matrix $\rho%(e,\bar{e}) = \langle e 
= \ket{\psi_m}\bra{\psi_m}%\bar{e}\rangle
$ with bra and ket states given by wavefunctionals  $\langle e\ket{\psi_m}$  and $\bra{\psi_m}\bar{e}\rangle$ 
of the respective components of  $A_+$. 
The basic observation that underlies our proposed 3D lift is that (i) this density matrix $\rho(e,\bar{e})$ can be identified with the partition function $\ishibra{m}\text{C}_+\rangle(e,\bar{e})$  
of the $\varphi_+$ Liouville theory with crosscap and Ishibashi boundary conditions with background metric $g=(e,\bar{e})$ and that (ii) performing the 2D path integral over $(e,\bar{e})$ amounts to taking the trace of this density matrix. From the 3D geometry perspective, taking this trace amounts to gluing the top and bottom half-spheres together, thus producing the expectation value $\langle W_m\rangle$ of the Wilson line on the closed non-orientable three manifold $\mathbb{RP}^3$. Our working assumption is that the $\varphi_-$ Liouville field is identified with the conformal mode of the metric $g = (e,\bar{e})$. Its role in this construction is to provide the inner product between the bra and ket state.

The conformal blocks form a natural orthonormal basis of the 3D gravity Hilbert space \cite{Verlinde:1989ua,Collier:2023fwi,Collier:2023cyw_Virasoro_Minimal_String}. On a closed surface without boundaries, conformal blocks $\chi_p(e,\bar{e})$ with given Liouville momentum factorize into a product $\chi_p(e)\bar{\chi}_p(\bar{e})$ of chiral blocks. With  boundaries, writing $\chi_p$ in this factorized form is not possible, as the Ishibashi conditions \eqref{ishicond} couple both sectors.  Note, however, that within the BRST cohomology, all Virasoro descendants are spurious and only the primary states are physical. The ground state with given Liouville momentum does factorize. This motivates the identification 
\bea
\label{chifact}
\chi^+_m\spc(e,\bar{e}) \ & \longleftrightarrow & \ 
\ket{\spc\hat{\chi}^+_m\spc} \equiv %\int\!\frac{dq}{2\pi} \, \frac{
\ket{\spc m\spc}\bra{\spc m\spc}
%} %{p^2-q^2 -\i \epsilon}  
\, \in \, {\cal{H}}_{\text{CS}_+}\otimes {\cal{H}}^*_{\text{CS}_-}.
\eea
We assume that the states $\ket{m}$ form an orthonormal basis $\langle n\ket{m} = \delta_{nm}$, c.f. \cite{Collier:2023fwi, Collier:2023cyw_Virasoro_Minimal_String}.

Taking inspiration from this identification, we  introduce another state in the dual Hilbert space
\bea
\ket{\spc \hat{Z}_-\spc }\; \in \; {\cal{H}}_{\text{CS}_-}\otimes {\cal{H}}^*_{\text{CS}_+} 
\eea
designed such that the full gravitational CLS partition function -- which via  \eqref{gintegral} amounts to coupling to the $\varphi_-$ theory, adding ghosts, and integrating over the $\tau$ modulus -- equals
\bea
\label{ztrace}
\int\![\d e\, \d\bar{e}]\, \psi_m^+\spc(e,\bar{e})  = \bra{\spc \hat{Z}_-\spc} \hat{\psi}_m^+\spc \rangle = {\rm tr}_{{\cal H}_{\rm CS}}\bigl(\spc \hat{Z}_- 
\spc {\hat{\psi}_m^+}\bigr) \spc .
\eea
%The state $\ket{\hat{Z}_-}$ is the Hilbert space vector associated with the  $\varphi_-$ partition function, c.f. equation \eqref{zminus3d}.\footnote{ In the 3D to 2D mapping, we should think of $(e,\bar{e})$ as sources that couple to the stress-tensor of the Virasoro edge modes living on the $S^2$ boundary of the three-ball $B_3$. Taking the inner product between the states involves projecting out the edge modes by gauging the Virasoro symmetry. The role of the $\varphi_-$-theory is to absorb the conformal anomaly and make this a well-defined procedure.}
%Thus far we have just given a suggestive 3D Hilbert space representation of the $\varphi_+$ partition function and of the integral over all 2D metrics $g=(e,\bar{e})$. 
What can we learn from this?  At this point, we make the \emph{Ansatz} that the partition function $Z_-$  defined by the integral over metrics \eqref{zminus3d} coincides with the $\varphi_-$-partition function given in equation \eqref{chifact}. We do not have a first-principles derivation of this Ansatz, but let us see what it looks like from the 3D perspective. As we will see, our choice \eqref{xzcap} of $\varphi_-$ boundary conditions is equivalent to postulating that the 2D gravity integral \eqref{gintegral} represents the trace in the 3D gravity Hilbert space. 

Applying the dual version of the map \eqref{chifact} to the $\varphi_-$ partition function $Z_-(\tau)$ given in \eqref{zminus},
we see that  $Z_-(\tau)$ becomes identified with the ``identity" operator in the dual Hilbert space 
\bea
\ket{\hat{Z}_-}  \ & \longleftrightarrow & \  \II\, = \; \int\! dq\,  \ket{\spc \hat{\chi}_q  \spc}
\spc = \int\! dq \spc \ket{q}\bra{{q}} \spc . %\, \in \, {\cal{H}}_{\text{CS}}\otimes {\cal{H}}^*_{\text{CS}}
\eea
Combining with \eqref{ztrace} and using the orthonormality of the basis states $\ket{q}$,   we see that integrating a matter partition function $\chi_m^+(g)$ over metrics is indeed equivalent to taking the trace of the corresponding operator $\hat{\chi}_m^+$ in \eqref{chifact}, leading to the simple formula 
\bea 
\int\![\d e\, \d\bar{e}]\,\chi^+{\!\!\!\!\smpc}_{m}(e,\bar{e}) \is %\tr_{{\cal H}_{\rm CS}}\bigl(\spc 
\langle\spc\II\spc \ket{\hat\chi_m} \spc = {\rm tr}_{{\cal H}_{\rm CS}}\bigl(\spc 
\spc {\hat{\chi}_m^+}\bigr) \, = \,
1\spc .
%= \int_0^\infty\!\! d\tau \; \; \frac{\chi^+_p(\tau\nspc+\nspc 1/2)}{\chi_0(\tau\nspc+\nspc1/2)^2}\int [d\varphi_-] \, e^{c_-S_L(\varphi_-)} 
%\eea
%Evaluating the CLS partition function $Z(s)$ now becomes a one line computation
%\bea
%Z(s)\, =\, \int\![dg] \, Z_+(s\spc|\spc g)\,  = \, \langle \spc \II \spc \ket{Z_+(s)} \is \int \! {dp}\, \psi_+(s,p)
%\\[2mm]{\mathcal{Z}}_{\rm spec}(s) = \int\![dg]\, Z^+_{\rm spec}(s\spc |\spc g) 
%{\mathcal{Z}}_{\rm spec}(s)\! \, = \, 
%\int\![\d g] \, Z^+_{\rm spec}(s\spc |\spc g)
%\nonumber\\[-2mm]\\[-2mm]\nonumber\is \, =\, 
%\,\langle\spc  \II\spc \ket{\hat{Z}{}^+_{\rm spec}(s)}\is \sum_{p_n}\;%=\i n\sfb/2}
% \cos(4\pi p_n s) \, \psi_{\rm initial}(p_n) 
\eea
Applying this rule to from \eqref{zplus}, we can now directly evaluate the Wilson line expectation value 
\bea
\label{ztrace2}
\bigl\langle W_m\bigr\rangle = \int\![\d e\, \d\bar{e}]\, \psi_m^+\spc(e,\bar{e})  = \bra{\spc \II} \hat{\psi}_m^+\spc \rangle = {\rm tr}_{{\cal H}_{\rm CS}}\bigl(\spc 
\spc {\hat{\psi}_m^+}\bigr) \, =\,
P_{\mathbb{1}m} T_m^{-1/2},
\eea
matching our result for $\psi_\text{initial}(m)$ obtained in section 5.

Summarizing, we see that the mixed crosscap boundary condition \eqref{xzcap} has the special property that it trivializes the contribution from the $\varphi_-$-sector. From the 3D perspective, it implies that one SU(1,1) CS fields $A_-$ effectively decouples, setting its partition function equal to one. %This decoupling is implemented by placing the 3D theory on a 3D geometry with a topological $A_+=A_-$ boundary condition \cite{Gaiotto_verlinde:2024kze, Collier:2025lux}, which can be unfolded into single SU(1,1) CS theory on the double cover. 
%We can compare this $\varphi_-$- state $\ket{Z_-}$ with the state $\ket{Z_-^{\spc\mbox{\tiny C}}}$ associated with the symmetric CLS crosscap boundary state $\ket{\text{C}_+}\otimes \ket{\text{C}_-}$. From equation \eqref{symmetric crosscap amplitude} in Appendix B, we read off that the state $\ket{Z_-^{\spc\mbox{\tiny C}}}$ and the scorresponding pectral amplitude $ {\mathcal{Z}}^{\mbox{\tiny CC}}_\text{spec}(\theta)$ take the form
%\bea & & \ket{Z_-^{\spc\mbox{\tiny C}}\,} \,=\, \int\! dq \, \frac{e^{-\pi q^2 }}{\sin(\pi \sfb q) \sinh(\pi q/\sfb)}\, \ket{\hat{\chi}_q}\\[2mm]
 %   {\mathcal{Z}}^{\mbox{\tiny CC}}_\text{spec}(\theta)\!\is\!
    %\sum_{n=1}^\infty\frac{\cos(2 n \theta)}{\tan(\pi b^2 n)}\, = \, \sum
%\!\!    \sum_{p_n=\i n\sfb/2}\!\!
 %\cos(4\pi p_n s) \, \psi_{\mbox{\tiny CC}}(p_n) , \qquad \ \ \psi_{\mbox{\tiny CC}}(p) = \frac{1}{\tanh(\pi p/ \sfb)}\label{6.1zspecCC}
%\eea
%This is exactly the same equation as for the crosscap wavefunction in the JT gravity limit, see equation (5.22) in \cite{Stanford:2019vob}. In that case has a simple group-theoretic interpretation \cite{witten1991quantum} with $p$ interpreted as related with the monodromy of a flat connection around the crosscap. 
%It seems likely that this equation has a similar explanation in SL(2,$\mathbb{C}$) Chern-Simons or dS$_3$ quantum gravity\cite{Collier:2025lux,Verlinde:2024zrh,Verlinde:2024znh,Tietto:2025oxn,Gaiotto:2024kze}. 

\def\taup{y_{\mbox{\tiny$P$}}}
\def\tauh{y_{\mbox{\tiny$H$}}}
\def\phip{x_{\mbox{\tiny$P$}}}
\def\phih{x_{\mbox{\tiny$H$}}}

\def\htaup{y_{\mbox{\tiny$P$}}}
\def\htauh{y_{\mbox{\tiny$H$}}}
\def\hphip{x_{\mbox{\tiny$P$}}}
\def\hphih{x_{\mbox{\tiny$H$}}}

\subsection{Relation with quantized dS$_3$ geometry}\label{sect:6.2surgery}
\vspace{-1mm}

The interpretation of the 2D partition function as an overlap between states of a covariant 3D theory gives a useful perspective on the physical observables of the 2D theory. We have seen in section \ref{sec: lightcone quantization} that the physical CLS phase space is parametrized by holonomy variables $\theta$ and $b_+$, related to the Liouville variables $s$ and $p_n$ via $
\theta = 2\pi \sfb\smpc s$ and $b = 4\pi \sfb\spc p$.\footnote{Recall that $\alpha_+=Q_+/2+\i P_+$ and $P_+=\sqrt{\i}p$ and with $s$ playing the same role as $p$.} The associated holonomy matrices 
\bea
\label{lhmonodromy2}
\hat\ll \, =\, \bigg(\! \begin{array}{cc} e^{\i \hat\theta} \! &\!  0\\[0mm] 0 \! &\! e^{-\i \hat\theta} \end{array}\!\bigg)\,,\quad
\hat\hh \, =\, \bigg(\! \begin{array}{cc} e^{ \i \hat{b}/2 } \! &\!  0\\[0mm] 0 \! &\! e^{-\i \hat{b}/2} \end{array}\!\bigg)
\eea
classically commute, but quantum mechanically they do not commute (hence the temporary hat notation). The eigen values of $\hat{L}$ and $\hat{M}$ satisfy the torus algebra 
% \AB{This assumes $[b,\theta]=\hbar$. In other words it is $i b$ which is quantized in units of $\hbar$, that makes more sense, because why would an angle get quantized.}
\bea
\label{torusalgebra}
 \hat{u} \hat{v} \! \is \! \mathsf{q} \hat{v} \hat{u}\,,\quad \hat{u} =  e^{\i \hat{\theta}}, \ \hat{v} = e^{\i \hat{b}/2}\,,\quad \mathsf{q} = e^{-\hbar/2} = e^{-\pi \sfb^2}\,.
 \eea
Via the 3D lift of complex Liouville gravity outlined in the previous section, the matrices $\hat{L}$ and $\hat{M}$ acquire a 3D interpretation as holonomies around the cycles inside the 3D geometry. We will outline this interpretation below.

The euclidean dS$_3$ metric can be written~as
\bea
\label{sds_metric}
\d s^2 = \d\tau^2 + \beta^2 \cos^2\!\tau\, \d y^2 + \alpha^2 \sin^2\!\tau \, \d x^2\,,\quad x\sim x+1\,,\quad y\sim y+1\,.
\eea
The north and south pode are at $\tau=0$ and $\tau=\pi$, and the cosmological horizon is at $\tau=\pi/2$.\footnote{Note however that in euclidean signature, both regions look identical and are interchanged under $\tau\leftrightarrow \pi/2 -\tau$.} The angles $\alpha$ and $\beta$ parametrize potential conical defects \cite{Hikida:2021ese,Hikida:2022ltr} at both locations with opening angles
\bea
\label{alphabeta}
\alpha\, = 2\theta\,,\quad \beta\, =b\,.
\eea
The $\tau$ coordinate can be viewed as describing Euclidean time evolution of a 2D torus geometry evolving from the north pode to the horizon. Allowing for more general such metrics, the spacetime looks like
\bea
\label{radialm}
\d s^2= \d\tau^2 + e^{2\sigma_-} \d s_2^2\,, \quad \d s_2^2 =  \Omega_2^{-1}{{|\d x + \Omega \d y|^2}}\,,
\eea
with $\sigma_-$ and $\Omega$ functions of $\tau$ only. We will now describe how our minisuperspace treatment of CLS in section \ref{sec: hamiltonian analysis} can be lifted into a minisuperspace description of this euclidean torus cosmology.

\begin{figure}[t]
\centering
\begin{tikzpicture}[rotate=0,xscale=.75,yscale=-.75] 
\tikzset{
    partial ellipse/.style args={#1:#2:#3}{
        insert path={+ (#1:#3) arc (#1:#2:#3)}
    }
}   
\draw[black,fill=darkgreen!12!cyan!10!white] (0,0) [partial ellipse=-180:180:2.5cm and 1.9cm];
\draw[black] (0,0) [partial ellipse=-180:180:2.5cm and 1.9cm];
\draw[white,fill=white] (0,0) [partial ellipse=-180:180:1cm and .5cm];
\draw[black] (0,0) [partial ellipse=-0:-180:1cm and .5cm];
\draw[black] (0,-.1) [partial ellipse=-0:180:1.2cm and .6cm];
 \draw[darkgreen, very thick] [partial ellipse=0:92:1.75cm and -1.39cm];
\draw[darkgreen, very thick] [partial ellipse=98:200:1.75cm and -1.39cm];
\draw[darkgreen, very thick] [partial ellipse=200:360:1.75cm and -1.39cm];
%\draw [thick,dotted,partial ellipse=-180:180:2.5cm and -2.5cm];
\draw[\darkred,thick,decorate, decoration={snake, segment length=1.5mm, amplitude=.1mm}](0.15,-1.25) [partial ellipse=13:353:.3cm and .7cm];
\draw[blue] (-2.5,2) node {\mbox{Torus A}};
%\draw[blue,very thick] (-1.95,-1.5) -- (-1.95,-.4);
 \draw[\darkred] (-.5,-.8) node {\mbox{\small $\hat{L}$}};
\draw[white,fill=white] (0,2.5) [partial ellipse=-180:180:.1cm and .05cm];
%\draw[thick] (-1.65,0) circle (.35mm);
%\draw[thick] (1.65,0) circle (.35mm);
\end{tikzpicture}~~
 \begin{tikzpicture}[scale=.5,baseline={([yshift=-2.3cm]current bounding box.center)}]
        \draw[<-, thick,darkgray] (60:1) arc (60:120:4) node[midway, below] {paste} node[midway, above] {\mbox{\small apply $(\hat{T}_\text{A})^{2}$\Large$\strut$}} ;
\end{tikzpicture}\hspace{-7mm}\begin{tikzpicture}[rotate=90,xscale=.75,yscale=-.75] 
\tikzset{
    partial ellipse/.style args={#1:#2:#3}{
        insert path={+ (#1:#3) arc (#1:#2:#3)}
    }
}   
\draw[black,fill=darkgreen!12!cyan!10!white] (0,0) [partial ellipse=-180:180:2.5cm and 1.9cm];
\draw[black] (0,0) [partial ellipse=-180:180:2.5cm and 1.9cm];
\draw[white,fill=white] (0,0) [partial ellipse=-180:180:1cm and .5cm];
\draw[black] (0,0) [partial ellipse=-0:-180:1cm and .5cm];
\draw[black] (0,-.1) [partial ellipse=-0:180:1.2cm and .6cm];
%\draw [thick,dotted,partial ellipse=-180:180:2.5cm and -2.5cm];
\draw[\darkgreen,thick,dotted](0.15,-1.2) [partial ellipse=13:353:.3cm and .7cm];
\draw[blue] (-2,-2.5) node {\mbox{Torus B}};
%\draw[blue,very thick] (-1.95,-1.5) -- (-1.95,-.4);
%\draw[thick] (-1.65,0) circle (.35mm);
%\draw[thick] (1.65,0) circle (.35mm);
\end{tikzpicture}~~~~~\raisebox{1.7cm}{\Large $=$}~~~~~
\raisebox{-.3cm}{\begin{tikzpicture}[rotate=0,xscale=.7,yscale=-.7] 
\tikzset{
    partial ellipse/.style args={#1:#2:#3}{
        insert path={+ (#1:#3) arc (#1:#2:#3)}
    }
}
 \draw[blue!07!white] [fill=darkgreen!12!cyan!10!white] [partial ellipse=0:360:3.2cm and 3.2cm]; 
 \draw[darkgreen, very thick] [partial ellipse=0:92:1.95cm and -1.2cm];
\draw[darkgreen, very thick] [partial ellipse=98:200:1.95cm and -1.2cm];
\draw[darkgreen, very thick] [partial ellipse=200:360:1.95cm and -1.2cm];
\draw[\darkred,thick,decorate, decoration={snake, segment length=1.5mm, amplitude=.1mm}](0.1,-1.2) [partial ellipse=13:353:.3cm and .8cm];
\draw[\darkblue] (-2,-1.6) node {\mbox{\large $S^3$}};
 \draw[\darkred] (-.55,-.3) node {\mbox{\small $\hat{L}$}};
\end{tikzpicture}}
\caption{The three-sphere $S^3$ with a circular line defect (green) can be obtained by gluing together two filled tori. The gluing identification implements a modular $S$-transform and may involve a non-trivial Dehn twist. This gluing prescription assigns to the 3D gravity theory a past and future state that live on 2D tori. The A-cycle line operator $\hat{L}$ measures the deficit angle of the defect.  }
\label{figtori}
\vspace{-2mm}
\end{figure}
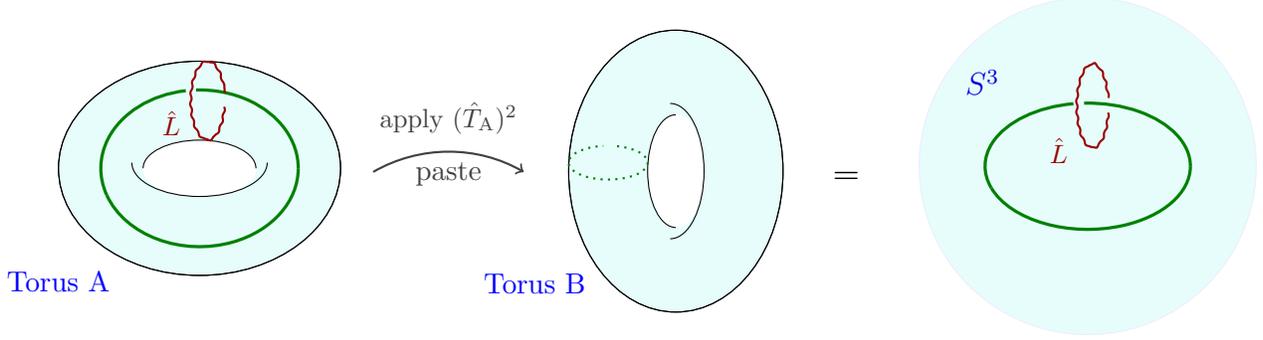

In the spirit of associating a wavefunction with the initial and final states, we momentarily divide the $\tau$ evolution into an initial and final state region representing the local regions near the pode (initial) and horizon (final), and introduce separate $x$ and $y$ coordinates for both regions
\begin{align}
\d s^2 &=  \d\tau^2 +\beta^2 \d y_\text{P}^2 +  \alpha^2\tau^2\,\d x_\text{P}^2\,,\quad \qquad\quad \;\; \text{near pode}\ (\tau=0)\\[2mm]
\d s^2 &=  \d\tau^2 +\beta_\text{H}^2\spc (\tau\!-\!\pi/2)^2 \d y_\text{H}^2 +  \alpha_\text{H}^2\spc\d x_\text{H}^2\,,\quad \text{near horizon} \; 
(\tau=\pi/2)
\end{align}
The $x_\text{P}$ circle shrinks to a point at the pode, whilst the $y_\text{H}$-circle shrinks to a point at the horizon. In general, we allow for conical singularities at both locations parameterized by $\alpha$, $\beta$ near the pode and $\alpha_\text{H}$, $\beta_\text{H}$ near the horizon . For pure dS$_3$ we have $\alpha=\alpha_\text{H} = 2\theta=2\pi$, $\beta=\beta_H = b=2\pi$ and $(y_\text{H},x_\text{H})=(y_\text{P},x_\text{P})$.\footnote{This corresponds to $p=1/2\sfb$ which classically implements indeed the Virasoro vacuum representation.} The pode and horizon patch describe two smooth filled-up tori (donuts) glued together via an $S$ modular transformation. Indeed, for the pode torus the A-cycle  is smoothly contractible while for the horizon torus the B-cycle is smoothly contractible. For $\theta\neq \pi$, the pode torus contains a conical singularity at $\tau=0$. This defect can be thought of as the consequence of a localized matter source, and the geometry is known as the SdS solution.\footnote{In spite of its name, the Schwarzschild-de Sitter spacetime has a cosmological horizon but no black hole horizon. } For $b\neq 2\pi$, the horizon torus also becomes singular at the horizon. As we will see, for the geometry that reproduces the SYK partition sum, the pode can be singular, while the horizon geometry remains smooth - but in a non-trivial way. 

In general we can allow for the possibility \cite{Castro:2011xb} that the two tori are glued together via a non-trivial SL(2,$\mathbb{Z}$) modular transformation $(x_\text{H},y_\text{H})= (a x_\text{P} + b y_\text{P},c x_\text{P} + d y_\text{P})$. From the formulas \eqref{psi_initial_3d} and \eqref{zspec_3d}, one can deduce that the geometry that reproduces the SYK partition function is obtained by gluing the two tori together with the transformation $\hat{S}\hat{T}_\text{A}^2$:
\bea
\label{hglue}
(x_\text{H},y_\text{H})\! \is  \!\bigl(x_\text{P}+2y_\text{P},y_\text{P})\,, \qquad 
(\alpha_\text{H},\beta_\text{H}) \spc=\spc \bigl(\alpha,\beta+2\alpha)
\eea 
This geometric procedure is schematically depicted in figure \ref{figtori}. The second equation, combined with \eqref{alphabeta}, identifies $\beta_\text{H}=b+4\theta$. So requiring smoothness of the horizon and setting $\beta_\text{H}=2\pi$, we recover the familiar identity \eqref{gsigmasol2.12} that specifies the crosscap boundary condition of the $G\Sigma$-theory. In this way, the 3D lift underscores the topological interpretation depicted in figure 6 of the relation \eqref{gsigmasol2.12}.

We now outline how the above geometric construction can be implemented at the quantum level via the CS formulation of dS$_3$ gravity.
%The quickest way to see that \eqref{hglue} does the trick is to note that it leads to the following relation between the opening angles at the podes and the horizon
A practical approach to quantizing dS$_3$ gravity is to use that its phase space coincides with (a subset of) the space of flat SU(1,1) connections. Defining $A_\pm = \omega \pm \i e$, with $\omega$ and $e$ the 3D spin connection and dreibein, the Einstein equations become flatness conditions $F(A_\pm) =0$ \cite{Witten:1989ip,Verlinde:1989ua}. Adopting a Hamiltonian approach with $\tau$ as time-direction, the classical phase space on a surface of torus topology is parametrized by holonomies around the A and B-cycle 
\bea
\label{hololoops}
\hat{L}_\pm = \text{P}\exp\Big(\oint_A  A_\pm\Big)\,,\quad
\hat{M}_\pm = \text{P}\exp\Big(\!\pm \oint_B A_\pm\Big)\,
\eea
evaluated in the spin 1/2 representation.\footnote{The eigen values of the open holonomy loops \eqref{hololoops} are gauge invariant, but the loops themselves are not. To get gauge invariant loops, one can either take the trace, or introduce a marked point with a preferred section $s$ of the flat SU(1,1) bundle and define open loop operators as in equation \eqref{yzholo} \cite{Gaiotto_verlinde:2024kze}. In quantum gravity on the torus, inserting a marking is a natural way of eliminating the zero mode due to translation invariance along the torus. A more precise definition of the loop operators \eqref{yzholo} is to represent the 3D wavefunctions as conformal blocks of the CLS CFT (that now lives on the 2D tori depicted in figure \ref{figtori}) and construct open Verlinde line operators connecting the marked point to itself\cite{Verlinde:1988sn_erik,gaiotto_open_2014}. The most practical Verlinde line operators are those associated with the (2,1) degenerate field. Insert a pair of (2,1) operators into a conformal block, starting from the identity OPE. Let $\chi_i(u)$, $i=1,2$, denote the two blocks created by the insertion of the (2,1) operator at $u$ (and another one at, say, the marked point). The null state equation implies that 
\begin{equation}
    (\de_u^2 - \sfb^2 T^\pm_{uu})\chi_i(u)=0\,,\quad (\de_v^2 - \sfb^2 T^\pm_{vv})\xi_i(v)=0\,.
\end{equation}
Here $\xi_i(v)$ is the analogous right-moving block. The quantum matrices $L$ and $M$ then describe the A- and B-cycle holonomy of the quantum uniformizing coordinates defined as the ratios $U = \chi_1/\chi_2\,,\ V = \xi_1/\xi_2\,$
of the blocks (see e.g. \cite{chen_quantum_2004}). These holonomies again classically commute but quantum mechanically do not: their quantum commutator follows from the well-known skein relations for $SL(2\mathbb{C})$ Wilson lines \cite{Verlinde:2024znh,Gaiotto_verlinde:2024kze}.}
Henceforth, to relate with a single copy of SYK, we restrict to real (non-spinning) SdS geometries.\footnote{Including a second copy of SYK one can describe the full phase space including spinning geometries.} These have 
\begin{equation}
    \hat{L}_+=\hat{L}_-\,,\quad \hat{M}_+=\hat{M}_-\,.
\end{equation}
In practice, this means this sector of the dS$_3$ phase space is captured by quantum mechanically by one SU$(1,1)$ CS theory (as opposed to SL$(2,\mathbb{C})$ CS for the full phase space).

The classical moduli space of flat SU(1,1) connections on the torus comes with a symplectic form that implies that the eigenvalues of $\hat{L}$ and $\hat{M}$ satisfy the torus algebra \eqref{torusalgebra}. So, while in the classical theory we can simultaneously diagonalize the holonomies $\hat{L}$ and $\hat{M}$, in the quantum theory we can not. Instead we have two sets of eigenstates:
\bea
\hat{L}\ket{\spc\theta\spc} = L(\theta) \ket{\spc\theta\spc}\,,\quad \hat{M}\ket{\spc b_n \spc} = M(b_n) \ket{\spc b_n\spc}\,.
\eea
Here we assume, motivated by the correspondence with SYK and the physical definition of $\theta$ as a deficit angle, or equivalently, as labeling an elliptic conjugate class of SU(1,1), that $\theta$ runs over a finite range between  0 and $\pi$. As a result, the eigenbasis of the B-cycle holonomy $\hat{M}$ becomes discrete, $b_n = 2\pi \i \sfb^2 \spc n$.
One can think of the states $\ket{\theta}$ as obtained from the 3D CS path integral over the handle body  obtained by filling in the torus such that the A-cycle is contractible, while inserting a line defect along $B$-cycle that creates the holonomy $L(\theta)$ around the defect. 

The state $\ket{b_n}$ is obtained in a similar way, except that the B-cycle is now contractible and the defect line wraps the A-cycle. We remark that real $n$ creates imaginary conical defects at the horizon, which are more like performing finite jumps in the Lorentzian time. We adopt the following normalization of the inner products between these eigen states
\bea
\langle \spc \theta_1 \spc \ket{\spc \theta_2} = \delta(\theta_1\!-\nspc\theta_2), \quad & & \quad \langle b_n | b_m \rangle % = \langle B: p_1 | B: p \rangle 
= \delta_{nm}, \qquad \ \ \langle\theta \ket{b_n} = \cos( n \theta)
\label{eq:normalization}
\eea
The third equation follows from the geometric fact that the  A- and B-cycle conformal blocks are related via the modular S-transformation, or simply from the fact that $\theta$ and $b$ are canonical conjugates
\bea
\ket{\spc \theta\spc }_\text{\nspc A} \is \sum_{b_n = \i \hbar n} \, S_{\theta n}  \ket{\spc b_n \spc}_\text{B}\,,\quad  S_{\theta n} = \cos( n\theta)\,. 
\label{eq:fzzt-block2}
\eea
The identity A- and B-cycle blocks $\ket{\spc \mathbb{1}\spc}_\text{A}  = \ket{\theta=\pi}$ and $\ket{\spc \mathbb{1}\spc}_\text{B}  = \ket{b=2\pi}$
represent the path integral over the empty torus handle-bodies, such that the corresponding holonomies are trivial
\bea
L \ket{\spc \mathbb{1}\spc}_\text{A} = 
\ket{\spc \mathbb{1}\spc}_\text{A} ,\quad & & \quad M \ket{\spc \mathbb{1}\spc}_\text{B} = 
\ket{\spc \mathbb{1}\spc}_\text{B} .
\eea

With this set-up in place, we propose the 3D lift of our 2D crosscap geometry and compute the associated partition function. As explained in section \ref{sect:skein} and illustrated in figure \ref{FIGURE mobius-figure}, the 2D crosscap has the property that a one-cycle encircling it acquires, when embedded in 3-dimensions, a self-linking number 2. In terms of the local geometry of the Wilson line defect itself, the self linking represents a double twist in the local framing.\footnote{Note that the framing represents a physical property of the line-defect because we are considering a Wilson line of single CS gauge field $A_+$. If instead we would consider a Wilson line of $A_+$ and $A_-$ combined, the framing anomaly would cancel. The same cancelation happens in the partition function of the double crosscap geometry in Appendix B.} 
Creating this  double twist can be done locally by acting with a double Dehn twist $\hat{T}_\text{A}^{2}$ on the state $\ket{\theta}$ representing the line-defect at the pode. Alternatively, we can let this double Dehn twist act on the vacuum state $\ket{\mathbb{1}}_B$ at the horizon. We thus propose to associate the 2D crosscap boundary condition with the definition of a ``crosscap horizon state'' $\ket{\text{C}}_\text{B}$ obtained by applying the double A-cycle Dehn twist $\hat{T}_A^{2}$ on the B-cycle identity state
\bea
\label{cdefg}
\ket{\spc \text{C}\spc }_\text{B} =\spc
\hat{T}_{\rm A}^2\spc  \ket{\spc \mathbb{1}\spc}_\text{B}\,. 
\eea
Using that $\hat{T}_\text{A} = S\spc \hat{T}_\text{B} S$, that $\hat{T}_B\ket{\mathbb{1}}_\text{B} = \ket{\mathbb{1}}_\text{B}$, and the expression  $\hat{P}=\sqrt{{T}} {S}\spc {T}^2 S \sqrt{{T}}$ for the modular $P$-matrix used in the definition \eqref{crosscap liouville} of the crosscap boundary state, we compute
\bea
\hat{T}_{\rm A}^2\spc  \ket{\spc \mathbb{1}\spc}_\text{B} \is \! \hat{S}\, \hat{T}_{\rm B}^2\, \hat{S}  \ket{\spc \mathbb{1}\spc}_\text{B} 
= \hat{T}_\text{B}^{-1/2}\hat{P} \ket{\spc \mathbb{1}\spc}_\text{B} 
\eea
Hence we find that the state $\ket{\text{C}}_\text{B}$ defined via \eqref{cdefg} indeed matches with the crosscap boundary state \eqref{crosscap liouville} in the complex Liouville CFT
\bea
\label{3dxcap}
\ket{\spc \text{C}\spc }_\text{B} = \sum_{n} \, P_{\mathbb{1}n} T_{n}^{-1/2} \ket{\spc b_n \spc}_\text{B}. 
\eea 
Although this state involves a sum over states $\ket{b_n}$ with non-trivial B-cycle holonomy, it is clear from its definition \eqref{cdefg} as a double Dehn twist of the B-cycle identity block that the crosscap state \eqref{3dxcap} defines a smooth 3D horizon geometry with a contractible cycle $B+2A$ with trivial holonomy $ML^2 = 1$! 

From the above expression \eqref{3dxcap} for the horizon state $\ket{C}_\text{B}$, the spectral partition function of the crosscap geometry with a deficit angle $\theta$ can be computed in dS$_3$ quantum mechanics from the overlap 
\begin{align}
\mathcal{Z}_{\text{spec}}(\theta) &= 
\mbox{\tiny ${\strut}$}_\text{B}\langle{\spc \text{C}}\ket{\spc \theta \spc}_\text{A} = \mbox{\tiny ${\strut}$}_\text{B}\!\bra{\mathbb{1}} \hat{T}_\text{A}^2\ket{\spc \theta\spc}_\text{A}
= \sum_{m \in 2\mathbb{Z}} S_{\theta m} \, P_{\mathbb{1}m}\,T^{-1/2}_{m} 
\label{eq:zspec}
\end{align}
As seen from the middle formula, this partition function indeed corresponds to the gluing prescription depicted in figure \ref{figtori}. Plugging in the expressions \eqref{sptmatrix}, or simply comparing with equation \eqref{psi_initial_3d}, we obtain a match with the 2D crosscap partition function defined and with the SYK spectrum \eqref{1.2}.\footnote{As a side comment: a simpler dS$_3$ amplitude without the double Dehn twist produces a partition function $_\text{B}\langle{\spc \mathbb{1}}\ket{\spc s \spc}_\text{A} = S_{\mathbb{1}s} =\sin(2\pi\sfb s) \sinh(2\pi s/\sfb) = \rho_\TT{MM}(s)$ which equals the spectral density of the matrix model dual to CLS \cite{Collier:2024lys_matrix}. We stress, however, that our logic is not to compute the simplest amplitudes in dS$_3$ quantum gravity, but to identify the boundary conditions in dS$_3$ quantum gravity that are provided by its candidate microscopic hologram, i.e.\,the SYK model.}

Another way that one may have arrived at this correspondence, with the knowledge of our description of the CLS minisuperspace in section \ref{sec: hamiltonian analysis}, is as follows. Using the parametrization \eqref{radialm} of SdS, we can formulate the radial evolution in $\tau$ by means of a 3D minisuperspace ADM Hamiltonian. Parametrizing the complex modular parameter $\Omega$ of the torus via
\bea
\Omega = \nu + \i\spc e^{\sigma_+}\,,
\eea
and setting up the canonical ADM formalism, one finds that the minisuperspace WDW constraint takes the following familiar form \cite{Carlip:1994ap, Carlip:2004ba}
\bea
H_{\rm WDW} = \pi_-^2 + e^{2\sigma_-}- \pi_+^2 - \mu^2 \spc e^{2\sigma_+} = 0\,,
\eea
where $\pi_\pm$ and $\mu$ denote the conjugate momenta to $\sigma_\pm$ and $\nu$. Since $H_\text{WDW}$ does not depend on $f$, its momentum $\mu$ is conserved. Hence we can set $\mu$ to a constant.\footnote{The $\sigma_+$ Hamiltonian (in the  subsector where $\mu$ is a given constant) is actually equivalent to that of Schwarzian QM \cite{Engelsoy:2016xyb}. The equation of motion of $\mu$ imposes a constraint $e^{2\sigma_+} = \dot{f}$ with $f = 2\mu \nu$. Eliminating $\sigma_+$ in favor of $f$, the Hamiltonian then indeed takes the form of the Schwarzian derivative $\{f,\tau\}$. } The WDW Hamiltonian then reduces to the minisuperspace Hamiltonian \eqref{H_WDW4.5} of the CLS, with $\varphi_- = \sigma_-$ and $\varphi_+ = \sigma_+$! This suggests that it may be possible to obtain the 3D gravity description via a trivial lift of the 2D CLS theory by adding an extra $y$ circle. We leave a more in-depth study of this 3D/2D/SYK dictionary for future work.

\section{Concluding remarks}\label{sect:disc}

In this paper we developed a precise correspondence between the $G\Sigma$ collective field theory of DSSYK and a gauge fixed version of the CLS worldsheet theory in which one of the two Liouville fields is used as a physical clock. The correspondence requires placing the CLS worldsheet theory on a disk with a crosscap. As an explicit check for the proposal, we have computed the CLS partition function on this geometry in section \ref{sec: liouville crosscap amplitude} and shown that it matches the SYK spectral density ${\mathcal{Z}}_\text{spec}(\theta)$.

The correspondence with SYK dictates an unusual choice of crosscap initial state \eqref{xzcap} in which  the $\varphi_+$ and $\varphi_-$ theory satisfy different boundary conditions. Schematically
\bea
Z_-\is \bra{\text{FZZT}} 
e^{\i {\pi}(\tau + \frac 1{2}) L_0^-}  \ket{\text{ZZ}}, 
\qquad \qquad 
Z_+\, = \,  \bra{\text{FZZT}}  
e^{\i {\pi} \tau L_0^-}  \ket{\text{C}}.
%= \int\! dq\,  \, \ishibra{q}e^{\i {\pi}(\tau + \frac 1{2}) L_0^-} \ishiket{q} %e^{- \pi \tau  L_0^-} 
\eea
$Z_+$ defines a proper M\"obius strip partition function. The $\varphi_-$ boundary conditions, on the other hand, do not look like a conventional crosscap. The boundary state $e^{\i \frac \pi 2 L_0^-}\ket{\text{ZZ}}$ satisfies the crosscap Ishibashi conditions, but it does not obviously satisfy the crosscap Cardy condition. Nevertheless,  as shown in section \ref{SECTION open channel}, we can write $Z_-$ as a proper trace over the open string Hilbert space, giving confidence that the $\varphi_-$ partition function can be given a proper path-integral representation. Finally, in section \ref{sect:threedee}, we developed the beginnings of a 1D/3D dictionary that identifies the partition function and amplitudes in the SYK model with expectation values of gravitational line operators in 3D de Sitter gravity. The 3D language appears to be very natural for capturing the geometric meaning of the boundary conditions and the algebraic structure of the physical operators of the 2D gravity theory.

The 2D gravity theory exhibits a striking difference between the closed and open string channels. In section 5 we found that its partition function takes the form of a discrete sum over Liouville momenta $p_n = \pi \sfb n$ and that (since the modular S-transformation amounts to a Fourier transform)  the integral over the Liouville momentum $q$ in the open string channel becomes a finite integral over a spectral angle $\theta$.
This brings us back to the question stated in the introduction: how can a covariant 2D gravity theory constructed with CFTs reproduce a bounded spectral density that looks figure 1? In our construction, there are two main elements responsible for the shape of the spectrum. The first is that the physical phase space of the complex Liouville theory is parametrized by SU(1,1) holonomies of the uniformizing coordinates that parametrize the solutions of the complex Liouville theory. Geometrically, one of these phase space variables is a periodic angle. In a dilaton gravity language, the periodicity of this angle is related to the presence of a shift symmetry in the value of the dilaton field \cite{Blommaert:2023opb, Blommaert:2023wad,Blommaert:2024whf,Blommaert:2024ydx}. This shift symmetry is unobstructed in our setting, since the Euler characteristic of the M\"obius strip vanishes.  In 3D, the SYK spectral angle $\theta$ acquires the interpretation of a conical deficit angle.

A second key feature of the spectrum is the presence of the gaussian suppression factor $T_p^{1/2} = e^{-\pi p^2}$. It arises due to the M\"obius twist in 2D or the framing twist of the circular defect in 3D. Crucially, in both the 2D and 3D settings the effect of the twist would cancel out in the symmetric case with the $\varphi_+$ and $\varphi_-$-sector both living on the same twisted geometry. Indeed, as shown in Appendix \ref{app:B} a CLS partition function with the symmetric crosscap state $\ket{\text{C}_+}\ket{\text{C}_-}$ does not contain this Gaussian factor, while in 3D the framing anomaly cancels between the CS$_+$ and CS$_-$ theory. To get the Gaussian factor, one needs a relative twist between the two sectors. 

In section 6, we summarized the connection between our 2D crosscap amplitude in CLS and the expectation value of circular Wilson line defects in 3D SU(1,1) CS theory. When applied to the finite temperature SYK partition function, the 3D dictionary leads to the suggestive identity
\bea
\label{suggestive_identity}
Z_{\rm SYK}(\beta) = \sum_m \, I_m(2\beta/\hbar) \, \bigl\langle W_m\bigr\rangle_{\mathbb{RP}^3}\,.
\eea
Geometrically, the right-hand side can be thought of as a 2D trumpet geometry (represented by the Bessel function) with outer boundary length $\beta$ and inner boundary length $b_m = \i\hbar m$, glued into 3D real projective space $\mathbb{RP}^3$, where the Wilson line $W_m$ is identified with the inner boundary of the trumpet. As
a spacetime geometry, $\mathbb{RP}_3$ describes elliptic 3D de Sitter space, the quotient of dS$_3$ by the antipodal map \cite{Parikh:2002py,Castro:2011xb}, while the physical phase space of 3D SU(1,1) CS theory with a line defect is identical to that of a non-rotating 3D Schwarschild-de Sitter spacetime. The identity \eqref{suggestive_identity} is one of many quantitative hints that point to a concrete holographic dictionary between SYK and the 3D de Sitter space. In section 6, we sketched some elements of this correspondence, but important open questions remain, even at the level of the partition function. 

One specific immediate task would be to recover the SYK partition function for small $\hbar = \lambda =\pi\sfb^2$ from a semi-classical expansion for small $G_N$ in 3D gravity. Here we just make a preliminary comment. The Einstein action $S(\alpha,\beta)$ of the classical euclidean 3D Schwarzschild-de Sitter geometry \eqref{sds_metric} with linked conical defects with opening angles $\alpha$ and $\beta$ evaluates to  \cite{Hikida:2021ese,Hikida:2022ltr}
\bea
S(\alpha,\beta) = -\frac{\alpha\beta}{8\pi G_N}.
\eea
In our setting, we impose a topological boundary condition $A_+=A_-$, or equivalently, consider a chiral version of 3D gravity corresponding to one SU(1,1) CS theory. In this chiral theory, the angles $\alpha = 2\theta $ and $\beta = b$ are canonically conjugate variables $[{b},\theta] = \frac{1}{8\pi G_N}$.\footnote{Here we use the standard relation between CS and 3D gravity \cite{Witten:1989ip} to equate $\hbar = \lambda = 8\pi G_N$ in 3D de Sitter units.} Moreover, as explained in section 6.2, the $\mathbb{Z}_2$ identification that projects $S^3$ to $\mathbb{RP}^3$ imposes the condition $b = 2\pi - 4\theta$. Incorporating these two facts into the computation of the gravitational action $S(\theta)$ results in
\bea
\label{gib-hawk}
\frac{dS(\theta)}{d\theta} = \frac{2\pi - 4\theta}{8\pi G_N} \quad & \rightarrow & \quad
S(\theta) = S_0-\frac{(\theta - \pi/2)^2}{4\pi G_N}
\eea
which matches the  $n=0$ leading order term in the expansion \eqref{ztot} of the SYK spectral density. We propose that equation \eqref{gib-hawk} is the correct  outcome of applying the semi-classical Gibbons-Hawking procedure to the geometric set-up described in section 6.2.\footnote{An unusual feature of \eqref{gib-hawk} is that it has a maximum at the opening angle $2\theta = \pi$ rather than $2\pi$. For a recent discussion of the relation between the DSSYK spectral entropy and the Gibbons-Hawking formula for 3D de Sitter, see \cite{Tietto:2025oxn}.}

% %%%%%%%%%%%
% % SECTION %
% %%%%%%%%%%%
\section*{Acknowledgments}

\vspace{-1mm}

We thank Davide Gaiotto, Akash Goel, Daniel Jafferis, Dongyeob Kim, Adam Levine, Henry Lin, Thomas Mertens, Beatrix M\"uhlmann, Vladimir Narovlansky, Liza Rosenburg, Douglas Stanford, Cynthia Yan and Mengyang Zhang for useful discussions. AB was supported by the Leinweber foundation, by the US DOE DE-SC0009988, and the Sivian fund.
%%%%%%%%%%%
% SECTION %
%%%%%%%%%%%

\medskip

\appendix

%%%%%%%%%%%%%%%%%%%%%
\section{Disk partition function}\label{sect:appendix}
\vspace{-1mm}

It is a natural question to ask how our CLS/SYK partition function with the crosscap initial condition relates to the more standard partition function of the CLS two matrix model. In CLS worldsheet language, the later describes the CLS partition function on the disk. Let us compare the two partition functions side by side
\begin{align}
\label{zcompare}
Z(\beta) &  = \int_0^\pi \!\! d\theta \, e^{\frac{\beta \spc {\cos(\theta)}} {2\pi \sfb^2}\spc} {\mathcal{Z}}_{\rm spec}(\theta),\qquad\qquad 
Z_{\rm disk}(\beta) = \int_0^{\theta_{\rm crit}}\!\! d\theta \; e^{\frac{\beta\spc {\cos(\theta)}} {2\pi \sfb^2}\spc }  \rho_{\rm MM}(\theta).
\end{align}
The spectral density of the CLS matrix model is given in terms of the modular $S$ matrix via 
\bea
\rho_{\rm MM}(\theta) = S_{\mathbb{1}s} = \sin(2\pi\sfb s) \sinh(2\pi s/\sfb), \qquad 2\pi \sfb \spc s=\theta
\eea
The upper bound $\theta_{\rm crit}$ is the point where the spectral density $\rho_{\rm MM}(\theta)$ turn negative. This occurs for $\theta_{\rm crit}=\pi$. So both integrals in fact run over the same range.

Both partition functions in \eqref{zcompare} are given by the overlap of a constant length boundary state and a  state that provides a smooth initial condition, either by inserting the crosscap or a regular disk. We can formally write
\bea
\label{zinnerproduct}
  Z(\beta) = \langle e^{\varphi_+}\!\nspc =\nspc \beta \ket{\text{C}_+}\qquad & & \qquad
  Z_{\rm disk}(\beta) = \langle e^{\varphi_+}\!\nspc =\nspc \beta \ishiket{\mathbb{1}}\; .
\eea
where $\ket{\text{C}_+}$ is the crosscap state and $\ishiket{\mathbb{1}}$ the identity Ishibashi state \cite{Collier:2024mlg_zz}
\footnote{Here we absorbed a factor of $ 1/\sqrt{S_{\mathbb{1} p} }$ into the definition of the Ishibashi state. So they are normalized according to $\ishibra{p}q\rangle\!\rangle = \delta(p-q)/S_{\mathbb{1}p}$.}
\bea
  \ket{\text{C}_+} =  \int\! \d p \, {P_{\mathbb{1} p} T_{p}^{-1/2} } \, e^{\i \pi L_{0}}{\ishiket{\spc p\spc }},\, \quad & & \quad 
  \ishiket{\mathbb{1}}\, = \, \sum_{p_{{}_{1\pm 1}}}
  \ishiket{p_{{}_{1\pm 1}}}\, .
\label{ishibashi-identity-state}  
\eea
Next, let us factorize the inner products \eqref{zinnerproduct} by inserting a complete Ishibashi basis. Defining the trumpet amplitude via 
\begin{align}
Z_\text{trumpet}\bigl(\beta,p\bigr)&= \langle e^{\varphi_+}\!\nspc =\nspc \beta \ishiket{p}\; = \int_0^\infty\!\! \d \theta\,\cos\bigl(\mbox{\large $\frac{2\i p\spc \theta}{\sfb}$}\bigr)\,e^{2{\beta {\cos(\theta)}}\spc }  = \sin(2\pi \sfb p) \, {I}_{2\sfb p}(2\beta)
    %= I_{m}\bigl(\mbox{\large $\frac{\beta}{2\pi\sfb^2}$}\bigr)\\[3mm] %\quad\ \hbar = 2\pi \sfb^2
\end{align}
we find that the two partition functions can be decomposed as a discrete sum of trumpet amplitudes
\begin{align}
Z(\beta)  = &\, \sum_{p_n}\, \psi_{\rm initial}(p_n) \spc Z_{\rm trumpet}(\beta,p_n),\qquad\ \ p_n\nspc=\nspc\textstyle \frac \i 2\sfb n \, \\[2mm]
Z_{\rm disk}(\beta)\; & 
 =\, \sum_{p_{{}_{1\pm 1}}}\, 
\psi_{\rm disk} Z_{\rm trumpet}(\beta,p_{{}_{1\pm 1}}),\qquad p_{{}_{1\pm 1}}\! =\nspc \textstyle \frac 1 2(\sfb^{-1}\!\nspc \pm\nspc\i\smpc \sfb) 
\end{align}
The disk wave function $\psi_{\rm disk}\! = \! \sfb^{-1}$ is just a trivial constant. Comparing with equation \eqref{ishibashi-identity-state}, we see that the initial wavefunction $\psi_\text{initial}(p_n)$ comes entirely from the crosscap state wavefunction \eqref{crosscap liouville}
\begin{equation} \label{psi initial from PT}
    \psi_\text{initial}\bigl(p_n\bigr) = P_{\mathbb{1}p_n} T_{p_n}^{-1/2}
    %= \cos(\pi\sfb p) \cosh(\pi p/\sfb) e^{\pi p^2} %\quad\quad 
    = \cos(\pi n/2)\cosh(\pi \sfb^2 n/2)e^{-\pi \sfb^2 n^2/4}\,.
\end{equation}
This expression matches with the wavefunction identified in \cite{blommaert_wormholes_2025} as the correct special boundary state for producing the exact match between the sine dilaton and DSSYK partition function. Via the above analysis, we have given this state a natural geometric interpretation. The close geometric and formal analogy between the disk and disk with crosscap partition function suggests that the latter may also admit a natural representation in the double scaled two matrix model. We leave this problem for future study.

\section{Symmetric crosscap amplitude}\label{app:B}

\par Our initial SYK crosscap state $\ket{\psi_\TT{initial}}$  is not equal to the standard crosscap state of CLS given by the overlap between the initial and final states
\bea 
\ket{\psi_\text{CC}} \is \ket{\text{C}_+}\otimes \ket{\text{C}_-} \\[3mm]
\ket{\psi_\text{final}(s)} \is \ket{\text{FZZT}(s)}\otimes \ket{\text{FZZT(0)}} \, .
\eea
Compared with \eqref{psi initial cft}, we replaced $e^{\i \pi L_{0-}} \ket{\TT{ZZ}}$ with the crosscap state $\ket{\text{C}_-}$. The latter solves both the crosscap reflecting boundary conditions \eqref{ishicond} and the crosscap Cardy conditions. It is instructive to compare the calculation of section \ref{sec: liouville crosscap amplitude} with the above symmetric crosscap amplitude of CLS\footnote{As in section \ref{sec: liouville crosscap amplitude}, we drop numerical prefactors to make the calculation more transparent.}.

The partition function we wish to compute reads:
\bea
\begin{split} \label{symmetric crosscap amplitude}
    Z_\TT{CC}(s) & =\int_{0}^{\i \infty}\!\!\mathrm{d} \tau \,Z_\text{ghost}\bra{\text{FZZT(0)}}e^{2 \pi \i \tau L_0^-} \ket{\text{C}_-} \bra{\text{FZZT}(s)}e^{2\pi \i \tau L_0^+}\ket{\text{C}_+} \\ 
    &=\int \d p\,\frac{\cosh(4\pi p s)}{\sin(\pi\sfb p)\sinh(\pi p/\sfb)}\int \d q\,\frac{1}{\sin(\pi\sfb q)\sinh(\pi q/\sfb)}\int_{0}^{\i \infty}\d \tau\,e^{-2\pi \tau (p^2-q^2)} \,.
\end{split}
\eea
To evaluate it, we proceed as in section \ref{sec: liouville crosscap amplitude}. We first pick up the poles at $p=\i\sfb n_1$ and $q=\i\sfb n_2$, analogously to the integral in \eqref{5.27}, for\footnote{Here the poles at $n_1 ,n_2 =0$ are excluded, as they are double poles.} $n_1, n_2 \neq 0 $. The moduli space integral over $\tau$ then imposes $n_1 = n_2$. Introducing $\theta = 2\pi \sfb s$, we get
\begin{equation}
    Z_\TT{CC}(\theta) = \sum_{n_1, n_2 = 1}^{+\infty} (-1)^{n_1 + n_2} \frac{\cos(2 n_1 \theta)}{\sinh(\pi \sfb^2 n_1)\sinh(\pi \sfb^2 n_2)} \int_{0}^{\i \infty} d\tau \, e^{2\pi \tau \sfb^2 (n_1^2 - n_2^2)}  = \infty \sum_{n  = 1}^{+\infty} \frac{1}{n} \frac{\cos(2 n \theta)}{\sinh(\pi \sfb^2 n)^2} \, .
\end{equation}
Note that the infinite prefactor arises from \eqref{chinner}, as for $n_1 = n_2$ (and hence $p = q$) we are left just with a divergent factor of $1/\epsilon = \infty$; for $n_1 \neq n_2$ we just get finite contributions, which can be neglected in comparison. Finally, to get the crosscap ``spectral amplitude'', we follow \eqref{5.3zpec}, adding a marked point on the boundary and taking the discontinuity $s \pm \i \sfb/2$. This effectively replaces $\cos(2 n \theta) \to n \sinh(2 \pi \sfb^2 n) \cos(2 n \theta)$ in the above equation, leading to the symmetric crosscap spectral amplitude:
\begin{equation} \label{true crosscap spectral amplitude}
    \ca{Z}_\TT{spec}^\TT{CC}(\theta) =  \sum_{n  = 1}^{+\infty}  \frac{\cos(2 n \theta)}{\tanh(\pi \sfb^2 n)} \, . 
\end{equation}

\par Let us now compare the crosscap amplitude \eqref{true crosscap spectral amplitude} with the SYK crosscap spectral amplitude \eqref{5.28}. We see that the symmetric crosscap boundary state results in very different physics than the SYK crosscap state. The former computes a geometric crosscap amplitude.  Indeed, rewriting \eqref{true crosscap spectral amplitude} as
\begin{equation}
    \ca{Z}_\TT{spec}^\TT{CC}(\theta) = \sum_{b_+ = 2 \i \hbar n} \psi_\TT{CC}(b_+) \cos(\frac{b_+ \theta}{\i \hbar})   \,, \quad\quad \psi_\TT{CC}(b_+) = \frac{1}{\tan(b_+/4)} \, ,
\end{equation}
we recognize that the wavefunction $\psi_\TT{CC}(b_+)$ associated to the state $\ket{\TT{C}_+}\otimes \ket{\TT{C}_-}$ is a discretized version of the JT gravity crosscap wavefunction \cite{Stanford:2019vob,Yan:2022nod}. The SYK crosscap boundary condition \eqref{xzcap}, on the other hand, reduces to the \emph{disk} amplitude in the JT limit, rather than the crosscap amplitude.

\bibliographystyle{ourbst}
\bibliography{biblio}

\end{document}